\documentclass[useAMS,usenatbib,fleqn]{mnras}

\usepackage[T1]{fontenc}
\usepackage{aecompl}
\usepackage{pslatex}
\usepackage{color}
\usepackage{amsmath}
\usepackage{url}
\usepackage{amssymb}
\usepackage{graphicx}
\definecolor{orange}{rgb}{1,0.3,0}
\definecolor{purple}{rgb}{1,0,1}

\newcommand*{\ditto}{\textquotedbl} 

\title[Evolution of the velocity dispersion -- temperature relation]{The \textit{XMM} Cluster
Survey: evolution of the velocity dispersion -- temperature relation over half a Hubble time}

\author[Wilson et al.]
{\parbox{\textwidth}{\raggedright Susan~Wilson,$^{1}$\thanks{E-mail: swilson072@gmail.com}
Matt~Hilton,$^{1}$\thanks{E-mail: hiltonm@ukzn.ac.za}
Philip~J.~Rooney,$^{2}$
Caroline~Caldwell,$^{3}$
Scott~T.~Kay,$^{4}$
Chris~A.~Collins,$^{3}$
Ian~G.~McCarthy,$^{3}$
A.~Kathy~Romer,$^{2}$
Alberto~Bermeo,$^{2}$
Rebecca~Bernstein,$^{5}$
Luiz~da~Costa,$^{6,7}$
Daniel~Gifford,$^{8}$ 
Devon~Hollowood,$^{9}$
Ben~Hoyle,$^{10}$
Tesla~Jeltema,$^{9}$
Andrew~R.~Liddle,$^{11}$
Marcio~A.~G~Maia,$^{6,7}$
Robert~G.~Mann,$^{11}$
Julian~A.~Mayers,$^{3}$
Nicola~Mehrtens,$^{12,13}$
Christopher~J.~Miller,$^{8}$
Robert~C.~Nichol,$^{14}$
Ricardo~Ogando,$^{6,7}$ 
Martin~Sahl\'en,$^{15}$
Benjamin ~Stahl,$^{9}$
John~P.~Stott,$^{16}$
Peter~A.~Thomas,$^{2}$
Pedro~T.~P.~Viana,$^{17,18}$
and
Harry~Wilcox$^{14}$
}\vspace{0.4cm}\\
\parbox{\textwidth}{\raggedright 
$^{1}$~Astrophysics \& Cosmology Research Unit, School of Mathematics, Statistics \& Computer Science, 
University of KwaZulu-Natal, Durban 4041, SA\\
$^{2}$~Astronomy Centre, University of Sussex, Falmer, Brighton, BN1 9QH, UK\\
$^{3}$~Astrophysics Research Institute, Liverpool John Moores University, IC2, Liverpool Science Park, 
146 Brownlow Hill, Liverpool, L3 5RF, UK\\
$^{4}$~Jodrell Bank Centre for Astrophysics, School of Physics and Astronomy, The University of Manchester, 
Manchester, M13 9PL, UK\\
$^{5}$~Observatories of the Carnegie Institution for Science, 813 Santa Barbara Street, Pasadena, CA 91101, USA\\
$^{6}$~Observat\'orio Nacional, Rua Gal. Jos\'e Cristino 77, Rio de Janeiro, RJ - 22460-040, Brazil\\
$^{7}$~Laborat\'orio Interinstitucional de e-Astronomia - LIneA, Rua Gal. Jos\'e Cristino 77, Rio de Janeiro, 
RJ - 20921-400, Brazil\\
$^{8}$~Astronomy Department, University of Michigan, Ann Arbor, MI 48109, USA \\
$^{9}$~Department of Physics and Santa Cruz Institute for Particle Physics, University of California, Santa Cruz, CA 95064, USA\\
$^{10}$~Universitaets-Sternwarte, Fakultaet fuer Physik, Ludwig-Maximilians Universitaet Muenchen, 
Scheinerstr. 1, D-81679 Muenchen, Germany\\
$^{11}$~Institute for Astronomy, University of Edinburgh, Royal Observatory, Blackford Hill, Edinburgh, EH9 
3HJ, UK\\
$^{12}$~George P. and Cynthia Woods Mitchell Institute for Fundamental Physics and Astronomy, Texas A \& M 
University, College Station,
TX, 77843-4242 USA\\
$^{13}$~Department of Physics and Astronomy, Texas A \& M University, College Station, TX, 77843-4242 USA\\
$^{14}$~Institute of Cosmology and Gravitation, University of Portsmouth, Dennis Sciama Building, Portsmouth, 
PO1 3FX, UK \\
$^{15}$~BIPAC, Department of Physics, University of Oxford, Denys Wilkinson Building, 1 Keble Road, Oxford OX1 3RH, UK\\
$^{16}$~Sub-department of Astrophysics, Department of Physics, University of Oxford, Denys Wilkinson Building, Keble Road,
Oxford OX1 3RH, UK\\
$^{17}$~Instituto de Astrof\'{\i}sica e Ci\^{e}ncias do Espa\c{c}o, Universidade do Porto, CAUP, Rua das 
Estrelas, 4150-762 Porto, Portugal \\
$^{18}$~Departamento de F\'isica e Astronomia, Faculdade de Ci\^encias, Universidade do Porto, Rua do Campo 
Alegre, 687, 4169-007 Porto, Portugal
}
}

\begin{document}

\date{Draft version: \today}
\date{Accepted for publication in MNRAS}

\maketitle

\begin{abstract}
We measure the evolution of the velocity dispersion--temperature ($\sigma_{\rm v}$--$T_{\rm X}$)
relation up to $z = 1$ using a sample of 38 galaxy clusters drawn from the \textit{XMM} Cluster Survey.
This work improves upon previous studies by the
use of a homogeneous cluster sample and in terms of the number of high redshift clusters included. We present
here new redshift and velocity dispersion measurements for 12 $z > 0.5$ clusters observed with the GMOS
instruments on the Gemini telescopes. Using an orthogonal regression method, we find that the slope of the 
relation is steeper than that expected
if clusters were self-similar, and that the evolution of the normalisation is slightly negative, but not 
significantly different from zero ($\sigma_{\rm v} \propto T^{0.86 \pm 0.14} E(z)^{-0.37 \pm 0.33}$).
We verify our results by applying our methods to cosmological hydrodynamical simulations. 
The lack of evolution seen in our data is consistent with simulations that include both feedback and radiative cooling.
\end{abstract}


\section{Introduction}
Clusters of galaxies are the largest coherent gravitationally bound objects in our Universe. By studying 
galaxy clusters, information can be gained about the formation of galaxies, and the effect of ongoing 
processes such as merging and AGN feedback. They can also be used as a probe of cosmology by studying 
the evolution of their number density with mass and redshift \citep[e.g.,][]{Vikhlinin2009, Hasselfield2013, 
Reichardt2013, Planck2015_XXIV}. However, the mass of galaxy clusters is not a 
quantity that can be directly measured, and therefore it needs to be determined using observable mass tracers 
such as X-ray properties (e.g., luminosity and temperature), the Sunyaev-Zel'dovich (SZ) effect signal, 
and optical properties, such as richness, line-of-sight velocity dispersion of member galaxies,
and shear due to gravitational lensing \citep[e.g.,][]{Ortiz-Gil2004, Vikhlinin2006, Rozo2009, Sifon2013, Nastasi2014, vonDerLinden2014, Hoekstra2015}. 

Scaling relations are power laws between galaxy cluster properties that have the potential to allow us to measure the mass of clusters using easily observable properties such as the X-ray Luminosity, X-ray Temperature and the velocity dispersion. These power laws can be predicted if we assume clusters are formed in the manner described by \citet{Kaiser1986}. In this model, known as the self-similar model, all galaxy clusters and groups are essentially identical objects which have been scaled up or down \citep{Maughan2012}. Strong self-similarity refers to when galaxy clusters have been scaled by mass and weak self-similarity refers to a scaling due to the changing density of the Universe with redshift \citep{Bower1997}. This model makes some key assumptions, as described by \cite{2012ARA&A..50..353K} and \cite{Maughan2012}. The first assumption is that we are in an Einstein-de-Sitter Universe, $\Omega_m =1$, so the clusters form via a single gravitational collapse at the observed redshift. Secondly, gravitational energy as a result of the collapse is the only source of energy to the intracluster medium (ICM). By introducing these assumptions we greatly simplify the problem so that properties of the density field depend on only two control parameters, the slope of the power spectrum of the initial perturbations and its normalisation. The strong self-similarity determines the slope and is not expected to evolve with redshift while the weak self-similarity is responsible for the evolution of the normalisation since in this simplified model it is due only to a change in density with redshift \citep{Bryan1998}.

The most commonly studied scaling relation is the luminosity -- temperature relation ($L_{\rm X} - T$), however there is still no consensus on how it evolves with redshift and if self-similarity holds. Some studies have found that the evolution of the normalisation of this relation is consistent with self-similarity \citep[e.g.,][]{Vikhlinin2002, Lumb2004, Maughan2006},
while other studies have found zero or negative evolution \citep[e.g.,][]{Ettori2004, Branchesi2007, Hilton2012, Clerc2012, Clerc2014}. \citet{Maughan2012} also found that the evolution of the $L_{\rm X} - T$
relation was not self-similar, but concluded that this could plausibly be explained by selection effects. 
 
In this paper we focus on the lesser studied relationship between the velocity dispersion of member galaxies
($\sigma_{\rm v}$) and the X-ray temperature ($T_{\rm X}$) of the ICM.
Since the velocity dispersion is a measure of the kinetic energy of the galaxies in the cluster, and 
temperature is related to the kinetic energy of the gas, both the gas and galaxies are tracers of the 
gravitational potential. One would expect a self-similar relationship of the form 
$\sigma_{\rm v} \propto T^{0.5}$, if clusters were formed purely due to the action of gravity 
\citep{Quintana1982, Kaiser1986, Voit2005}. However, almost all previous studies of the relation have found a steeper
power-law slope than this (see Table~\ref{previous}). The relation is also not expected to evolve with
redshift. To date this has been tested only by \cite{Wu1998} and \citet{Nastasi2014}. Even then, all but 
four clusters in the \citet{Wu1998} sample are at $z < 0.5$. \citet{Nastasi2014} made a measurement of the 
relation at $0.6 < z < 1.5$ using a sample of 12 clusters, obtaining results consistent with previous studies
at low redshift.

One may expect evolution in cluster scaling relations due to the increase of star formation and AGN activity 
at high redshift \citep[e.g.,][]{Silverman2005, Magnelli2009}, or due to the increase in frequency of 
galaxy cluster mergers with increasing redshift \citep[e.g.,][]{Cohn2005, Kay2007, Mann2012}. Galaxy cluster 
mergers are among the most energetic events in the Universe, and simulations have shown that these could 
result in the boosting of cluster X-ray temperatures\citep[e.g.,][]{Ritchie2002, Randall2002, Poole2007}. Figure 9 in \cite{Ritchie2002} shows how the temperature is boosted when two equal mass systems have a head on collision with varying initial distances between their centres.  
All of these processes add energy into the ICM, and so we might expect to see an overall increase in the 
average temperatures of galaxy clusters above that expected from the self-similar case at a given redshift.

\begin{table*}
\caption{Previous measurements of the velocity dispersion--temperature relation. Here the relation is in the 
form $\sigma_{\rm v} = 10^A T^B$, where $\sigma_{\rm v}$ is measured in km\,s$^{-1}$ and  $T$ 
is measured in keV.}
\begin{tabular*}{\textwidth}{l @{\extracolsep{\fill}} cccllc}
\hline
Paper     & Number of clusters   &$A$      & $B$ & Redshift range & Fitting method \\
 \hline
\citet{Edge1991}	&	\phantom{0}23	&	2.60 $\pm$ 0.08	&	0.46  $\pm$ 0.12	&	$z<0.1$ & Least squares\\
\citet{Lubin1993}	&	\phantom{0}41	&	2.52 $\pm$ 0.07	&	0.60  $\pm$ 0.11	&	$z<0.2$ & $\chi^2$\\
\citet{Bird1995}	&	\phantom{0}22	&	2.50 $\pm$ 0.09	&	0.61  $\pm$ 0.13	&	$z<0.1$ & Bisector\\
\citet{Girardi1996}	&	\phantom{0}37	&	2.53 $\pm$ 0.04	&	0.61  $\pm$ 0.05	&	$z<0.2$ & Bisector\\
\citet{Ponman1996}	&	\phantom{0}27	&	2.54 $\pm$ 0.04	&	0.55  $\pm$ 0.05	&	$z<0.15$ & Bisector\\
\citet{White1997}	&	\phantom{0}35	&	2.53 $\pm$ 0.08	&	0.60  $\pm$ 0.10	&	$z<0.2$ & Orthogonal\\
\citet{Wu1998}	&	\phantom{0}94	&	2.47 $\pm$ 0.06	&	0.67  $\pm$ 0.09	&	$z<0.9$ & Orthogonal\\
\citet{Wu1998}	&	110	&	2.57 $\pm$ 0.03	&	0.49  $\pm$ 0.05	&	$z<0.1$ & Orthogonal\\
\citet{Wu1998}	&	\phantom{0}39	&	2.57 $\pm$ 0.08	&	0.56  $\pm$ 0.09	&	$0.1<z<0.9$ & Orthogonal\\
\citet{Wu1999}	&	\phantom{0}92	&	2.49 $\pm$ 0.03	&	0.64  $\pm$ 0.02	&	$z<0.45$ & Orthogonal\\
\citet{Xue2000}	&	109	&	2.53 $\pm$ 0.03	&	0.58  $\pm$ 0.05	&	$z<0.2$ & Orthogonal\\
\citet{Nastasi2014}	&	\phantom{0}12	& 2.47 $\pm$ 0.19  &	0.64  $\pm$ 0.34	&	$0.64<z<1.46$ & Bisector\\
\hline
\end{tabular*}
\label{previous}  
\end{table*} 

In this paper, we study a sample of 38 $z < 1.0$ galaxy clusters drawn from the \textit{XMM} Cluster Survey 
\citep[XCS;][]{Mehrtens2012}. We divide the sample into two groups: a low redshift sample ($0.0 < z < 0.5$), and a high redshift sample ($0.5 < z < 1.0$), such that each group has an equal number of clusters in each, and then proceed to test for evolution in the $\sigma_{\rm v}-T$ relation. We describe the sample and processing
of the optical and X-ray data in Section~\ref{Sample}. Section~\ref{Method} discusses the method used to 
determine cluster membership and for measuring the velocity dispersion and describes the methods used 
for fitting the $\sigma_{\rm v}$--$T$ relation, and we present our results
in Section~\ref{Results}. We discuss our findings in Section~\ref{Discussion} and present our conclusions in 
Section~\ref{Conclusions}.

We assume a cosmology with $\Omega_{m} = 0.27$, $\Omega_{\Lambda} = 0.73$, and 
$H_0=\,70$\,km\,s$^{-1}$\,Mpc$^{-1}$ throughout.

\section{Sample and Data Reduction}
\label{Sample}

The cluster sample for this work is drawn from XCS, a serendipitous X-ray cluster survey being conducted
using archival \textit{XMM-Newton} data. Data Release 1 (DR1) of the XCS is described in \citet{Mehrtens2012}.
The overall aims of the XCS project are to measure cosmological parameters through the evolution of the cluster mass
function with redshift \citep{Sahlen2009}, study the evolution of galaxies in clusters 
\citep{Collins2009, Hilton2009, Hilton2010, Stott2010} and investigate the X-ray scaling relations as a way 
to study the evolution of the cluster gas with redshift \citep{Hilton2012}. 

The XCS Automated Pipeline Algorithm (XAPA) described in \citet{LloydDavies2011} was used to search the XMM archive for cluster candidates. 
\citet{Mehrtens2012} describes confirmation of a subset of these candidates as clusters using the combination of data 
from the literature and optical follow-up observations. This left a final sample of 503 
X-ray confirmed galaxy clusters, 255 which were previously unknown and 356 of which were new X-ray detections. 
Of these, 464 have redshift estimates, and 402 have temperature measurements. 

For XCS-DR1 the cluster-averaged X-ray temperatures (${\it T}_{\rm X}$) were measured using an automated pipeline described in detail in \citet{LloydDavies2011}. In summary this pipeline operates as follows: spectra were generated in the 0.3-7.9 keV band using photons in the XAPA source ellipse, (where the XAPA ellipse corresponds 0.08 -- 0.56 of R500, with a median value of to 0.36 R$_{500}$, where R$_{500}$ is calculated using Equation 2 and Table 2 from \cite{Arnaud2005}); an in-field background subtraction method was used; and model fitting was done inside XSPEC \citep{Schafer1991} using an absorbed MEKAL \citep{Mewe1986} model and Cash statistics \citep{Cash1979}. In the fit,  the hydrogen column density was fixed to the \citet{Dickey1990} value and the metal abundance to 0.3 times the Solar value. For this paper we have updated the ${\it T}_{\rm X}$ values compared to \citet{Mehrtens2012}. The pipeline is very similar to that described in \citet{LloydDavies2011}, but using updated versions of the XMM calibration and XSPEC.
The median X-ray count for all the clusters in our final sample was 1919 with a minimum count of 220. We note that for only 1 of the clusters in our sample are the X-ray counts used for the spectral analysis less than 300. This is the minimum threshold defined by \citet{LloydDavies2011} for reliable, i.e. with a fractional error of $<$0.4, Tx measurements at T$_x>$5keV (See Figure 16 from \citet{LloydDavies2011}). The remaining cluster was fit using 220 counts, but has a temperature of 3.5keV (so still has an expected fractional error of ~0.4)

\begin{figure*}
\includegraphics[trim={0 0.6cm 0 0},clip,width=\columnwidth]{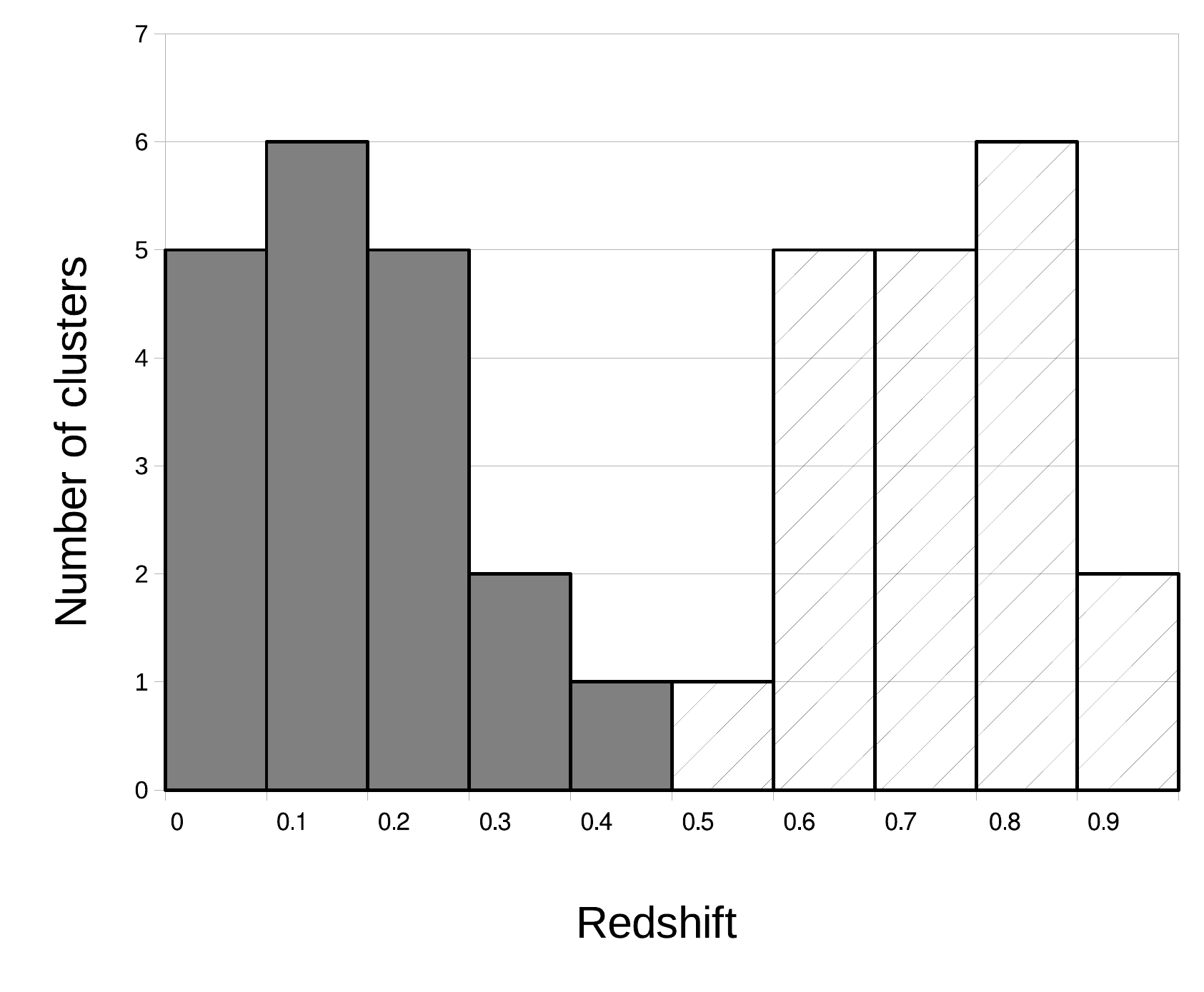}
\includegraphics[trim={0 0.6cm 0 0},clip,width=\columnwidth]{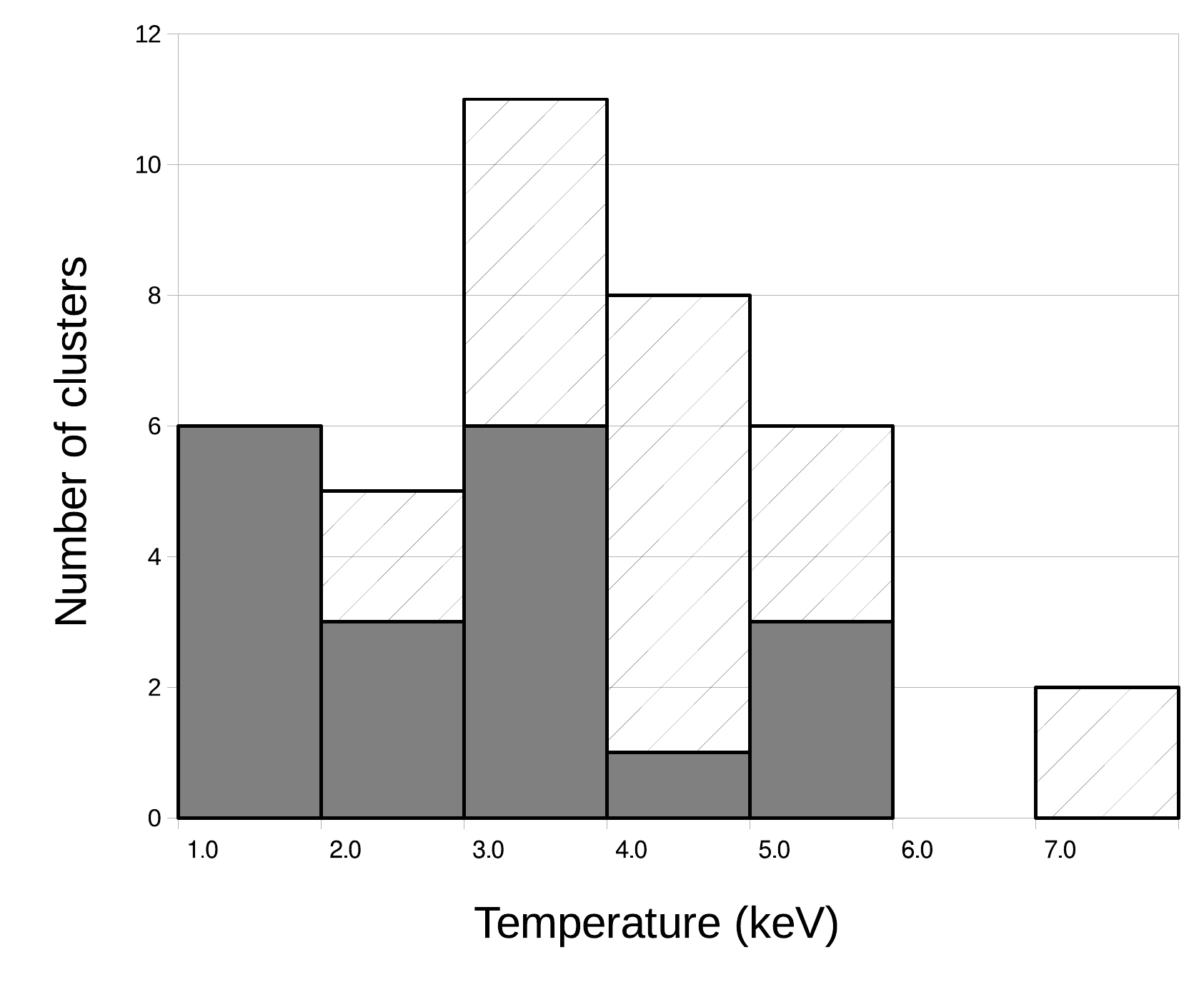}
\caption{The redshift and temperature distributions of the low and high redshift cluster samples used in this
work. The solid grey marks the low redshift sample ($z<0.5$) and the shading with diagonal lines marks the 
high redshift sample ($z>0.5$). Note that the high redshift sample (T$_{median}=$ 4.5 keV) contains more high temperature clusters than the low redshift sample (T$_{median}=$ 3.0 keV).}
\label{hist}
\end{figure*}

For this paper both the samples were constructed from XCS DR1, except for one of the 
high redshift sample clusters (XMMXCS J113602.9-032943.2) which is a previously unreported XCS cluster 
detection. Fig.~\ref{hist} shows the redshift and temperature distributions of the two samples.
The high redshift sample contains more high temperature clusters 
than the low redshift sample, which may be due to selection effects which result in higher luminosity and hence higher temperature clusters being chosen at higher redshift.

%
%

\subsection{Low redshift sample}

The low redshift sample contains 19 clusters whose properties can be found in Table~\ref{DR1}. In order to 
obtain this sample we excluded all clusters from the DR1 sample that did not have temperatures or which had a 
redshift $z>0.5$, leaving us with a sample of 320 clusters. We performed a search in NED\footnote{This 
research has made use of the NASA/IPAC Extragalactic Database (NED) which is operated by the Jet Propulsion 
Laboratory, California Institute of Technology, under contract with the National Aeronautics and Space 
Administration.} for galaxies surrounding each cluster. 
We included only clusters which had spectroscopic redshifts resulting in our sample size being decreased from 320 to 296. Since NED collects data from many different sources the reliability of the redshifts can not be guaranteed. Hence, where possible we use only one source of redshifts per cluster to ensure homogeneity. These redshifts are specified to the 4th decimal place but unfortunately for most an uncertainty is not included in the original sample and therefore we assumed an accuracy of 1\%.

We excluded galaxies located at a projected radial distance $>R_{200}$ (the radius within which the mean density is 200 
times the critical density of the Universe at the cluster redshift) as such galaxies are unlikely to be 
cluster members. To ensure we did not exclude possible members, for this initial step $R_{200}$ was 
calculated using a fiducial velocity dispersion of 2000\,km\,s$^{-1}$ following \citet{Finn2005},

\begin{equation}
R_{200}\,({\rm Mpc}) = 
2.47\frac{\sigma_{\rm v}}{1000\,\rm{km\,s^{-1}}}\frac{1}{\sqrt{\Omega_{\Lambda}+ \Omega_0(1+z)^3}}. 
\label{R200}
\end{equation}
Here $\sigma_{\rm v}$ is the line of sight velocity dispersion (see Section \ref{veldisp}) and $z$ is 
the redshift of the cluster. Equation~\ref{R200} assumes that the galaxy velocity distribution follows an 
isothermal sphere dark matter profile. The fiducial $R_{200}$ values span the range 2--4\,Mpc. Section~\ref{membership} below describes how this initial cluster membership selection was refined to give the final cluster members. 
We then excluded all clusters which had less than 10 galaxies as this would provide us with two few members for accurate velocity dispersion calculation leaving us with a sample size of 19 clusters.
\subsection{High redshift sample}

The high redshift sample is made up of 19 clusters whose properties can be found in Table~\ref{Gemini}.
Member redshifts were determined from observations using the Gemini telescopes for 12 of these clusters 
(see Section \ref{obs}). The other seven clusters used data obtained from \citet{Nastasi2014}. They drew 
both on new observations and on existing data. For example, the observations of three of the \citet{Nastasi2014}
clusters we have used in this paper (XMMXCS J105659.5-033728.0, XMMXCS J113602.9-032943.2 and XMMXCS J182132.9+682755.0 
in Table \ref{Gemini}) were presented in \citet{Tran1999} respectively. The observations of the other 4 
clusters we have used in this paper were presented for the first time in \citet{Nastasi2014}.
These four were discovered independently (to XCS) by the XMM Newton Distant Cluster Project 
\citep[XDCP;][]{Fassbender2011}. \citet{Nastasi2014} also presented galaxy redshift data for another six XDCP clusters, 
however we have not used those in this paper because there are insufficient galaxies to derive an accurate velocity 
dispersion\footnote{The methodology described in Section \ref{Method} was applied to these six clusters before they were excluded from our study.}. 
For the seven clusters that relied on \citet{Nastasi2014} data, new temperatures were 
obtained using XCS pipelines and the velocity dispersion was recalculated using the \citet{Nastasi2014} cluster redshift together with the method 
described in Section \ref{Method}. 

\subsubsection{Observations} 
\label{obs}
Observations of 12 $z > 0.5$ clusters were obtained using the Gemini Multi Object Spectographs (GMOS) on both
the Gemini telescopes from 2010 to 2012. The nod-and-shuffle mode \citep{Glazebrook2001} was used to allow 
better sky subtraction and shorter slit lengths when compared to conventional techniques. For all 
observations the R400 grating and OG515 order blocking filter were used, giving wavelength coverage of 5400 
-- 9700 {\AA}. The GMOS field of view samples out to R200 at the the redshifts of our sample \cite{Sifon2013}. A total of 30 masks were observed with a varying number of target slitlets. Each slitlet had length 
3$^{\prime \prime}$ and width 1$^{\prime \prime}$. 
Target galaxies were selected to be fainter than the 
brightest cluster galaxy (which was also targeted in the slit masks), on the basis of $i$-band pre-imaging 
obtained from Gemini. We also used colour or photo-$z$ information, where available,
to maximise our efficiency in targeting cluster members. For five clusters which had $r$, $z$-band photometry 
from the National Optical Astronomy Observatory--XMM Cluster Survey 
\citep[NXS; described in][]{Mehrtens2012}, we preferentially selected galaxies with $r-z$ colours expected for
passively evolving galaxies at the cluster redshift \citep[see][for details]{Mehrtens2012}. For four clusters,
we used photometric redshifts for galaxies from SDSS DR7 \citep{Abazajian2009}. For XMMXCS J113602.9-032943.2, 
we used galaxy photo-$z$s that were measured from our own $riz$ photometry obtained at the William Herschel
Telescope (WHT) on 2011 May 5.
Observations at three different central wavelengths (7500, 7550 and 7600 {\AA}) were used 
to obtain coverage over the gaps between the GMOS CCDs. For all observations an 85 percentile image quality 
and 50 percentile sky transparency were requested. The details of the individual observations are given in 
Table~\ref{obslog}.

\subsubsection{Spectroscopic data reduction}

The data were reduced in a similar manner to \cite{Hilton2010}, using \textsc{Pyraf} and the 
Gemini IRAF\footnote[2]{IRAF is distributed by the National Optical Astronomy Observatories,which are operated 
by the Association of Universities for Research in Astronomy, Inc., under cooperative agreement with the 
National Science Foundation.} package. We used the tools from this package 
to subtract bias frames; make flat fields; apply flat field corrections and create mosaic images. We then 
applied nod-and-shuffle sky subtraction using the \texttt{gnsskysub} task. Wavelength calibration was 
determined from arc frames taken between the science frames, using standard IRAF tasks. All data were then 
combined using a median, rejecting bad pixels using a mask constructed from the nod-and-shuffle dark frames. 
Finally, we combined the pairs of spectra corresponding to each nod position, and extracted one-dimensional 
spectra using a simple boxcar algorithm.

\subsubsection{Galaxy redshift measurements}
\label{zmeasurements}

\begin{figure*}
\includegraphics[width=\textwidth]{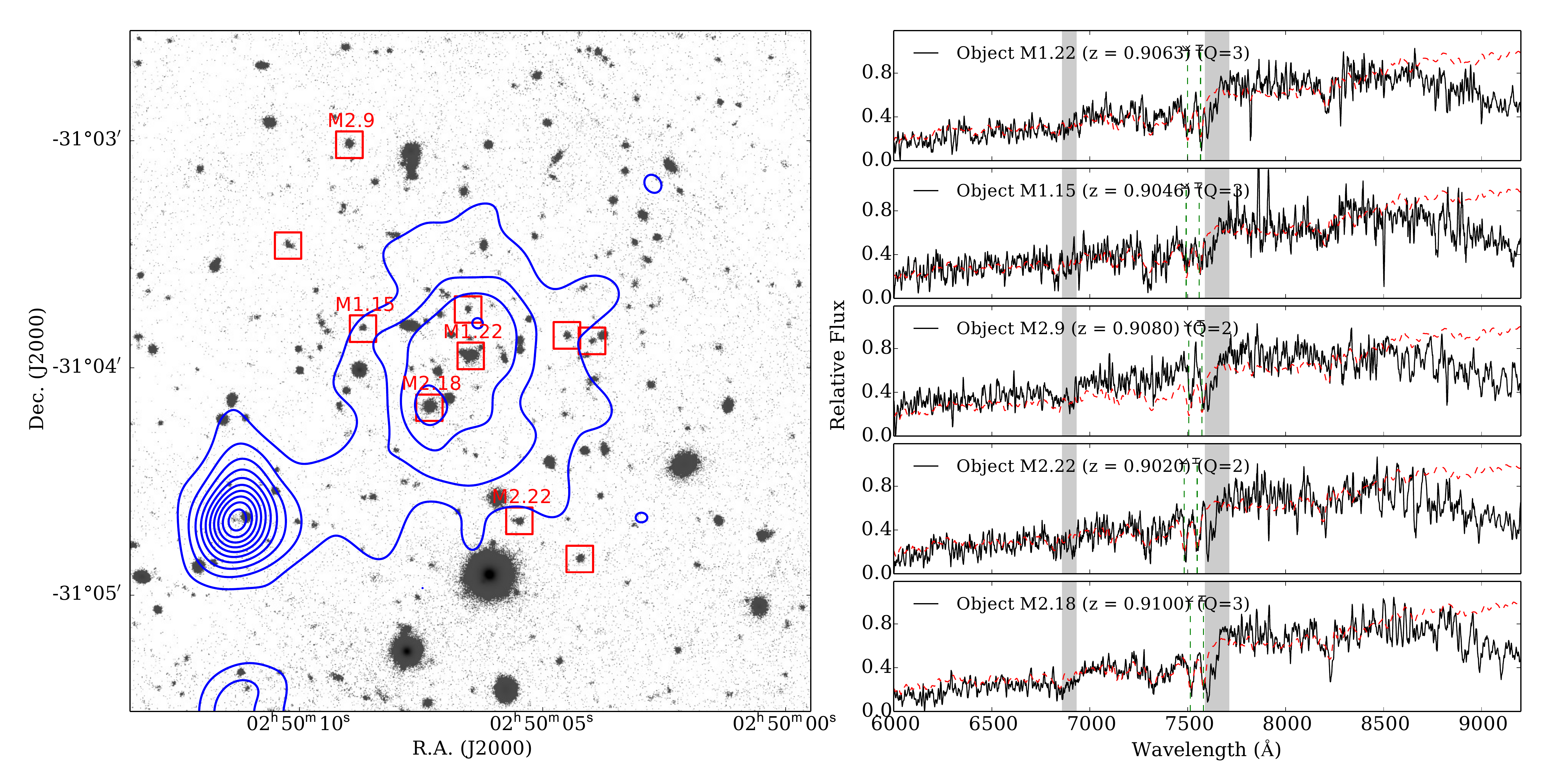}
\caption{
The $z$=0.91 cluster XMMXCS J$025006.4-310400.8$. The left hand panel shows the Gemini $i$-band image overlayed with the X--ray contours in blue. The red squares represent possible galaxy cluster members. Each possible member is labeled Mx.y, where x is the mask number and y is the object ID. The right hand panel shows 
the Gemini spectra (black lines) for a subset of these galaxies. The grey bands indicate regions affected by telluric absorption lines. The red line is the best fit SDSS template. The green dotted vertical lines show the positions of the H and K lines at the galaxy redshift.
}  
\label{spectra}
\end{figure*}

We measured galaxy redshifts from the spectra by cross-correlation with SDSS spectral 
templates\footnote[3]{http://www.sdss.org/dr7/algorithms/spectemplates/index.
html} using the RVSAO/XCSAO package for IRAF \citep{Kurtz1998}. XCSAO implements the method described by 
\cite{Tonry1979}. The spectra were compared to six different templates over varying redshifts with the final 
redshift measurement being determined after visual inspection. Redshifts were assigned a quality flag 
according to the following scheme: $Q=3$ corresponds to two or 
more strongly detected features; $Q=2$ refers to one strongly detected or two weakly detected 
features; $Q=1$ one weakly detected feature and $Q=0$ when no features could be 
identified.
The features used were spectral lines, with the most commonly identified being \textsc{[Oii]} 3727\,\AA{}, H, K, H$\beta$, and the \textsc{[Oiii]} 4959, 5007\,\AA{} lines. 
Only galaxies with a quality rating of $Q\geq2$ were used in this study because these 
have reasonably secure redshifts. Fig.\ref{spectra} shows spectra of some member galaxies of the 
cluster XMMXCS J025006.4-310400.8 as an example. Tables of redshifts for galaxies in each cluster 
field as well as histograms depicting the included/excluded members and the best-fit Gaussian $^{16}$ can be found in Appendix~\ref{RedshiftCatalogue}.
 
Table \ref{apptable} is shown, as an example, below for cluster XMMXCS J025006.4-310400.8.

\begin{table*}
\centering

 \begin{tabular*}{\textwidth}{c @{\extracolsep{\fill}} ccccccc}
  \hline
   ID & Mask     & RA(J2000)         & Dec(J2000)  & z & Quality & Member\\
 \hline
 
1	&	GS-2010B-Q-46-06	&	02$^\mathrm{h}$50$^\mathrm{m}$22.92$^\mathrm{s}$	&	-31$^\circ$03$^\prime$53.0$^{\prime \prime}$	&	0.8337	&	3	&		\\
8	&	GS-2010B-Q-46-06	&	02$^\mathrm{h}$50$^\mathrm{m}$15.25$^\mathrm{s}$	&	-31$^\circ$03$^\prime$33.5$^{\prime \prime}$	&	0.7263	&	3	&		\\
9	&	GS-2010B-Q-46-06	&	02$^\mathrm{h}$50$^\mathrm{m}$13.27$^\mathrm{s}$	&	-31$^\circ$03$^\prime$35.7$^{\prime \prime}$	&	0.6168	&	3	&		\\
10	&	GS-2010B-Q-46-06	&	02$^\mathrm{h}$50$^\mathrm{m}$12.05$^\mathrm{s}$	&	-31$^\circ$03$^\prime$08.0$^{\prime \prime}$	&	0.7146	&	3	&		\\
14	&	GS-2010B-Q-46-06	&	02$^\mathrm{h}$50$^\mathrm{m}$06.63$^\mathrm{s}$	&	-31$^\circ$03$^\prime$13.7$^{\prime \prime}$	&	0.8496	&	3	&		\\
15	&	GS-2010B-Q-46-06	&	02$^\mathrm{h}$50$^\mathrm{m}$08.70$^\mathrm{s}$	&	-31$^\circ$03$^\prime$49.7$^{\prime \prime}$	&	0.9052	&	3	&	$\surd$	\\
16	&	GS-2010B-Q-46-06	&	02$^\mathrm{h}$50$^\mathrm{m}$03.78$^\mathrm{s}$	&	-31$^\circ$03$^\prime$51.5$^{\prime \prime}$	&	0.3533	&	3	&		\\
17	&	GS-2010B-Q-46-06	&	02$^\mathrm{h}$50$^\mathrm{m}$06.89$^\mathrm{s}$	&	-31$^\circ$03$^\prime$51.5$^{\prime \prime}$	&	0.9217	&	3	&	$\surd$	\\
18	&	GS-2010B-Q-46-06	&	02$^\mathrm{h}$50$^\mathrm{m}$05.48$^\mathrm{s}$	&	-31$^\circ$03$^\prime$53.0$^{\prime \prime}$	&	0.6972	&	3	&		\\
19	&	GS-2010B-Q-46-06	&	02$^\mathrm{h}$50$^\mathrm{m}$04.50$^\mathrm{s}$	&	-31$^\circ$03$^\prime$51.5$^{\prime \prime}$	&	0.9149	&	3	&	$\surd$	\\
21	&	GS-2010B-Q-46-06	&	02$^\mathrm{h}$50$^\mathrm{m}$02.79$^\mathrm{s}$	&	-31$^\circ$04$^\prime$04.9$^{\prime \prime}$	&	0.9831	&	3	&		\\
22	&	GS-2010B-Q-46-06	&	02$^\mathrm{h}$50$^\mathrm{m}$06.48$^\mathrm{s}$	&	-31$^\circ$03$^\prime$56.9$^{\prime \prime}$	&	0.9069	&	3	&	$\surd$	\\
23	&	GS-2010B-Q-46-06	&	02$^\mathrm{h}$50$^\mathrm{m}$09.04$^\mathrm{s}$	&	-31$^\circ$04$^\prime$06.3$^{\prime \prime}$	&	0.7567	&	3	&		\\
25	&	GS-2010B-Q-46-06	&	02$^\mathrm{h}$50$^\mathrm{m}$03.83$^\mathrm{s}$	&	-31$^\circ$04$^\prime$34.0$^{\prime \prime}$	&	0.8988	&	3	&	$\surd$	\\
26	&	GS-2010B-Q-46-06	&	02$^\mathrm{h}$50$^\mathrm{m}$04.24$^\mathrm{s}$	&	-31$^\circ$04$^\prime$50.6$^{\prime \prime}$	&	0.9095	&	3	&	$\surd$	\\
28	&	GS-2010B-Q-46-06	&	02$^\mathrm{h}$50$^\mathrm{m}$04.59$^\mathrm{s}$	&	-31$^\circ$05$^\prime$41.7$^{\prime \prime}$	&	0.6197	&	3	&		\\
34	&	GS-2010B-Q-46-06	&	02$^\mathrm{h}$50$^\mathrm{m}$04.26$^\mathrm{s}$	&	-31$^\circ$07$^\prime$05.2$^{\prime \prime}$	&	0.1261	&	3	&		\\
1	&	GS-2012B-Q-011-09	&	02$^\mathrm{h}$50$^\mathrm{m}$22.92$^\mathrm{s}$	&	-31$^\circ$03$^\prime$53.0$^{\prime \prime}$	&	0.5924	&	2	&		\\
2	&	GS-2012B-Q-011-09	&	02$^\mathrm{h}$50$^\mathrm{m}$18.11$^\mathrm{s}$	&	-31$^\circ$03$^\prime$10.5$^{\prime \prime}$	&	0.9326	&	2	&		\\
4	&	GS-2012B-Q-011-09	&	02$^\mathrm{h}$50$^\mathrm{m}$16.05$^\mathrm{s}$	&	-31$^\circ$03$^\prime$23.1$^{\prime \prime}$	&	1.0077	&	2	&		\\
5	&	GS-2012B-Q-011-09	&	02$^\mathrm{h}$50$^\mathrm{m}$14.89$^\mathrm{s}$	&	-31$^\circ$03$^\prime$32.1$^{\prime \prime}$	&	0.7245	&	3	&		\\
6	&	GS-2012B-Q-011-09	&	02$^\mathrm{h}$50$^\mathrm{m}$14.94$^\mathrm{s}$	&	-31$^\circ$02$^\prime$56.8$^{\prime \prime}$	&	0.9927	&	2	&		\\
9	&	GS-2012B-Q-011-09	&	02$^\mathrm{h}$50$^\mathrm{m}$08.98$^\mathrm{s}$	&	-31$^\circ$03$^\prime$01.1$^{\prime \prime}$	&	0.9086	&	3	&	$\surd$	\\
10	&	GS-2012B-Q-011-09	&	02$^\mathrm{h}$50$^\mathrm{m}$10.24$^\mathrm{s}$	&	-31$^\circ$03$^\prime$27.8$^{\prime \prime}$	&	0.9056	&	3	&	$\surd$	\\
11	&	GS-2012B-Q-011-09	&	02$^\mathrm{h}$50$^\mathrm{m}$07.01$^\mathrm{s}$	&	-31$^\circ$01$^\prime$00.9$^{\prime \prime}$	&	0.5204	&	3	&		\\
13	&	GS-2012B-Q-011-09	&	02$^\mathrm{h}$50$^\mathrm{m}$06.54$^\mathrm{s}$	&	-31$^\circ$03$^\prime$44.7$^{\prime \prime}$	&	0.9126	&	3	&	$\surd$	\\
17	&	GS-2012B-Q-011-09	&	02$^\mathrm{h}$50$^\mathrm{m}$03.99$^\mathrm{s}$	&	-31$^\circ$03$^\prime$53.0$^{\prime \prime}$	&	0.9176	&	3	&	$\surd$	\\
18	&	GS-2012B-Q-011-09	&	02$^\mathrm{h}$50$^\mathrm{m}$07.33$^\mathrm{s}$	&	-31$^\circ$04$^\prime$10.6$^{\prime \prime}$	&	0.9106	&	3	&	$\surd$	\\
19	&	GS-2012B-Q-011-09	&	02$^\mathrm{h}$50$^\mathrm{m}$02.67$^\mathrm{s}$	&	-31$^\circ$03$^\prime$26.0$^{\prime \prime}$	&	0.9797	&	3	&		\\
20	&	GS-2012B-Q-011-09	&	02$^\mathrm{h}$50$^\mathrm{m}$07.38$^\mathrm{s}$	&	-31$^\circ$05$^\prime$28.7$^{\prime \prime}$	&	0.6494	&	2	&		\\
21	&	GS-2012B-Q-011-09	&	02$^\mathrm{h}$49$^\mathrm{m}$58.10$^\mathrm{s}$	&	-31$^\circ$03$^\prime$39.6$^{\prime \prime}$	&	0.8696	&	3	&	$\surd$	\\
22	&	GS-2012B-Q-011-09	&	02$^\mathrm{h}$50$^\mathrm{m}$05.48$^\mathrm{s}$	&	-31$^\circ$04$^\prime$40.5$^{\prime \prime}$	&	0.9026	&	2	&		\\
27	&	GS-2012B-Q-011-09	&	02$^\mathrm{h}$49$^\mathrm{m}$58.68$^\mathrm{s}$	&	-31$^\circ$05$^\prime$25.8$^{\prime \prime}$	&	0.6274	&	3	&		\\
28	&	GS-2012B-Q-011-09	&	02$^\mathrm{h}$49$^\mathrm{m}$59.99$^\mathrm{s}$	&	-31$^\circ$05$^\prime$07.8$^{\prime \prime}$	&	0.8816	&	2	&		\\
29	&	GS-2012B-Q-011-09	&	02$^\mathrm{h}$49$^\mathrm{m}$58.60$^\mathrm{s}$	&	-31$^\circ$04$^\prime$59.2$^{\prime \prime}$	&	0.9216	&	3	&	$\surd$	\\
30	&	GS-2012B-Q-011-09	&	02$^\mathrm{h}$50$^\mathrm{m}$00.51$^\mathrm{s}$	&	-31$^\circ$04$^\prime$44.5$^{\prime \prime}$	&	0.5654	&	3	&		\\
31	&	GS-2012B-Q-011-09	&	02$^\mathrm{h}$50$^\mathrm{m}$04.26$^\mathrm{s}$	&	-31$^\circ$07$^\prime$05.2$^{\prime \prime}$	&	0.9827	&	3	&		\\

\hline

\end{tabular*}
\caption{We depict the galaxy redshifts for the cluster XMMXCS J025006.4-310400.8. Column 1 gives an arbitrary ID for each galaxy, column 2 and 3 give the right ascension and declination respectively, and column 4 gives the redshift of the galaxy. Column 5 gives the quality flag as explained in Section \ref{zmeasurements} and column 6 shows whether or not the galaxy was included as a member for the determination of the velocity dispersion.}
\label{apptable}
\end{table*}


\begin{table*}
\caption{ 
Low redshift sample ($0.0 < z < 0.5$): column 1 gives the name of the XCS Cluster, columns 2 and 3 
give its J2000 right ascension and declination. Column 4 gives the redshift,with the uncertainty found using bootstrapping and column 5 gives the redshift from literature $^{15}$. Column 6 gives the 
temperature with its positive and negative 1$\sigma$ uncertainty. Column 7 gives the number of confirmed members and 
columns 8 and 9 give the calculated velocity dispersion and R$_{200}$ respectively. The references for the redshifts are as follows: 1. \citet{Mehrtens2012} 2. \citet{Cappi1998} 3. \citet{Yoon2008} 4. \citet{Mulchaey2006} 5. \citet{Vikhlinin1998} 6.\citet{Takey2013} 7. \citet{Finoguenov2007} 8. \citet{Takey2011} 9. \citet{Hao2010} 10. \citet{Hennawi2008} 11. \citet{Struble1999} 12. \citet{Burenin2007} 13. \citet{Koester2007} 14. \citet{Mullis2003} and 15. \citet{Sakelliou1998}}.
\begin{tabular*}{\textwidth}{c @{\extracolsep{\fill}} lccccccccc}
\hline
Name     & & RA        & Dec      & $z$ & $z_{lit}$ &  $T$   & Members         & $\sigma_{\rm v}$  & $R_{200}$\\
         & & (J2000)   & (J2000)  &   & & (keV) &                 & (km\,s$^{-1}$)    & (Mpc)\\
 \hline
XMMXCS J000013.9-251052.1	& &	00$^\mathrm{h}$00$^\mathrm{m}$13.9$^\mathrm{s}$	&	
-25$^\circ$10$^\prime$52.1$^{\prime \prime}$	&	0.0845 $\pm$ 0.0004	& 0.08$^1$ &	1.80	$_{-	0.20	} ^{+	0.40}$	&	19	& 	410	$\	\pm\		\phantom{0}80	$	&	1.11	\\
XMMXCS J003430.1-431905.6	& &	00$^\mathrm{h}$34$^\mathrm{m}$30.1$^\mathrm{s}$	&	
-43$^\circ$19$^\prime$05.6$^{\prime \prime}$	&	0.3958 $\pm$ 0.0010	& 0.40$^1$ &	3.50	$_{-	0.20	} ^{+	0.20}$	&	22	& 	920	$\	\pm\		150	$	&	1.96	\\
XMMXCS J005603.0-373248.0	& &	00$^\mathrm{h}$56$^\mathrm{m}$03.0$^\mathrm{s}$	&	
-37$^\circ$32$^\prime$48.0$^{\prime \prime}$	&	0.1659 $\pm$ 0.0009	& 0.16$^2$ &	5.20	$_{-	0.20	} ^{+	0.30}$	
&	22	&	900	$\	\pm\		140	$	&	2.06	\\
XMMXCS J015315.0+010214.2	& &	01$^\mathrm{h}$53$^\mathrm{m}$15.0$^\mathrm{s}$	&	
+01$^\circ$02$^\prime$14.2$^{\prime \prime}$	&	0.0593 $\pm$ 0.0002 & 0.06$^3$	&	1.08	$_{-	0.02	} ^{+	0.02}$	
&	12	&	240	$\	\pm\		\phantom{0}80	$	&	0.55	\\
XMMXCS J072054.3+710900.5	& &	07$^\mathrm{h}$20$^\mathrm{m}$54.3$^\mathrm{s}$	&	
+71$^\circ$09$^\prime$00.5$^{\prime \prime}$	&	0.2309 $\pm$ 0.0005  & 0.23$^4$	&	2.90	$_{-	0.40	} ^{+	0.50}$	
&	29	&	550	$\	\pm\		\phantom{0}60	$	&	1.20	\\
XMMXCS J081918.6+705457.5	& &	08$^\mathrm{h}$19$^\mathrm{m}$18.6$^\mathrm{s}$	&	
+70$^\circ$54$^\prime$57.5$^{\prime \prime}$	&	0.2298 $\pm$ 0.0005 & 0.23$^5$	&	3.00	$_{-	0.60	} ^{+	0.80}$	
&	19	&	410	$\	\pm\		\phantom{0}70	$	&	0.83	\\
XMMXCS J094358.2+164120.7	& &	09$^\mathrm{h}$43$^\mathrm{m}$58.2$^\mathrm{s}$	&	
+16$^\circ$41$^\prime$20.7$^{\prime \prime}$	&	0.2539 $\pm$ 0.0005 & 0.25$^1$	&	1.50	$_{-	0.20	} ^{+	0.40}$	
&	27	&	590	$\	\pm\		\phantom{0}90	$	&	1.54	\\
XMMXCS J095957.6+251629.0 	& &	09$^\mathrm{h}$59$^\mathrm{m}$57.6$^\mathrm{s}$	&	
+25$^\circ$16$^\prime$29.0$^{\prime \prime}$	&	0.0523 $\pm$ 0.0005 & 0.08$^6$	&	1.40	$_{-	0.05	} ^{+	0.05}$	
&	15	&	510	$\	\pm\		220	$	&	1.79	\\
XMMXCS J100047.4+013926.9	& &	10$^\mathrm{h}$00$^\mathrm{m}$47.4$^\mathrm{s}$	&	
+01$^\circ$39$^\prime$26.9$^{\prime \prime}$	&	0.2202 $\pm$ 0.0006 & 0.22$^7$	&	3.30	$_{-	0.20	} ^{+	0.20}$	
&	16	&	560	$\	\pm\		140	$	&	1.41	\\
XMMXCS J100141.7+022539.8	& &	10$^\mathrm{h}$01$^\mathrm{m}$41.7$^\mathrm{s}$	&	
+02$^\circ$25$^\prime$39.8$^{\prime \prime}$	&	0.1233 $\pm$ 0.0005 & 0.12$^8$	&	1.43	$_{-	0.03	} ^{+	0.06}$	
&	26	&	590	$\	\pm\		130	$	&	1.05	\\
XMMXCS J104044.4+395710.4	& &	10$^\mathrm{h}$40$^\mathrm{m}$44.4$^\mathrm{s}$	&	
+39$^\circ$57$^\prime$10.4$^{\prime \prime}$	&	0.1389 $\pm$ 0.0007 & 0.16$^9$	&	3.54	$_{-	0.03	} ^{+	0.03}$	
&	17	&	860	$\	\pm\		150	$	&	2.12	\\
XMMXCS J111515.6+531949.5	& &	11$^\mathrm{h}$15$^\mathrm{m}$15.6$^\mathrm{s}$	&	
+53$^\circ$19$^\prime$49.5$^{\prime \prime}$	&	0.4663 $\pm$ 0.0010 & 0.47$^{10}$	&	5.40	$_{-	0.90	} ^{+	1.50}$	
&	16	&	910	$\	\pm\		310	$	&	1.75	\\
XMMXCS J115112.0+550655.5	& &	11$^\mathrm{h}$51$^\mathrm{m}$12.0$^\mathrm{s}$	&	
+55$^\circ$06$^\prime$55.5$^{\prime \prime}$	&	0.0791 $\pm$ 0.0003 &  0.08$^{11}$	&	1.66	$_{-	0.04	} ^{+	0.04}$	
&	16	&	330	$\	\pm\		100	$	&	1.50	\\
XMMXCS J123144.4+413732.0	& &	12$^\mathrm{h}$31$^\mathrm{m}$44.4$^\mathrm{s}$	&	
+41$^\circ$37$^\prime$32.0$^{\prime \prime}$	&	0.1735 $\pm$ 0.0009 & 0.18$^{12}$	&	2.70	$_{-	0.40	} ^{+	0.60}$	
&	10	&	480	$\	\pm\		100	$	&	1.26	\\
XMMXCS J151618.6+000531.3	& &	15$^\mathrm{h}$16$^\mathrm{m}$18.6$^\mathrm{s}$	&	
+00$^\circ$05$^\prime$31.3$^{\prime \prime}$	&	0.1200 $\pm$ 0.0005 & 0.13$^{13}$	&	5.40	$_{-	0.10	} ^{+	0.10}$	
&	35	&	870	$\	\pm\		220	$	&	2.01	\\
XMMXCS J161132.7+541628.3	& &	16$^\mathrm{h}$11$^\mathrm{m}$32.7$^\mathrm{s}$	&	
+54$^\circ$16$^\prime$28.3$^{\prime \prime}$	&	0.3372 $\pm$ 0.0013 & 0.33$^{8}$	&	4.60	$_{-	0.80	} ^{+	1.20}$	
&	12	&	790	$\	\pm\		150	$	&	1.69	\\
XMMXCS J163015.6+243423.2	& &	16$^\mathrm{h}$30$^\mathrm{m}$15.6$^\mathrm{s}$	&	
+24$^\circ$34$^\prime$23.2$^{\prime \prime}$	&	0.0625 $\pm$ 0.0003 & 0.07$^{14}$	&	3.50	$_{-	0.40	} ^{+	0.60}$	
&	62	&	710	$\	\pm\		130	$	&	2.20	\\
XMMXCS J223939.3-054327.4	& &	22$^\mathrm{h}$39$^\mathrm{m}$39.3$^\mathrm{s}$	&	
-05$^\circ$43$^\prime$27.4$^{\prime \prime}$	&	0.2451 $\pm$ 0.0003 & 0.24$^{14}$	&	2.80	$_{-	0.20	} ^{+	0.20}$	
&	68	&	560	$\	\pm\		\phantom{0}70	$	&	1.32	\\
XMMXCS J233757.0+271121.0	& &	23$^\mathrm{h}$37$^\mathrm{m}$57.0$^\mathrm{s}$	&	
+27$^\circ$11$^\prime$21.0$^{\prime \prime}$	&	0.1237 $\pm$ 0.0007 & 0.12$^{15}$	&	3.40	$_{-	0.40	} ^{+	0.60}$	
&	12	&	460	$\	\pm\		110	$	&	1.49	\\
\hline
\end{tabular*}
\label{DR1}
\end{table*}

\begin{table*}
\caption{High redshift sample ($0.5 < z < 1.0$): all columns are as explained in Table~\ref{DR1}. The 
superscripts in column one indicate the origin of redshift data when it did not come from our own 
observations. $^{1}$ also known as MS1054-03 was observed with Keck for 8.6 hours \citep{Tran1999}. $^{2}$ was observed with Keck 
\citep{Donahue1999}. $^{3}$ also known as RXJ1821.6+6827 was observed with CFHT, Keck and the 2.2\,m telescope at the University of Hawaii 
\citep{Gioia2004}. $^{4}$ are all clusters taken from the XDCP survey and were observed with the VLT--FORS2 
spectograph \citep{Nastasi2014}. The references for the redshifts are as follows: 1. \citet{Nastasi2014} 2.\citet{Scharf1997} 3.\citet{Mehrtens2012} 4.\citet{Adami2011} 5.\citet{Suhada2011} 6.\citet{Bellagamba2011} 7.\citet{Gioia1994} 8.\citet{Basilakos2004} 9.\citet{Gioia2004} 10.\citet{Perlman2002}}
\begin{tabular*}{\textwidth}{c @{\extracolsep{\fill}} lccccccccc}
\hline
Name     & & RA        & Dec & $z$ & $z_{lit}$ & $T$ & Members         & $\sigma_{\rm v}$  & $R_{200}$ \\
& & (J2000) &  (J2000) & & & (keV) & & (km\,s$^{-1}$) & (Mpc) \\
 \hline
\footnotemark[4] XMMXCS J000216.1-355633.8	& &	00$^\mathrm{h}$02$^\mathrm{m}$16.1$^\mathrm{s}$	&	
-35$^\circ$56$^\prime$33.8$^{\prime \prime}$	&	0.7709 $\pm$ 0.0021 & 0.77$^{1}$	&	4.83	$_{-	0.76	} ^{+	1.01	5
}$	&	13	&	1100	$\	\pm\	190	$	&	1.77	&	\\
\phantom{0} XMMXCS J005656.6-274031.9	& &	00$^\mathrm{h}$56$^\mathrm{m}$56.6$^\mathrm{s}$	&	
-27$^\circ$40$^\prime$31.9$^{\prime \prime}$	&	0.5601 $\pm$ 0.0007 & 0.56$^{2}$	&	3.30	$_{-	0.63	} ^{+	0.94	
}$	&	15	&	\phantom{0}380	$\	\pm\	\phantom{0}60	$	&	0.66	&	
\\
\phantom{0} XMMXCS J015241.1-133855.9	& &	01$^\mathrm{h}$52$^\mathrm{m}$41.1$^\mathrm{s}$	&	
-13$^\circ$38$^\prime$55.9$^{\prime \prime}$ 	&	0.8268 $\pm$ 0.0010 & 0.82$^{3}$	&	3.23	$_{-	0.31	} ^{+	0.38	
}$	&	29	&	\phantom{0}840	$\	\pm\	150	$	&	1.33	&	\\
\phantom{0} XMMXCS J021734.7-051326.9	& &	02$^\mathrm{h}$17$^\mathrm{m}$34.7$^\mathrm{s}$	&	
-05$^\circ$13$^\prime$26.9$^{\prime \prime}$ 	&	0.6467 $\pm$ 0.0012 & 0.65$^{4}$	&	2.23	$_{-	0.44	} ^{+	0.90	
}$	&	12	&	\phantom{0}620	$\	\pm\	210	$	&	1.11	&	\\
\phantom{0} XMMXCS J025006.4-310400.8	& &	02$^\mathrm{h}$50$^\mathrm{m}$06.4$^\mathrm{s}$	&	
-31$^\circ$04$^\prime$00.8$^{\prime \prime}$ 	&	0.9100 $\pm$ 0.0024 & 0.90$^{2}$	&	4.50	$_{-	0.88	} ^{+	1.33	
}$	&	13	&	1120	$\	\pm\	260	$	&	1.66	&	\\
\phantom{0} XMMXCS J030205.1-000003.6	& &	03$^\mathrm{h}$02$^\mathrm{m}$05.1$^\mathrm{s}$	&	
-00$^\circ$00$^\prime$03.6$^{\prime \prime}$ 	&	0.6450 $\pm$ 0.0007 & 0.65$^{5}$	&	5.82	$_{-	1.32	} ^{+	2.09	
}$	&	16	&	\phantom{0}610	$\	\pm\	180	$	&	1.04	&	\\
\footnotemark[4] XMMXCS J095417.1+173805.9	& &	09$^\mathrm{h}$54$^\mathrm{m}$17.1$^\mathrm{s}$	&	
17$^\circ$38$^\prime$05.9$^{\prime \prime}$	&	0.8272 $\pm$ 0.0017 & 0.82$^{1}$	&	3.65	$_{-	0.51	} ^{+	0.62	
}$	&	10	&	\phantom{0}940	$\	\pm\	310	$	&	1.42	&	\\
\phantom{0} XMMXCS J095940.7+023113.4	& &	09$^\mathrm{h}$59$^\mathrm{m}$40.7$^\mathrm{s}$	&	
+02$^\circ$31$^\prime$13.4$^{\prime \prime}$ 	&	0.7291 $\pm$ 0.0005 & 0.72$^{6}$	&	5.02	$_{-	0.55	} ^{+	0.68	
}$	&	25	&	\phantom{0}470	$\	\pm\	\phantom{0}90	$	&	0.88	&	
\\
\footnotemark[1] XMMXCS J105659.5-033728.0	& &	10$^\mathrm{h}$56$^\mathrm{m}$59.5$^\mathrm{s}$	&	
-03$^\circ$37$^\prime$28.0$^{\prime \prime}$	&	0.8336 $\pm$ 0.0013 & 0.82$^{7}$	&	7.57	$_{-	0.40	} ^{+	0.43	
}$	&	29	&	1010	$\	\pm\	120	$	&	1.57	&	\\
\phantom{0} XMMXCS J112349.4+052955.1	& &	11$^\mathrm{h}$23$^\mathrm{m}$49.4$^\mathrm{s}$	&	
+05$^\circ$29$^\prime$55.1$^{\prime \prime}$ 	&	0.6550 $\pm$ 0.0007 & 0.65$^{3}$	&	4.62	$_{-	0.95	} ^{+	1.55	
}$	&	17	&	\phantom{0}600	$\	\pm\	210	$	&	1.05	&	\\
\phantom{0} XMMXCS J113602.9-032943.2	& &	11$^\mathrm{h}$36$^\mathrm{m}$02.9$^\mathrm{s}$	&	
-03$^\circ$29$^\prime$43.2$^{\prime \prime}$ 	&	0.8297 $\pm$ 0.0011 & 	&	3.32	$_{-	0.78	} ^{+	1.20	
}$	&	21	&	\phantom{0}700	$\	\pm\	110	$	&	1.06	&	\\
\footnotemark[2] XMMXCS J114023.0+660819.0	& &	11$^\mathrm{h}$40$^\mathrm{m}$23.9$^\mathrm{s}$	&	
+66$^\circ$08$^\prime$19.0$^{\prime \prime}$	&	0.7855 $\pm$ 0.0015 & 0.78$^{7}$	&	7.47	$_{-	0.77	} ^{+	0.92	
}$	&	22	&	\phantom{0}950	$\	\pm\	100	$	&	1.51	&	\\
\footnotemark[4] XMMXCS J124312.2-131307.2	& &	12$^\mathrm{h}$43$^\mathrm{m}$12.2$^\mathrm{s}$	&	
-13$^\circ$13$^\prime$07.2$^{\prime \prime}$	&	0.7910 $\pm$ 0.0014 & 0.80$^{1}$&	4.92	$_{-	1.54	} ^{+	2.93	
}$	&	11	&	\phantom{0}790	$\	\pm\	460	$	&	1.19	&	\\
\phantom{0} XMMXCS J134305.1-000056.8	& &	13$^\mathrm{h}$43$^\mathrm{m}$05.1$^\mathrm{s}$	&	
-00$^\circ$00$^\prime$56.8$^{\prime \prime}$ 	&	0.6894 $\pm$ 0.0011 & 0.67$^{8}$	&	4.49	$_{-	0.57	} ^{+	0.72	
}$	&	23	&	\phantom{0}920	$\	\pm\	170	$	&	1.72	&	\\
\phantom{0} XMMXCS J145009.3+090428.8	& &	14$^\mathrm{h}$50$^\mathrm{m}$09.3$^\mathrm{s}$	&	
+09$^\circ$04$^\prime$28.8$^{\prime \prime}$ 	&	0.6412 $\pm$ 0.0007 & 0.60$^{3}$	&	3.84	$_{-	0.55	} ^{+	0.66	
}$	&	22	&	\phantom{0}630	$\	\pm\	\phantom{0}90	$	&	1.07	&	
\\
\footnotemark[3] XMMXCS J182132.9+682755.0	& &	18$^\mathrm{h}$21$^\mathrm{m}$32.9$^\mathrm{s}$	&	
+68$^\circ$27$^\prime$55.0$^{\prime \prime}$	&	0.8166 $\pm$ 0.0011 & 0.82$^{9}$	&	4.49	$_{-	0.56	} ^{+	0.79	
}$	&	19	&	\phantom{0}860	$\	\pm\	130	$	&	1.34	&	\\
\phantom{0} XMMXCS J215221.0-273022.6	& &	21$^\mathrm{h}$52$^\mathrm{m}$21.0$^\mathrm{s}$	&	
-27$^\circ$30$^\prime$22.6$^{\prime \prime}$ 	&	0.8276 $\pm$ 0.0011 & 0.82$^{3}$	&	2.18	$_{-	0.45	} ^{+	0.67	
}$	&	15	&	\phantom{0}530	$\	\pm\	150	$	&	0.86	&	\\
\phantom{0} XMMXCS J230247.7+084355.9	& &	23$^\mathrm{h}$02$^\mathrm{m}$47.7$^\mathrm{s}$	&	
+08$^\circ$43$^\prime$55.9$^{\prime \prime}$	&	0.7187 $\pm$ 0.0014 & 0.72$^{10}$	&	5.29	$_{-	0.50	} ^{+	0.59	
}$	&	22	&	1010	$\	\pm\	130	$	&	1.60	&	\\
\footnotemark[4] XMMXCS J235616.4-344144.3	& &	23$^\mathrm{h}$56$^\mathrm{m}$16.4$^\mathrm{s}$	&	
-34$^\circ$41$^\prime$44.3$^{\prime \prime}$	&	0.9391 $\pm$ 0.0012 & 0.94$^{1}$	&	4.57	$_{-	0.41	} ^{+	0.48	
}$	&	10	&	\phantom{0}670	$\	\pm\	260	$	&	0.91	&	\\
\hline
\end{tabular*}
\label{Gemini}
\end{table*}

\section{Analysis}
\label{Method}

\subsection{Membership determination and velocity dispersion measurements}
In this section we describe the methodology used to determine cluster membership and calculate the velocity 
dispersion of each cluster.

\subsubsection{Cluster redshifts}

For all of the clusters an estimate of the redshift is known either from the literature or from previous 
observations and this is used as a starting point. The peculiar velocity of each of the galaxies is calculated 
relative to this redshift estimate using
\begin{equation}
 v_i=c \times \frac{z_i - \bar{z}}{1+\bar{z}},
 \label{pecvel}
\end{equation}
where $v_i$ is the peculiar velocity of the $i$th galaxy, $z_i$ is the redshift of the $i$th galaxy, 
$\bar{z}$ is the redshift of the cluster and $c$ is the speed of light. Extreme foreground and background 
sources were removed by applying a 3000\,km\,s$^{-1}$ cut with respect to 
the cluster redshift and then the redshift was recalculated using the biweight location method described by 
\cite{Beers1990}. This process was iterated until the redshift converged. 

\subsubsection{Cluster membership}
\label{membership}
A fixed gapper method, similar to that of \cite{Fadda1996} and \cite{Crawford2014}, was applied to 
determine which galaxies are cluster members. The reasoning behind this method is that by studying a histogram 
of the redshifts of possible members there should be a clear distinction between the cluster and the 
fore/background galaxies. Therefore we can exclude interlopers by finding the velocity difference between 
adjacent galaxies and setting a fixed gap that should not be exceeded. \cite{DePropris2002} found this optimum 
gap to be 1000\,km\,s$^{-1}$, which avoids the merging of subclusters but also prevents the breaking up of 
real systems into smaller groups. Therefore all our galaxies were sorted by peculiar velocity and the 
difference between all adjacent pairs was calculated. Any galaxies which had a difference between adjacent 
galaxies of greater than 1000\,km\,s$^{-1}$ were considered interlopers and were removed. This process was 
iterated until the number of galaxies converged.

\subsubsection{Velocity dispersion}
\label{veldisp}
We used our confirmed galaxy cluster members to calculate an initial estimate of the velocity dispersion of
each cluster using the biweight scale method described in \cite{Beers1990}. 
We then calculated $R_{200}$ using Equation \ref{R200}, and excluded all galaxies located at projected 
cluster-centric radial distances outside $R_{200}$. The velocity dispersion of each cluster was then 
recalculated. This final radial cut did not remove more than two galaxies from the 
final sample for each cluster. Tables~\ref{DR1} and \ref{Gemini} list the final redshifts, 
velocity dispersions, and $R_{200}$ values for the low and high redshift samples respectively.


\subsection{Fitting the velocity dispersion -- temperature relation}
\label{Fitting}
To determine the scaling relation between the velocity dispersion and temperature, we fitted a power law of 
the form
\begin{equation}
\label{model}
\log \left(\frac{\sigma_{\rm v}}{1000 \,{\rm km \,s}^{-1}}\right) = A+ B\, \log \left(\frac{T}{5 \, 
\rm{keV}}\right) + C\,\log E(z).
\end{equation}
Here, 5\,keV and 1000\,km s$^{-1}$ are the pivot temperature and velocity 
dispersion respectively for our fit. These were chosen to reduce the covariance between the normalisation $A$
and the slope $B$, and for ease of comparison to previous studies. In the above, evolution of the 
normalisation is parametrised as $E(z)^C$, where $E(z) = \sqrt{\Omega_m(1+z)^3 + \Omega_\Lambda}$ 
describes the redshift evolution of the Hubble parameter. For the self similar case, $B = 0.5$ and $C = 0$
are expected.

Similarly to \citet{Hilton2012}, the best fit values for these parameters were found using Markov Chain 
Monte-Carlo (MCMC) with the Metropolis algorithm. Both orthogonal and bisector regression 
methods were used. For the orthogonal method, the probability for a given cluster to be drawn from the model
scaling relation is
\begin{equation} 
P_{\rm{model}}=\frac{1}{\sqrt{2\pi(\Delta r^2 + \Delta 
S^2)}}\exp\left[\frac{-(r-r_{\rm{model}})^2}{2(\Delta r^2 + S^2)}\right],
\label{orthmodel}
\end{equation}

where $r - r_{\rm model}$ is the orthogonal distance of the cluster from the model relation, $\Delta 
r$ is the error on the orthogonal distance and $S$ is the intrinsic scatter orthogonal to the model relation. 
$\Delta r$ is calculated from the projection in the direction orthogonal to the model line of the ellipse 
defined by the errors on $\log \sigma_{\rm v}$ and $\log T$, chosen according to the position of a given point relative to the model  fit line. For the bisector method, the intrinsic scatter and measurement errors are treated independently for each axis. 
Therefore in the equation for $P_{\rm model}$, $r_{\rm model}$ is replaced by
\begin{equation}
y_{\rm{model}}=\log \left(\frac{\sigma_{\rm v}}{1000 \,{\rm km \, s}^{-1}}\right) - \left[ A + 
B\,\log \left(\frac{T}{5 \, \rm{keV}}\right) + C \log E(z) \right],
\end{equation}
and
\begin{equation}
 x_{\rm{model}} = \log \left( \frac{T}{5 \, \rm{keV}} \right) - 
 \left[ 
\frac{ \log \left( \frac{\sigma_{\rm v}} {1000 \,{\rm km \,s}^{-1}} \right)  - A - C \log E(z)}{B} \right], 
\end{equation}
where $r$ and $\Delta r$ are replaced by $x$, $\Delta x$ or $y$, $\Delta y$ as appropriate. The intrinsic 
scatter $S$ is replaced by two parameters $S_x$ and $S_y$. 

For both methods, the likelihood $\mathcal{L}$ of a given model is simply the product of $P_{\rm model}$ for each
cluster in the sample, i.e., in the orthogonal case
\begin{equation}
\mathcal{L}(\sigma_{\rm v}, T | A, B, C, S) \propto P_{\rm prior}(A, B, C, S) \prod_i{P_{\rm model, \emph i}},
\label{likelihood}
\end{equation}
where we assume generous, uniform priors on each parameter, as listed in Table \ref{priors}.

\begin{table}
\caption{Priors on $\sigma_{\rm v}$ -- $T$ relation fit parameters}
\label{priors}
\begin{tabular}{ccc}
\hline
Parameter & Uniform Prior & Notes\\
\hline
$A$ & (-5.0,5.0)& -\\
$B$ & (0.0,2.0)& -\\
$C$ & (-1.0,1.0)& - \\
$S$ & (0.01,1.0)& Orthogonal method only\\
$S_x$ & (0.01,1.0)& Bisector method only\\
$S_y$ & (0.01,1.0)& Bisector method only\\
\hline
\end{tabular}
\end{table}


\section{Results}
\label{Results}

\subsection{Evolution of the slope and intrinsic scatter}
\label{slope}

\begin{figure*}
\includegraphics[width=\columnwidth]{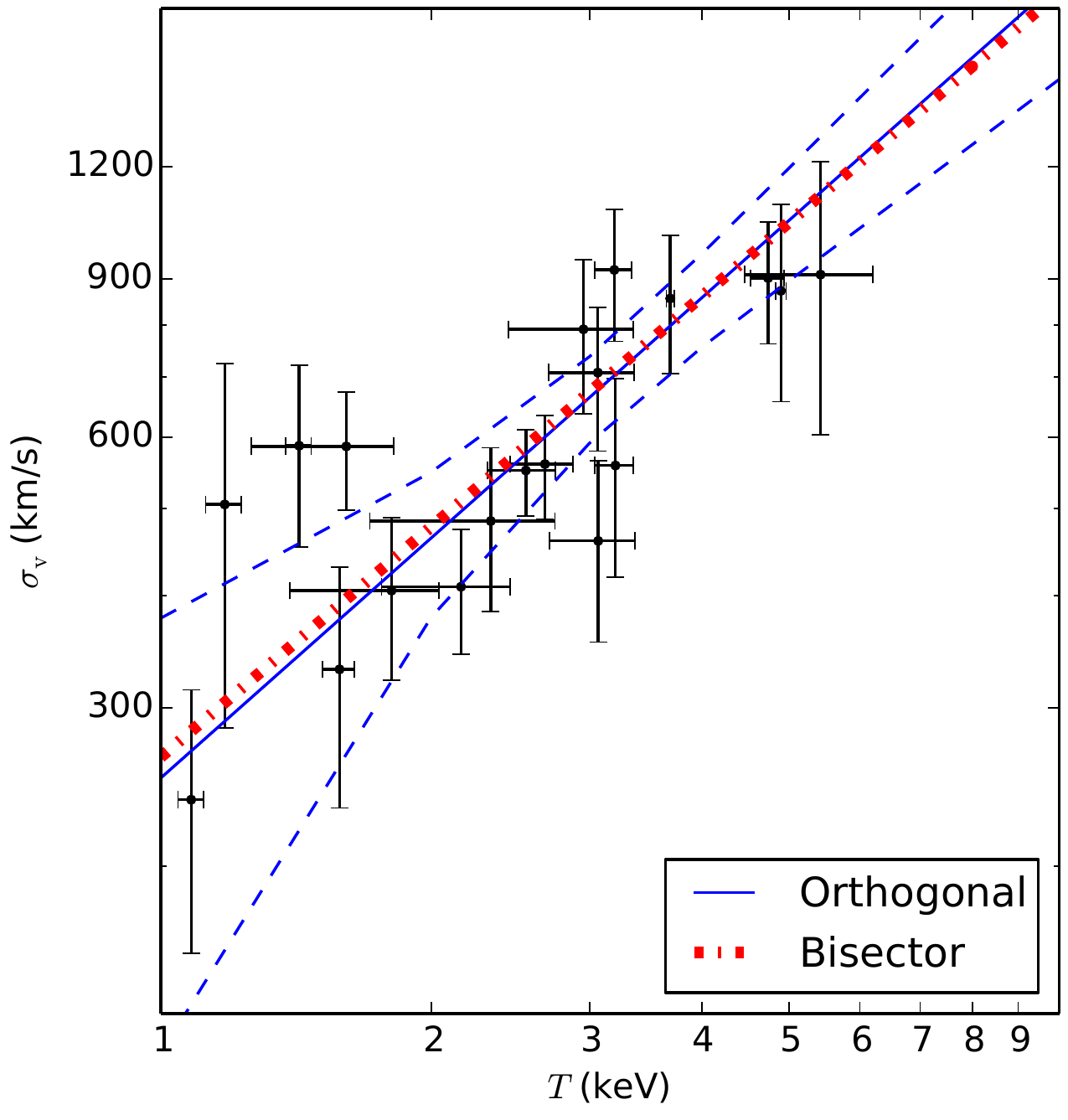}
\includegraphics[width=\columnwidth]{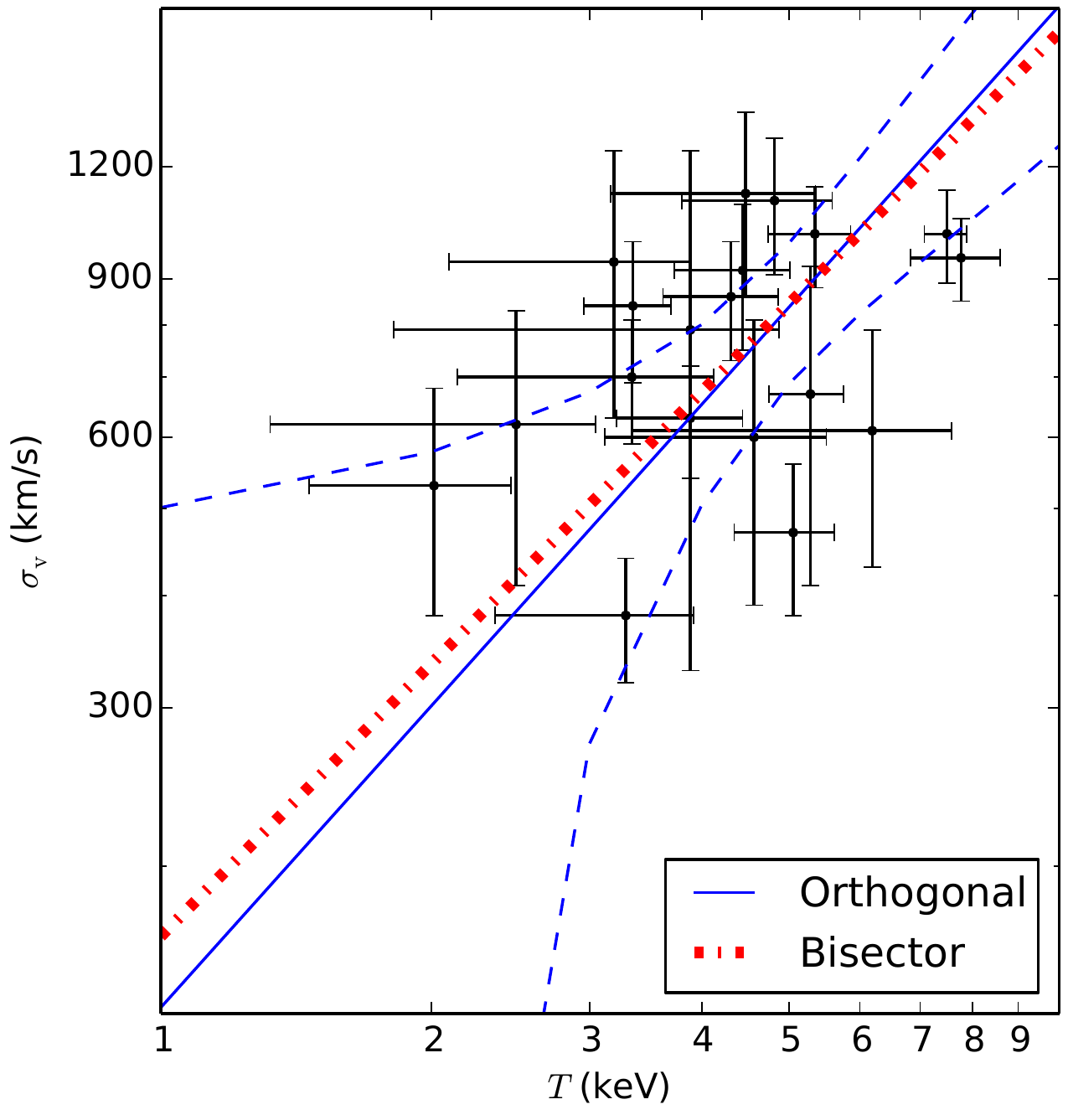}
\caption{The $\sigma_{\rm v}$--$T$ relation assuming no evolution, i.e., $C=0$ in 
Equation \ref{model}, for low (left - $0.0 < z < 0.5$) and high (right- $0.5 < z < 0.9$) redshift samples. The solid blue line shows an orthogonal regression fit to the data with the dashed line representing the 95 \% confidence interval. 
The dot-dashed line shows a bisector regression fit to the data (see Section \ref{Fitting}). A model of the
form seen in Equation \ref{model} was used in the Metropolis algorithm to determine a line of best-fit.
(see Section \ref{slope}). It is interesting to note that the two best-measured systems (XMMXCS J105659.5-033728.0 and XMMXCS J114023.0+660819.0) in the high-redshift subsample are relatively far off the best-fit relation, with a higher than predicted temperature. Our current observations do not provide good enough spatial resolution or deep enough multi-colour photometry to determine the exact reason for this and require further study and re-observations. }
\label{plots1}
\end{figure*}

\begin{figure*}
\includegraphics[width=\columnwidth]{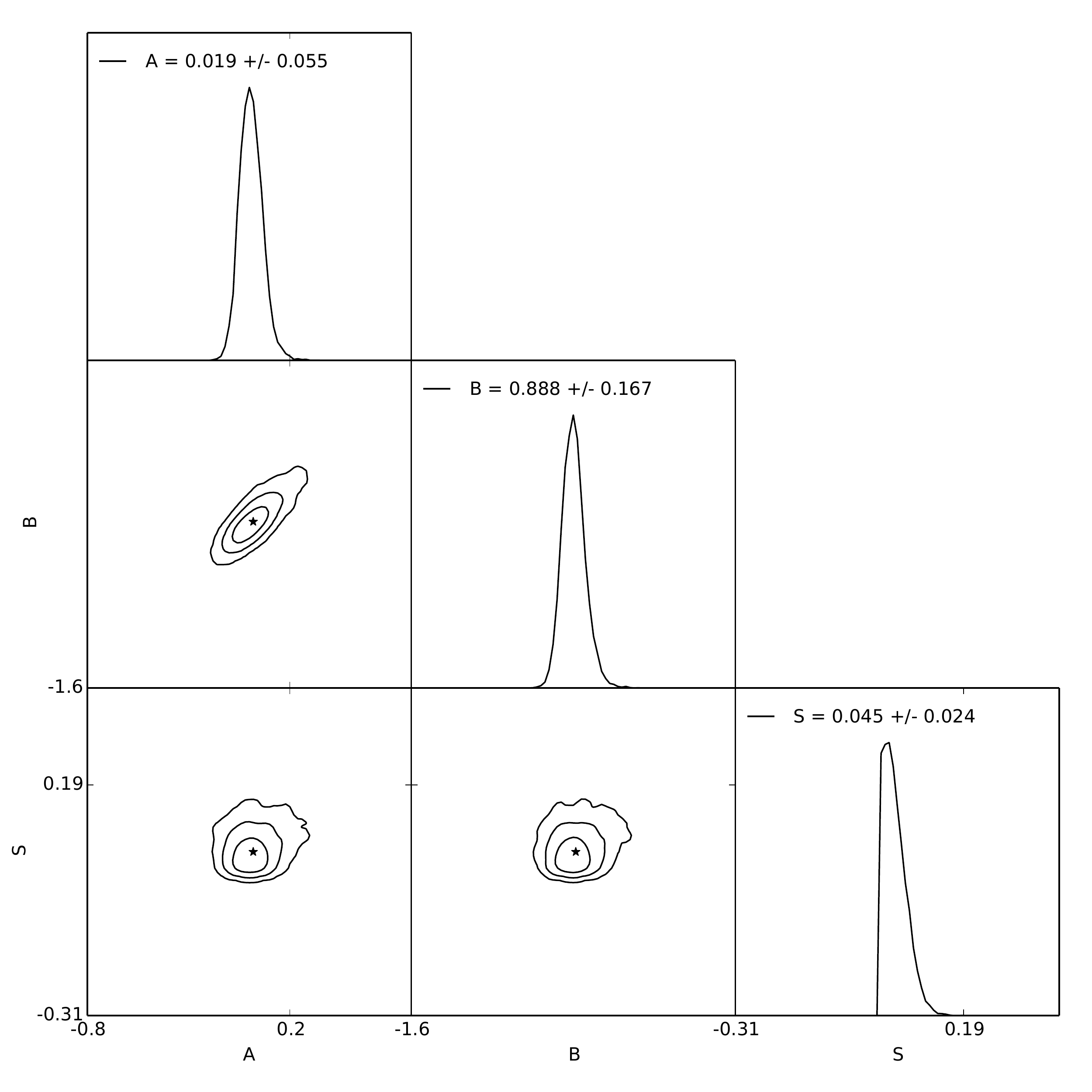}
\includegraphics[width=\columnwidth]{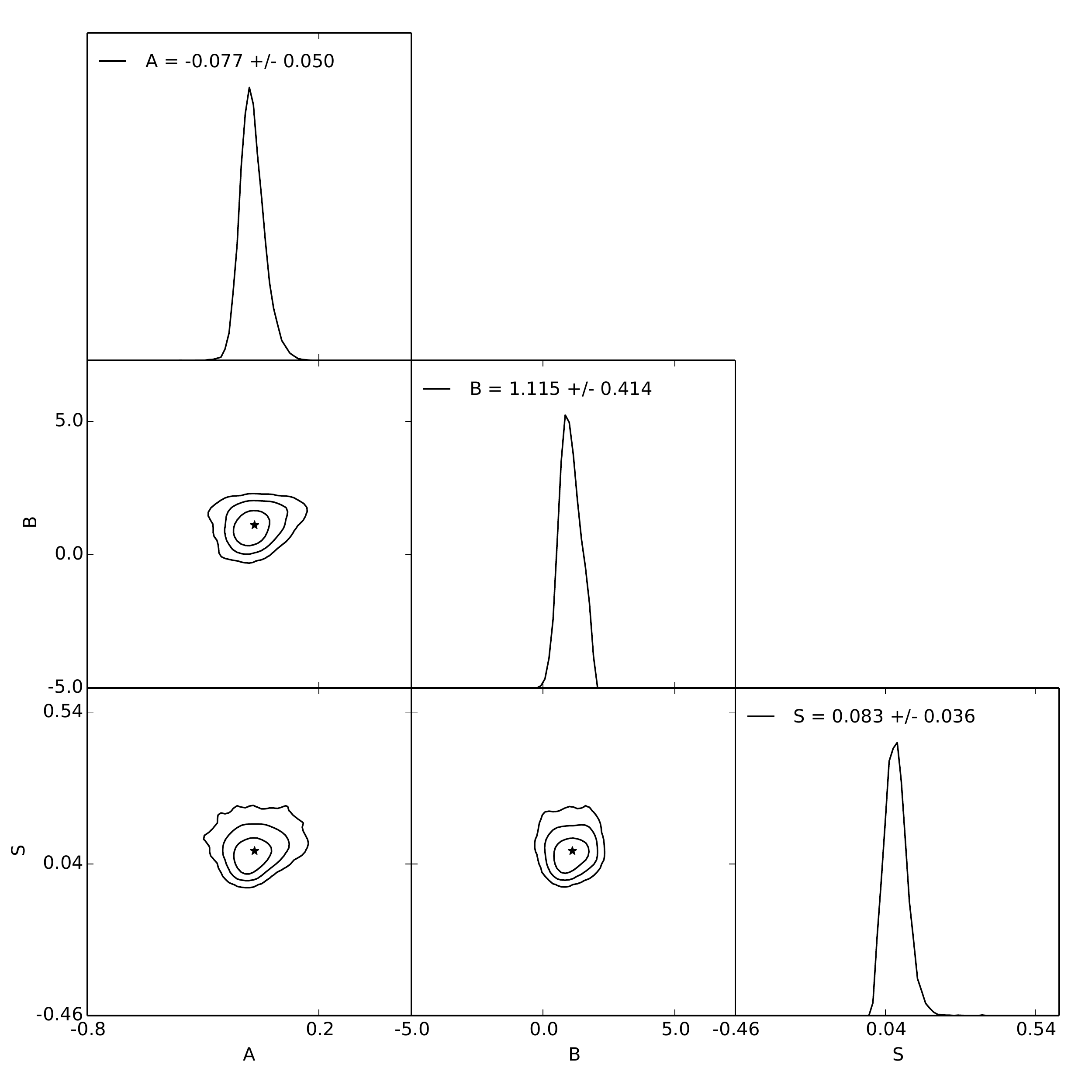}
\caption{Corner plots for the low (left - $0.0 < z < 0.5$) and high (right- $0.5 < z < 0.9$) redshift sample showing all the one and two dimensional projections of the posterior probability distributions of the three parameters when using the orthogonal method. The values of each parameter are given in the top centre. The histograms show the one dimensional marginalised distribution for each parameter and the other plots show the 2 dimensional version, where the contours show 1,2 and 3 $\sigma$.}
\label{2dplots}
\end{figure*}

For the model given in Equation \ref{model}, it is assumed that the slope (parameter $B$) is not evolving with redshift. To 
test this, the $\sigma_{\rm v}$--$T$ relation was fitted with $C = 0$ in two redshift bins, 
$0.0 < z < 0.5$ and $0.5 < z <0.9$, with 19 clusters in each bin. The parameters $A$, $B$ and $S$ were 
obtained using the MCMC method described above for the high and low redshift samples individually. 
The results for this are shown in Fig.~\ref{plots1} and Fig.~\ref{2dplots}. 

Using the orthogonal method we found $B= 1.12 \pm 0.41$ for the high redshift sample and $B=0.89 \pm 0.16$
for the low redshift sample. However, we found that the slope of the relation for the high redshift sample
is unconstrained if the prior on $B$ is relaxed further. We assume for the remainder of this paper that
the slope does not evolve with redshift, though clearly either a larger sample or more accurate 
measurements of individual clusters are needed to confirm that this is true.

The intrinsic scatter is $S = 0.05 \pm 0.02$ for the low redshift sample and $S = 0.08 \pm 0.04$ for the high
redshift sample. Therefore there is no evidence that the intrinsic scatter varies with redshift.

\begin{table*}
 \caption{Best-fit $\sigma_{\rm v}$--$T$ scaling relation parameters using both the orthognal and bisector
 regression methods (see Section~\ref{Fitting}).}
\centering 
   \begin{tabular*}{\textwidth}{c @{\extracolsep{\fill}} cllll}
  \hline
  Method & Parameter & Low Redshift & High Redshift & Combined  & Combined \\	
         &           &              &               & (no evolution) & (with evolution) \\
  \hline
		
Orthogonal &	$A$	&	0.02	$\pm$	0.06	&	-0.08	$\pm$	0.05	&	-0.05	$\pm$	
0.04	&	0.02	$\pm$	0.05	\\
&		$B$	&	0.89	$\pm$	0.16	&	1.12	$\pm$	0.41	&	0.72	$\pm$	
0.12	&	0.86	$\pm$	0.14	 \\
&		$S$	&	0.05	$\pm$	0.03	&	0.08	$\pm$	0.04	&	0.06	$\pm$	
0.03	&	0.06	$\pm$	0.03	 \\
&		$C$	& 0 & 0 & 0 &	 -0.37 $\pm$ 0.33	 \\
\hline
Bisector &	$A$	&	0.02	$\pm$	0.04	&	-0.07	$\pm$	0.03	&	-0.04	$\pm$	
0.02	&	0.02	$\pm$	0.04	\\
&	$B$	&	0.85	$\pm$	0.11	&	1.01	$\pm$	0.17	&	0.77	$\pm$	0.08	
&	0.86	$\pm$	0.09	\\
&	$S_x$	&	0.15	$\pm$	0.03	&	0.19	$\pm$	0.03	&	0.15	$\pm$	0.02	
&	0.14	$\pm$	0.02	\\
&	$S_y$	&	0.07	$\pm$	0.03	&	0.12	$\pm$	0.04	&	0.09	$\pm$	0.02	
&	0.09	$\pm$	0.02	\\
&	$C$	&	0 & 0 & 0 & -0.49	$\pm$	0.25	\\

\hline

\end{tabular*}

\label{ResultsTable}
\end{table*}

\subsection{Evolution of the normalisation}
\label{EvN}

To test for the evolution of the normalisation (parameter $A$ in Equation~\ref{model}), the low and high redshift 
samples were combined and $C$ was allowed to vary in the MCMC analysis. The results obtained are 
shown by the the scaling relation plot in Fig.~\ref{EvolutionPlot}. We found $C=-0.53 \pm 0.27$, 
meaning that for a given $\sigma_{\rm v}$, a higher $T$ is obtained at higher redshift. 
However, the no evolution relation falls within the 95 per cent confidence interval and therefore we conclude
that there is no significant evidence in favour of evolution.

\begin{figure}
\includegraphics[width=\columnwidth]{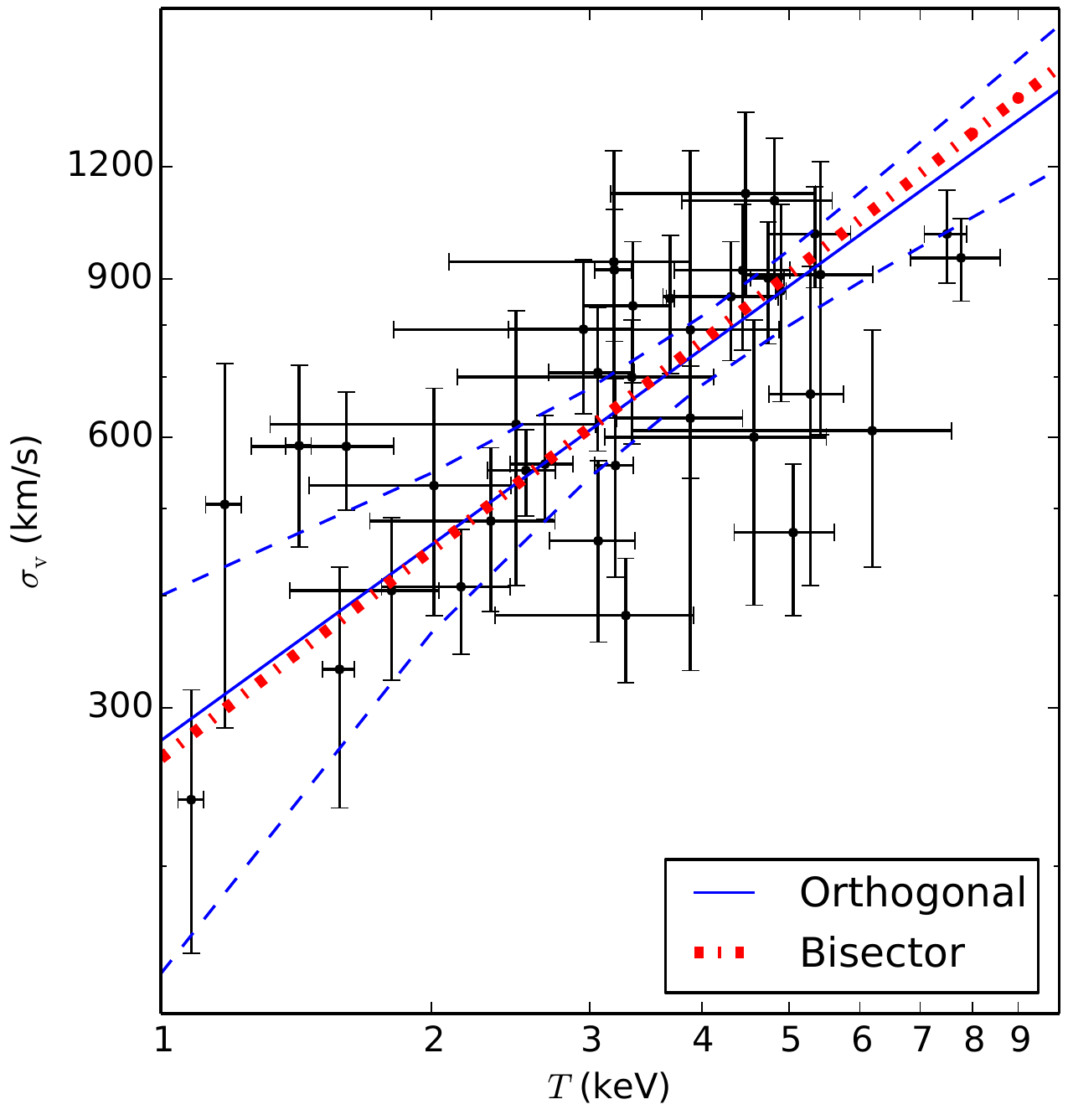}
\caption{The $\sigma_{\rm v}$--$T$ relation assuming no evolution, i,e. $C=0$ in Equation 
\ref{model}, for the combined sample. All lines are as explained in Fig. \ref{plots1}. }
\label{noEvolution}
\end{figure}

\begin{figure}
\includegraphics[width=\columnwidth,keepaspectratio=true]{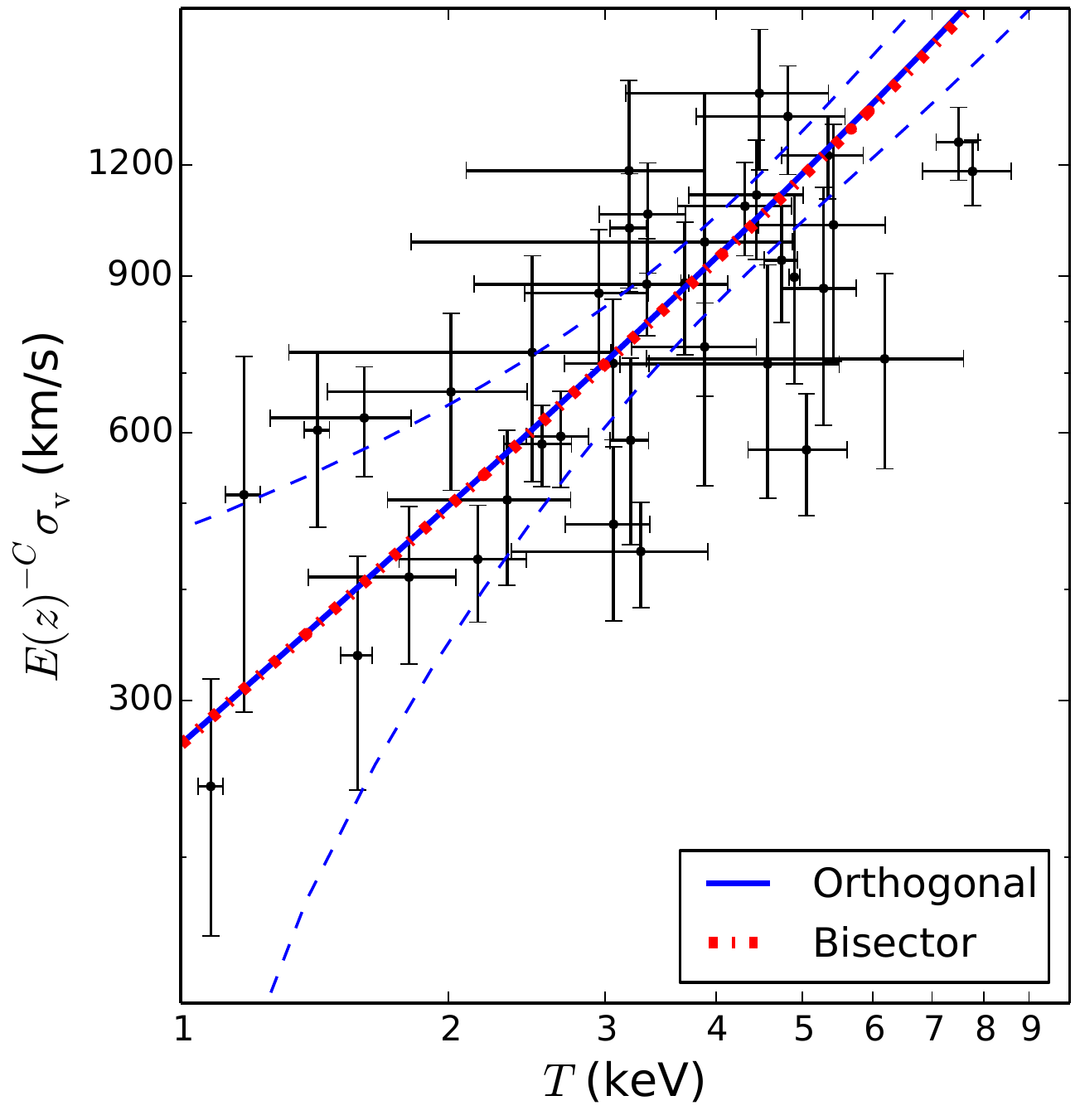}
\caption{Plots showing the $\sigma_{\rm v}$--$T$ relation for the combined sample with varying evolution, i.e. 
$C$ is a free parameter in Equation \ref{model}. The velocity dispersion is scaled to take into the account 
the evolution by multiplying by $E(z)^{-C}$. All lines are as explained in Fig. \ref{plots1}.}
\label{EvolutionPlot}
\end{figure}

\begin{figure}
\label{ev}
\includegraphics[width=\columnwidth]{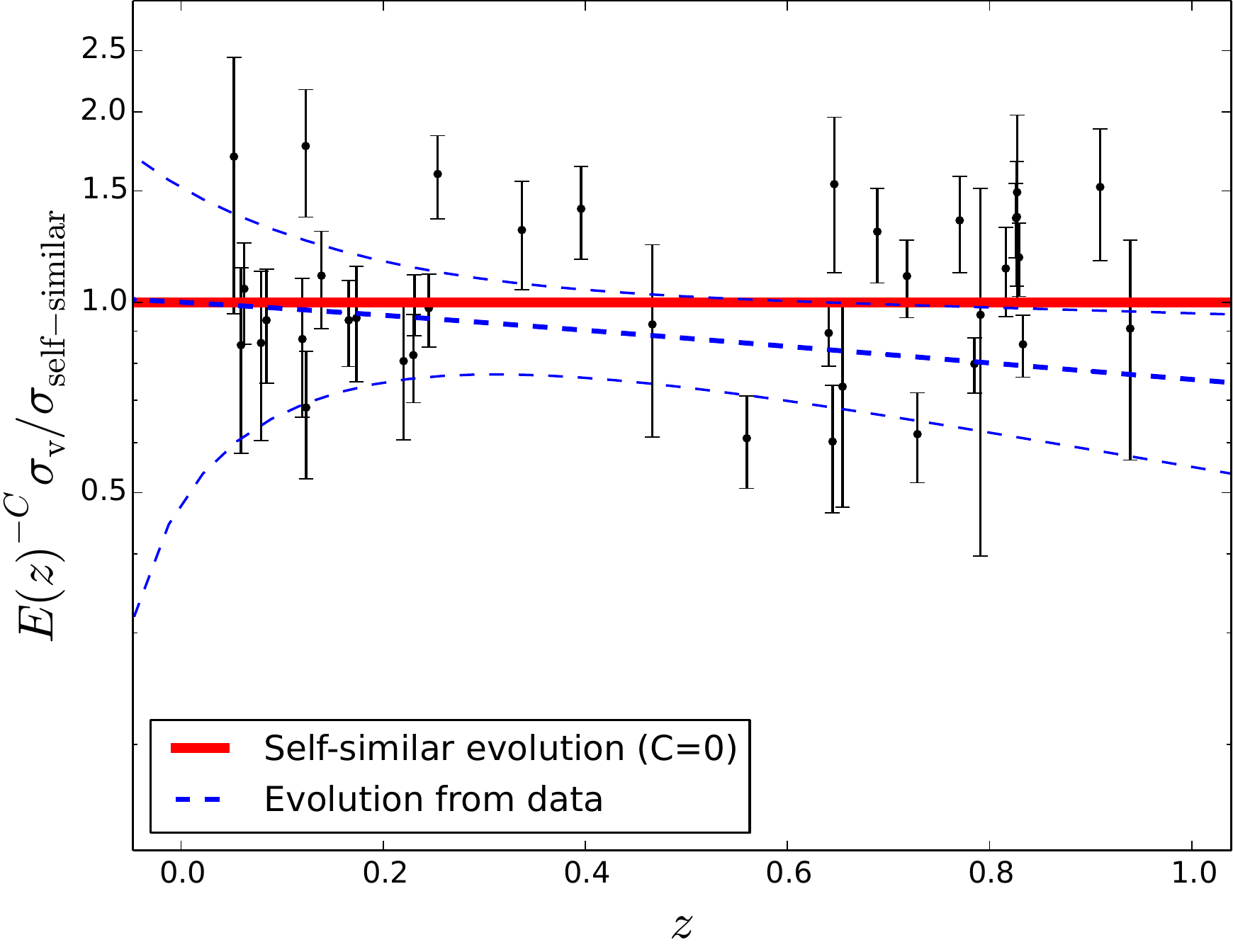}
\caption{Plot showing the evolution of the normalisation of the $\sigma_{\rm v}- T$ relation  
obtained for the data with the 95$\%$ confidence intervals in the dashed lines as compared to the self similar 
relation which predicts no evolution shown as the solid line. The black points show our sample data.}
\end{figure}

We also applied a statistical test known as the Akaike information criterion (AIC) to determine whether the 
model with or without evolution (Fig. \ref{noEvolution}) was preferred. The AIC estimates the quality of each 
model relative to each other and is therefore a means of model selection. It is defined by \citep{AIC} as
\begin{equation}
 \mathrm{AIC}= 2k - 2\mathrm{ln}(\mathcal{L}),
\end{equation}
where $\mathcal{L}$ is the maximised likelihood function (Equation~\ref{likelihood}) and $k$ is the number of 
free parameters. The AIC includes a penalty for using extra parameters as a way to discourage overfitting and 
rewards goodness of fit based on the likelihood function. Therefore the lower the value of the AIC, the better 
the fit. For the combined sample with the no evolution model the AIC value was -64.6 and when the fourth 
parameter for evolution ($C$) was included this increased to -62.1. Therefore, combining this with the results
from the $\sigma_{\rm v}$--$T$ relation fit, it can be concluded that the preferred model is the one with no 
evolution of the normalisation of the scaling relation.


\section{Discussion}
\label{Discussion}
%
%
%
%

\subsection{Comparison with previous results}
\label{Discussion_previous}
Table~\ref{ResultsTable} and Figs.\ref{plots1}--\ref{EvolutionPlot} present  the results of applying the orthogonal and 
bisector fitting methods to the low redshift, high redshift, and combined samples. We see that the bisector 
and orthogonal method give very similar results especially for our total sample without evolution. 
\cite{Hogg2010} suggests that the bisector method should be avoided, as by simply finding the difference 
between a forward and reverse fitting method large systematic errors will be introduced, but it has been 
widely used for scaling relation measurements in the past and is therefore included for completeness.

Results from previous studies of the $\sigma_{\rm v}$--$T$ relation are collected in Table~\ref{previous}. 
All of these studies, except for \cite{Edge1991} and the low redshift sample of \cite{Wu1998}, 
obtained a slope steeper than the expected self-similar slope of $\sigma_{\rm v} \propto T^{0.5}$. 
We measured $B=0.72 \pm 0.12$ using the orthogonal fitting method and $B=0.77 \pm 
0.08$ using the bisector fitting method for our combined sample. Therefore both the orthogonal and bisector 
slopes are in agreement with each other and the previous values in the literature, except for the result obtained by 
\cite{Edge1991} which is only consistent with the orthogonal result.

Except for work done by \cite{Wu1998} and \cite{Nastasi2014}, all the previous results were obtained for low 
redshift samples and no test for evolution was performed. \cite{Wu1998} divided their sample into two groups, 
$z<0.1$ and $z\geq 0.1$, and found no significant evolution, however their sample included only four 
clusters in the redshift range $0.5 < z < 1.0$. \cite{Nastasi2014} had a sample of 12 galaxy clusters and found a very 
large error of more than 50 per cent on their slope. They concluded that their sample size was too small to 
accurately measure evolution. We conclude that the data presented in this paper -- a homogeneous cluster
sample that is larger than those used in previous studies at $z > 0.5$ -- are consistent with previous 
results.
\subsection{Description of simulations}
\label{Sims}

Comparison to simulations are important for two main reasons. Firstly, we can determine if there is any bias due to
sample selection as the simulations provide both a bigger temperature and redshift range. It also allows us to compare 
different simulation models and learn about the nature of the non-gravitational physics through their effect on the gas temperature.

The Millennium Gas Project is a set of hydrodynamical simulations described in \cite{Short2010} which uses the 
same initial perturbations as the Millennium Simulation \citep{Springel2005}. These simulations include a 
variety of models, including gravity only; energy injection with radiative cooling; and feedback only. For 
comparison to the data presented in this paper, the feedback only model (FO) in a volume of 250\,$h^{-1}$ 
Mpc$^3$ was used. This model includes supernova and AGN feedback using a semi-analytic galaxy formation model. 
Heating due to supernovae and AGN and the star formation rate are obtained using the model of 
\cite{DeLucia2007}. The AGN feedback model used is described in \cite{Bower2008}, which is dependent on the 
matter accreted by the central black hole and the efficiency with which the matter is converted to energy near 
the event horizon, with the upper limit being at two per cent of the Eddington rate.

As a comparison to the velocity dispersion of the cluster, two proxies were considered, the velocity 
dispersion of the stars ($\sigma_{\rm Stars}$) and that obtained from the dark matter particles ($\sigma_{DM}$). 
The temperatures used from the simulation were spectroscopic-like temperatures 
\citep[$T_{\rm sl}$;][]{Mazzotta2004}. To ensure that only 
clusters similar to those in our sample were included we excluded all groups from the simulation with a mass 
less than 10$^{14}$ M$_{\sun}$. We also included a temperature cut, $2 < T (\rm keV) < 11$, and a redshift cut, 
$0 < z < 1$, to match our sample.

We also compared to the results of the BAHAMAS hydrodynamical simulation (McCarthy et al., in prep. and Caldwell et 
al., in prep.). Here, a 400\,$h^{-1}$ Mpc$^3$ box is used, with initial conditions based on Planck 2013 cosmological
parameters \citep{Planck2013}, and both AGN and supernovae feedback models as described by \cite{LeBrun2014}. A galaxy mass 
lower limit of $ 5 \times 10^9$ M$_{\sun}$ and a cluster mass lower limit of $10^{14}$\,M$_{\sun}$ were implemented. 
This simulation reproduces a large number of X-ray, SZ, and optical scaling relations of groups and clusters. 
However, unlike previous simulations, the new simulation also reproduces the observed galaxy stellar mass function 
remarkably well over a wide range of stellar masses. The velocity dispersion is traced by galaxies and is 
calculated using the gapper technique described by \cite{Beers1990}. The temperatures used from the simulation were spectroscopic 
($T_{\rm S}$).

We note that spectroscopic-like temperatures, as used in both simulations, are most robust at $T > 2$\, keV, where
the bremsstrahlung mechanism dominates \citep{Short2010}. Therefore, while we have applied mass, temperature, and redshift
cuts to the simulated cluster catalogues that are a reasonable match to our observed sample, the correspondance is not exact,
as 6 of the observed clusters have $T < 2$\,keV. This is a compromise aimed at limiting the potential impact of low mass
clusters with less reliable temperature measurements in the simulations.


Matching the velocity dispersions to the simulations, however, is not as straight-forward, and is outside the scope of this paper, but for completeness we briefly discuss causes of bias identified in previous studies. \cite{Old2013} studied the recovery of velocity dispersions from simulation data and explored how sample selection can impact the measurements and cause a bias. They introduced $I$-band magnitude limits and found that the velocity dispersion recovered from the halos was systematically higher than that from the galaxies. When this sample was further limited to just the brightest galaxies, this discrepancy was enhanced. They suggest that the reason for this is that dynamical friction greatly affects the velocity of the galaxies, and therefore to reduce this bias a strictly magnitude-limited sample should be avoided. 

\cite{Old2013} also calculated the velocity dispersion over different radial distances to see how it varied as a function of distance from the cluster centre. They found that the velocity dispersion was sensitive to the radius in which it was measured with a difference of 10 per cent in measurements being found between 0.5--1\,$R_{200}$. \cite{Sifon2016} also studied the impact of the choice of radius on velocity dispersion measurements, and found that the bias is negligible for measurements that sample beyond 0.7\,$R_{200}$ (see their Figure 4). Since our observations sample out to at least 0.7 $R_{200}$ for all clusters, and beyond $R_{200}$ for more than half the sample, we expect our velocity dispersion measurements to be unaffected by this source of bias.

We used these two sets of simulations to test our orthogonal fitting methods both with and without evolution.

\subsection{Comparison with simulations - Fitting with no evolution}
\begin{table*}
 \caption{Best fit values for the parameters in Equation~\ref{orthmodel} (slope, intercept and scatter) for the 
various models obtained from simulations without evolution. For the Millennium Gas Project we use dark matter 
(DM) and stars as the tracers for the velocity dispersion. The BAHAMAS simulation uses galaxies. The 
Millennium Gas simulations use spectroscopic like temperatures ($T_{sl}$) and the BAHAMAS simulation use 
spectroscopic temperatures ($T_{s}$). Caldwell et al. in prep present a different method for determining the $\sigma_{\rm v}- T$ relation as discussed in Section \ref{Sims}, the results of which are also shown below.}
   \begin{tabular*}{\textwidth}{c @{\extracolsep{\fill}} ccccccc}
  \hline
   Simulation & $\sigma_{\rm {tracer}}$ & T$_{model}$    & $A$        & $B$ & $S$\\  
 \hline
 Millennium Gas & DM & $T_{sl}$ & -0.011 $\pm$ 0.002 & 0.553 $\pm$ 0.008 & 0.028 $\pm$ 0.001  \\
 Millennium Gas & Stars & $T_{sl}$ & -0.034 $\pm$ 0.003 & 0.621 $\pm$ 0.010 & 0.034 $\pm$ 0.001  \\
 BAHAMAS  & Galaxies & $T_{s}$ & -0.055 $\pm$ 0.003 & 0.848 $\pm$ 0.012 & 0.055 $\pm$ 0.001  \\
  BAHAMAS (Caldwell et al., in prep) & Galaxies & T$_{s}$ & -0.133 & 0.545  &  \\
\hline
\end{tabular*}
\label{Simulationsno}
\end{table*}

\begin{figure*}
\includegraphics[width=\columnwidth]{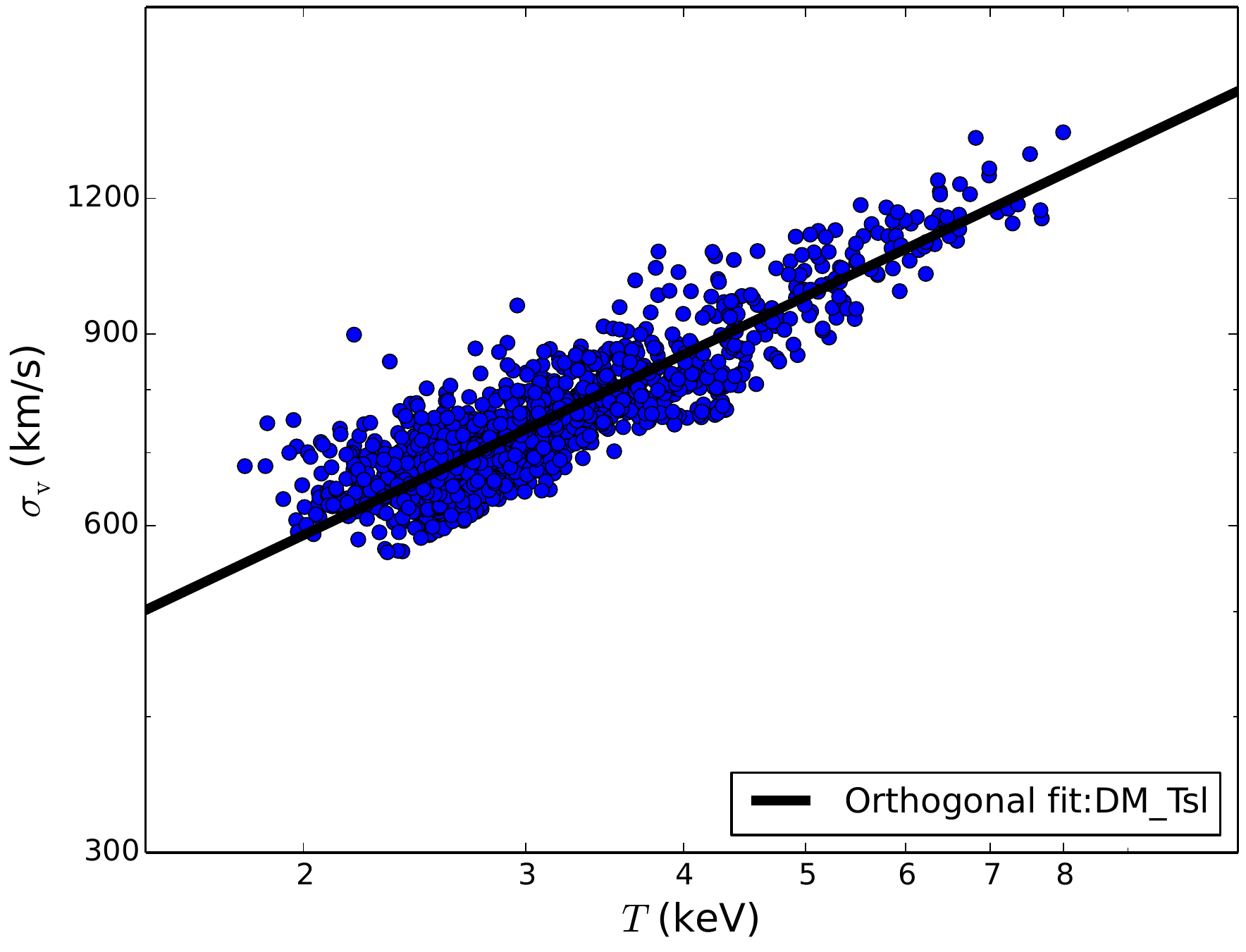}
\includegraphics[width=\columnwidth]{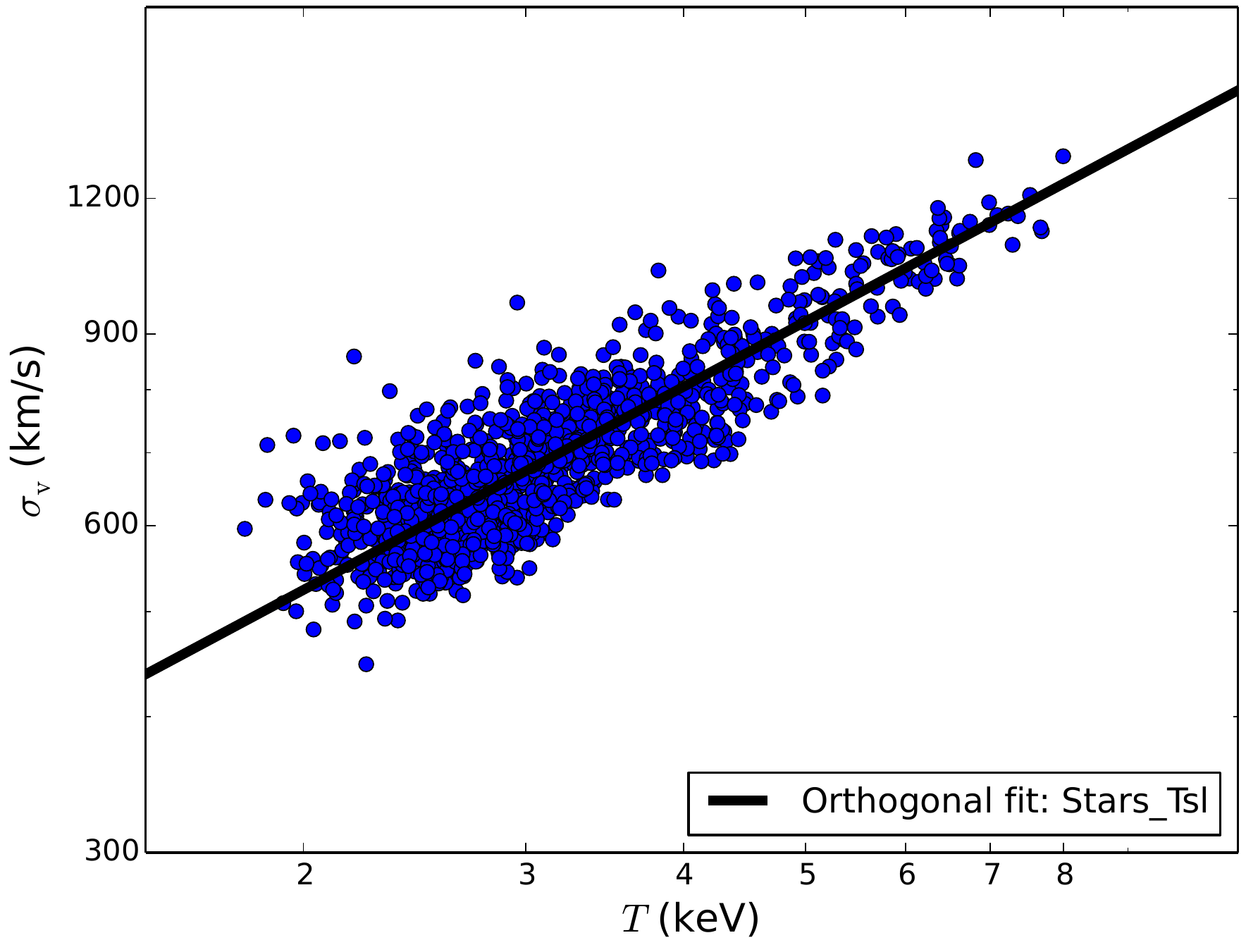}
\caption{The $\sigma_{\rm v}$--$T$ relation for the Millennium Gas Project simulations using dark 
matter and stars as proxies for the velocity dispersion. The blue dots are the data obtained from the 
simulation and the solid black line shows the fit using the orthogonal regression method. The slope is 
slightly steeper for the stars ($B$=0.62 $\pm$ 0.01) than for the dark matter ($B$=0.55 $\pm$ 0.08) but both are 
consistent with previous studies of the $\sigma_{\rm v}- T$ relation and the results obtained from our 
data.}
\label{Millennium}
\end{figure*}

The orthogonal fitting method described in Section \ref{Fitting} was applied to both sets of simulations with
$C = 0$. The parameters $A$, $B$ and $S$ for both the Millennium Gas Project and BAHAMAS simulations are shown in 
Table~\ref{Simulationsno}. The $\sigma_{\rm v}$--$T$ relation for the Millennium Gas Project with the two 
different $\sigma_{\rm v}$ proxies are shown in Fig.~\ref{Millennium}. The slope is 
slightly steeper for the stars ($B$=0.62 $\pm$ 0.01) than for the dark matter ($B$=0.55 $\pm$ 0.08) but both are 
consistent with previous studies of the $\sigma_{\rm v}- T$ relation and the results obtained from our 
data.

\begin{figure}
\includegraphics[width=\columnwidth]{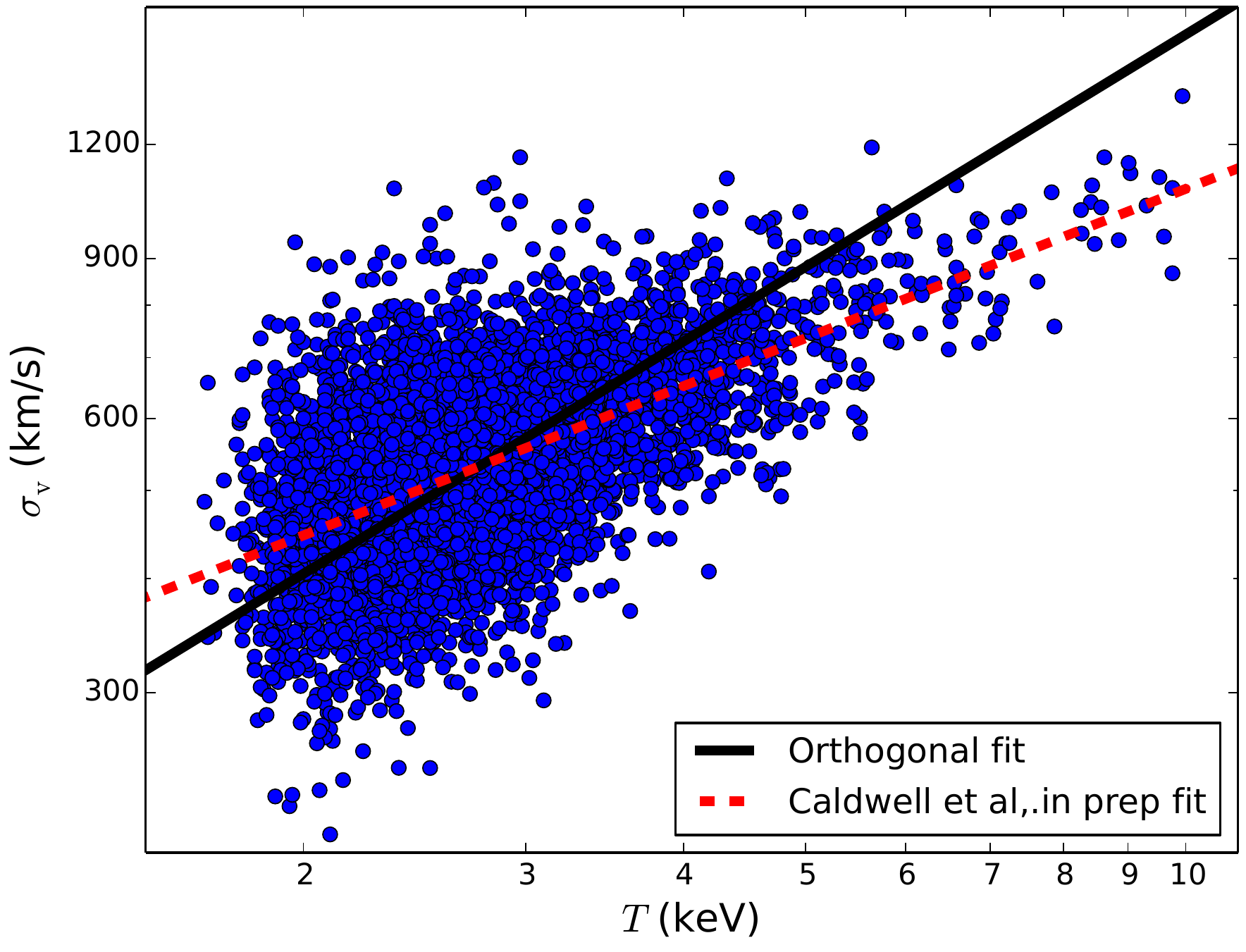}
 \caption{
Plot showing the $\sigma_{\rm v}$--$T$ relation for the BAHAMAS simulation data. The blue 
circles represent the data points from the simulation. The solid line is the fit obtained using the orthogonal 
 method and the dashed line is the fit obtained using the method described in Section \ref{Sims} and performed by Caldwell et al., in prep. From this it can be seen that the orthogonal fit over-estimates the velocity dispersion at $T_{\rm 
 X}>$5 keV (see Section \ref{Sims}
%
}
\label{Caroline}
\end{figure}

Fig.~\ref{Caroline} shows that the orthogonal fit to the full BAHAMAS sample systematically overestimates the 
average velocity dispersion at $T > 5$\,keV. This may be due, in part, to the model not being a complete 
description of the data, as the scatter appears to vary with temperature. This is not captured in our orthogonal regression model (Equation \ref{orthmodel}), i.e., $S$ is assumed to be constant with both $T$ and $z$. A comparison was made to a fit performed by Caldwell et al. (in prep.) to the BAHAMAS data. In this method the $\sigma^2 - kT$ relation was derived by first parametrically determining the mean functions and redshift evolution of velocity dispersion and temperature, separately, with respect to mass ($M_{500}$ critical). The velocity dispersion measurements are averaged in 0.25 log$_{10}(M_{500c})$ mass bins, to avoid biases from high cluster counts and the large scatter seen at low $T$ (Fig.~\ref{Caroline}). These mean values of velocity dispersion and mass are fit with a power law, and the temperature relation is derived with the same method. This was converted into the $\sigma_{v}-T$ relation (with slope $B=0.545$) that is plotted in Fig. \ref{Caroline}. This method provides a better fit to BAHAMAS data over the full temperature range than the orthogonal method. We use the slope obtained using the Caldwell et al. method for further studies and comparisons. We examine the effect of bias in the recovered slope on our results in Section \ref{SlopeBias}. 

We note that there is no single method which gives the underlying `true' scaling relation in the presence of errors on both variables and intrinsic 
scatter: the recovered slope and  normalisation depend upon the details of the method used. The fitting of a scaling relation is also affected by the selection processes used when determining your sample and for this study this has not been corrected for. Since our clusters are selected on X-ray luminosity rather than temperature or velocity dispersion we believe that our values are not biased and therefore selection effects will not have as big an impact on our results.

\subsection{The effect of biased slope measurements on the evolution of the normalisation}
\label{SlopeBias}

Having seen, using the BAHAMAS simulation, that the slope recovered using the orthogonal regression method may be biased high, we now discuss the potential impact of a biased slope measurement on our conclusions regarding the observed
cluster sample in Section~\ref{Discussion_previous}. 
\begin{figure}
 \centering

 \includegraphics[width=1.1\linewidth,keepaspectratio=true]{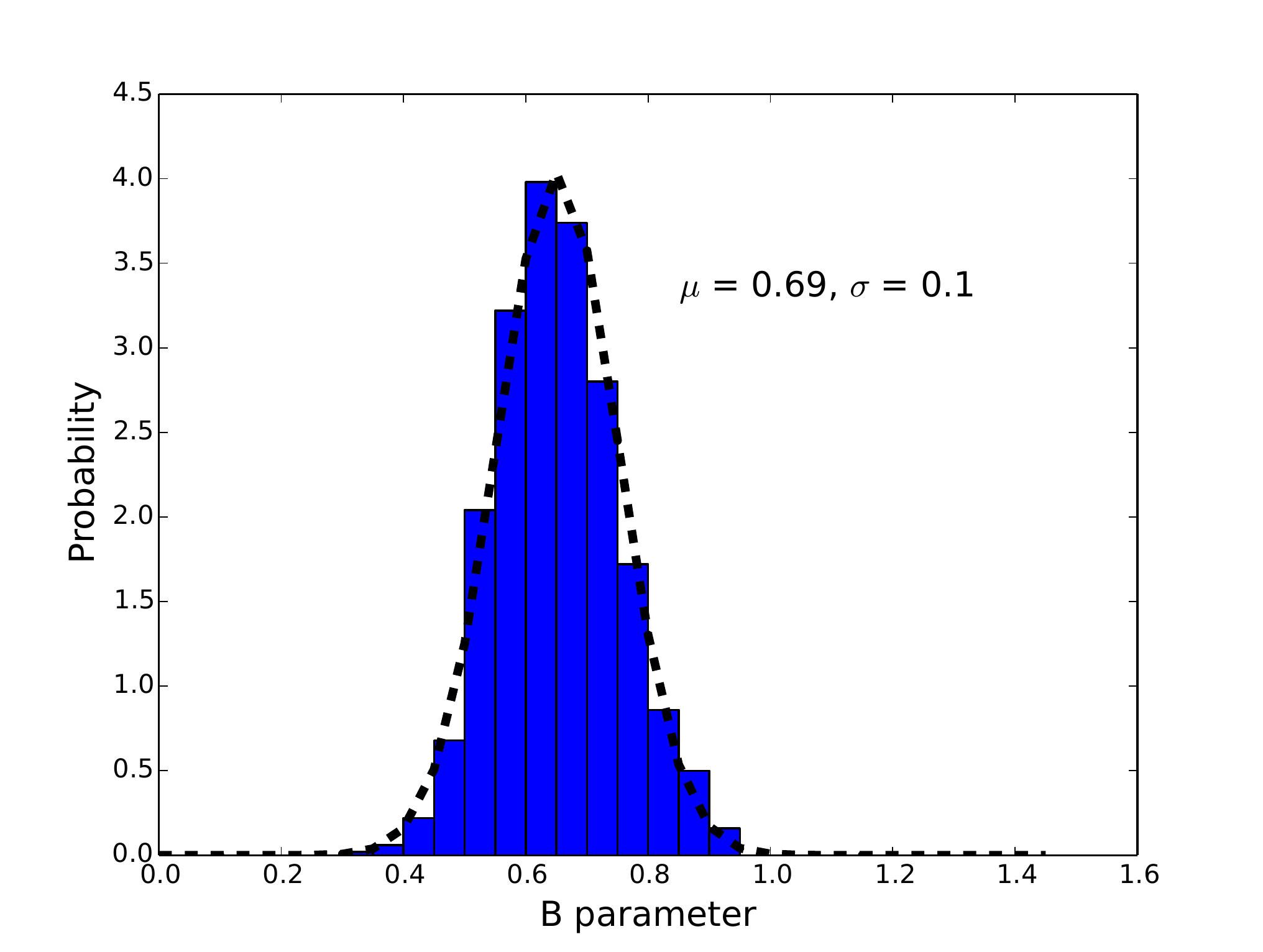}
 \caption{This histogram shows the probability of getting a specific value for the slope of the best-fit $\sigma-T$ relation given a set T distribution. We chose various subsamples from the BAHAMAS simulation which had the same T distribution as our sample and calculated the slope for each. The mean slope obtained is B=0.69$\pm$ 0.13, which is within 2 sigma of the value obtained from Caldwell et al, in prep, so there is a slight bias from the distribution of the sample.}
 \label{SlopeHist}
\end{figure}

To investigate this, we generated 1000 mock samples (each containing
38 clusters) from the BAHAMAS simulation with the same temperature distribution as the observed sample, and applied the 
orthogonal regression method. Fig.~\ref{SlopeHist} shows the 
distribution of recoverd slope values. The average is $B=0.69 \pm 0.13$, which is 2$\sigma$ higher than the slope 
obtained from the fit performed by Caldwell et al. in prep (Section~\ref{Sims}). Therefore, \textit{if}
the BAHAMAS sample is representative of the real cluster population, then we would conclude that the slope 
we have measured for the observed cluster sample is biased high.

\begin{figure}
\includegraphics[width=\columnwidth]{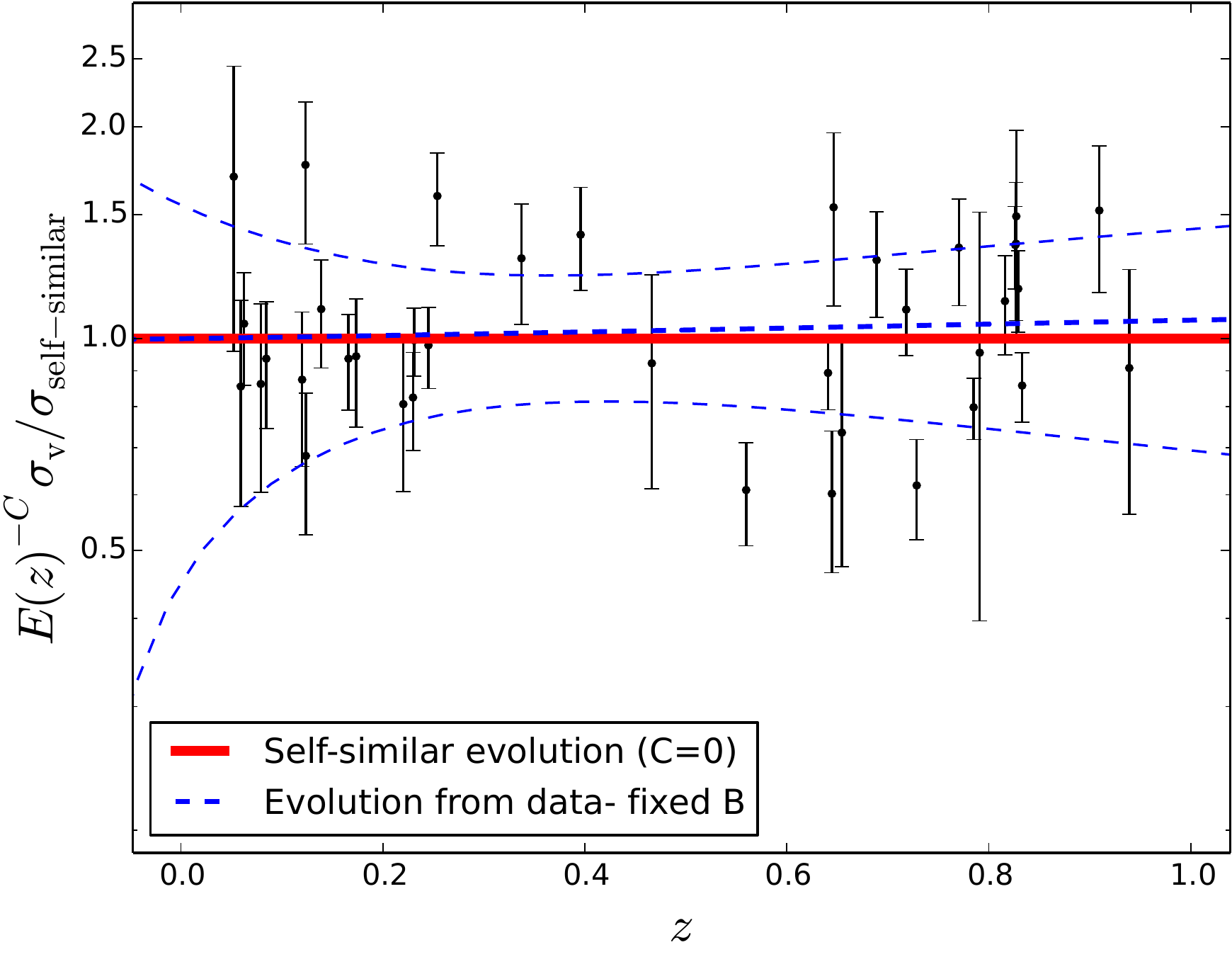}
\caption{Plot showing the evolution of the normalisation of the $\sigma_{\rm v}- T$ relation, with 
$B=0.545$, obtained for the observed cluster sample with the 95$\%$ confidence intervals in the dashed lines, as compared to the 
self similar relation which predicts no evolution shown as the solid line. The black points show the measurements for the clusters in our sample.}
\label{fixedB}
\end{figure}

To check if a biased slope estimate affects our conclusions regarding the lack of significant evidence
for evolution of the normalisation of the relation (Section~\ref{EvN}), we fixed the slope to $B = 0.545$ 
and re-ran the orthogonal fit for the observed cluster sample. We found $C=0.15 \pm 0.28$, which is consistent
with no evolution (Fig.~\ref{fixedB}). Therefore, even if the slope value of $B = 0.86 \pm 0.14$ that we 
measured was biased high for any reason, this does not affect our conclusion that we do not see significant 
evidence in favour of evolution.

\subsection{Comparison with simulations - Fitting with evolution}

\begin{figure}
\includegraphics[width=\columnwidth]{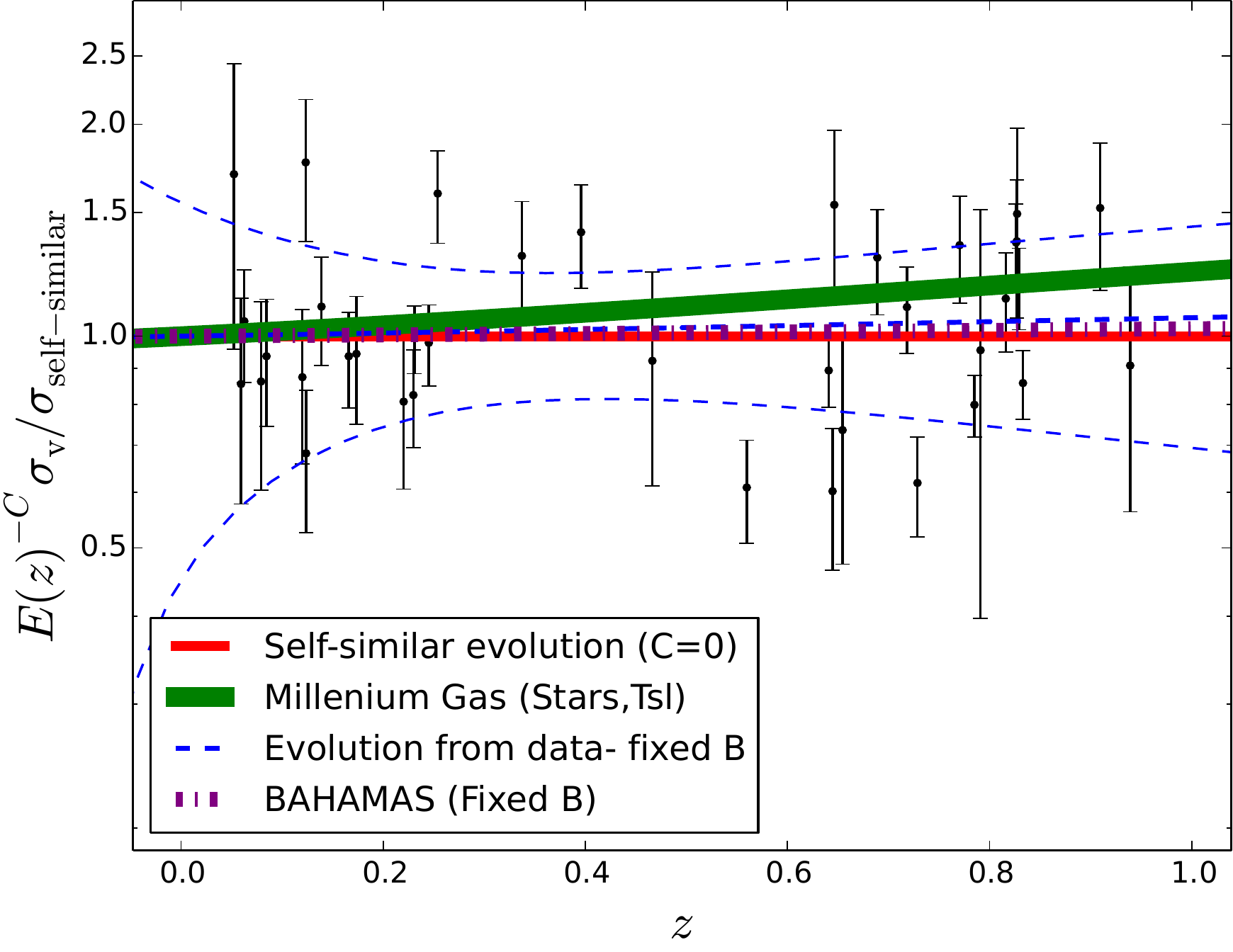}
\caption{We compared the evolution of the normalisation of $\sigma_{\rm v}- T$ relation of the 
Millennium Gas and BAHAMAS simulations with the self similar relation and that found from our data using a 
fixed slope. The solid line shows the line representing the self-similar relation i.e. $C=0$, the dot-dashed 
line represents the BAHAMAS simulation results with a fixed $B=0.545$ to avoid bias and 
the vertically dashed line represents the result from the Millennium Gas simulation. The blue dashed line and 
black points are our orthogonal fit and observed sample respectively.}
\label{ev_sim}
\end{figure}

\begin{table*}

 \caption{Best fit values for the parameters in Equation~\ref{orthmodel} (slope, intercept, scatter and evolution) for the 
various models obtained from simulations. All abbreviations are as in Table \ref{Simulationsno}.}
\centering
  
   \begin{tabular*}{\textwidth}{c @{\extracolsep{\fill}} ccccccc}
  \hline
   Simulation & $\sigma_{\rm{tracer}}$ & $T_{model}$    & A        & B & S & C\\  
 \hline
 Millennium Gas & DM & $T_{sl}$ & -0.031 $\pm$ 0.002 & 0.551 $\pm$ 0.006 & 0.0220 $\pm$ 0.0010 & 0.371 $\pm$ 0.014\\
 Millennium Gas & Stars & $T_{sl}$ & -0.056 $\pm$ 0.002 & 0.619 $\pm$ 0.009 & 0.0295 $\pm$ 0.0010 & 0.397 $\pm$ 0.019 \\
 BAHAMAS (fixed B) & Galaxies & $T_{s}$ & -0.135 $\pm$ 0.002 & 0.545 & 0.0390 $\pm$ 0.0010 & 0.046 $\pm$ 0.016 \\
BAHAMAS (varying B) & Galaxies & $T_{s}$ & -0.071 $\pm$ 0.005 & 0.779 $\pm$ 0.014 & 0.0570 $\pm$ 0.0010 & -0.029 $\pm$ 0.024 \\
 
\hline

\end{tabular*}

\label{Simulations}
\end{table*}

We now investigate evolution in the normalisation of the $\sigma_{\rm v}$--$T$ relation in the 
simulations by fitting for the value of $C$, as we did for the observed sample (see Section \ref{EvN}). The 
results are shown in Table \ref{Simulations} and graphically in Fig.~\ref{ev_sim}. The BAHAMAS simulation was tested both with a slope that was allowed to vary and a fixed slope. Although, in Section ~\ref{SlopeBias} we showed that the over-estimated slope does not affect evolution, we included a fixed slope for further comparison. Both were found to be consistent with zero evolution, further proving that the biased slope does not affect evolution. However, the simulations from the Millennium Gas Project show small but 
significant positive evolution ($C = 0.273 \pm 0.013$ for $\sigma_{Stars}$ and $T_{\rm sl}$). 
To see the reason for this, we can re-write the $\sigma_{\rm v}$--$T$ relation in terms of the 
$\sigma_{\rm v}$--$M$ and $T$--$M$ relations, where $M$ is the cluster mass 
\citep[see, e.g.,][]{Maughan2014}. We define
\begin{equation}
 \sigma_{\rm v}= 10^{A_{\sigma_{\rm v} T}} \, \left(\frac{T}{5\,\rm{keV}}\right)^{B_{\sigma_{\rm v} 
T}} \, E(z)^{C_{\sigma_{\rm v} T}}, 
\end{equation}
where 
\begin{equation}
\begin{aligned}
&B_{\sigma_{\rm v} T} = B_{\sigma_{\rm v} M}/B_{TM} \,,\\
&A_{\sigma_{\rm v} T} = A_{\sigma_{\rm v} M} - A_{TM}B_{\sigma_{\rm v} T},\, \rm{and}\\
&C_{\sigma_{\rm v} T} = C_{\sigma_{\rm v} M} - C_{TM}B_{\sigma_{\rm v} T}.
\end{aligned}
\end{equation}
Here, $A$, $B$ and $C$ have the same meaning as before, and the subscripts indicate the corresponding
relation (e.g., $B_{TM}$ indicates the slope of the $T$--$M$ relation). If we set  
$C_{\sigma_{\rm v} M}=1/3$, $C_{TM}=2/3$ and $B_{\sigma_{\rm v} T}=1/2$ as predicted by the self-similar relation,
then we obtain $C_{\sigma_{\rm v} T}=0$ as expected.
We performed fits to determine the values of $C_{\sigma_{\rm v} M}$, $C_{TM}$ and $B_{\sigma_{\rm v} T}$
in the Millennium Gas simulation at $z=0$ and $z=0.5$. We found that $C_{\sigma_{\rm v} M} = 1/3$
when using either $\sigma_{\rm Stars}$ or $\sigma_{\rm DM}$ as the measure of $\sigma_{\rm v}$, and that
$B_{\sigma_{\rm v} T}$ varied from 0.55--0.6 (depending on whether spectroscopic-like or mass-weighted
temperature estimates were used), which is slightly higher than the self-similar value, but not by enough to 
explain the positive evolution measured in the $\sigma_{\rm v}$--$T$ relation. This leads to the conclusion
that the evolution is driven by the value of $C_{TM}$, and it was found that the measured value for the dark
matter was $C_{TM}=2/3$ as expected, but that this decreased to values between 0--0.2 for the gas. 
Therefore in the Millennium Gas simulation, the lack of redshift evolution in the $T-M$ relation drives the
positive evolution in the $\sigma_{\rm v}$--$T$ relation.

The most likely explanation for the lack of redshift evolution in the $T-M$ relation in the Millennium Gas 
simulation is the absence of radiative cooling. When both cooling and feedback are included in simulations
(as in BAHAMAS), the feedback acts as a regulation mechanism, heating the surrounding dense gas and expelling
it from the cluster core. This in turn leads to higher-entropy gas flowing inwards. In the Millennium Gas 
simulation, the feedback model heats the gas and directly increases its entropy, which is eventually distributed 
throughout the cluster. This builds up over time as more and more energy is pumped into the gas from the 
growing black holes, and has the effect of slowing down the evolution of the $T-M$ relation (compared to
the evolution expected due to the decreasing background density with redshift). This in turn leads to the
positive evolution of the $\sigma_{\rm v}$--$T$ relation. It is likely that the more sophisticated 
feedback model used in BAHAMAS, where the entropy evolution is driven by radiative cooling, is the more 
realistic of the two.



\section{Conclusions}
\label{Conclusions}

We have studied the evolution of the velocity dispersion--temperature ($\sigma_{\rm v}- T$) relation 
using a cluster sample spanning the range $0.0 < z < 1.0$ drawn from XCS. This work improves upon previous 
studies in terms of the use of a homogeneous cluster sample and the number of $z > 0.5$ clusters included.
We present new redshift and velocity dispersion measurements based on Gemini data for 12 such $z > 0.5$ XCS 
clusters.

We used an orthogonal regression method to measure the normalisation, slope and intrinsic scatter of the 
$\sigma_{\rm v}$--$T$ relation for two subsamples: 19 clusters at $z<0.5$, and 19 clusters with $z>0.5$. 
In both cases, we found the slope of the relation to be consistent with the findings of previous studies, i.e., 
slightly steeper than expected from self-similarity.
Under the assumption that the slope of the relation does not evolve with redshift, we measured the evolution 
of the normalisation of the relation using the complete sample of 38 clusters. We found this to be slightly 
negative but not significantly different from the self-similar solution 
($\sigma_{\rm v} \propto T^{0.86 \pm 0.14} E(z)^{-0.37 \pm 0.33}$). Moreover, 
a no evolution model is the preferred choice when considering the Akaike Information Criterion.

We applied the same scaling relation analysis methods to the BAHAMAS and Millennium Gas cosmological 
hydrodynamical simulations. The $\sigma_{\rm v}$--$T$ relation does not evolve in BAHAMAS,
in agreement with our findings for the observed cluster sample. However, positive evolution is seen
in the Millennium Gas simulation. The difference is most likely due to the inclusion of self-consistent 
modelling of radiative cooling in BAHAMAS, which is absent in the Millennium Gas simulation. This leads to
a very slowly evolving $T$--$M$ relation in the Millennium Gas simulation, which in turn drives the positive 
evolution of the $\sigma_{\rm v}$--$T$ relation.
While this work has improved upon previous studies in terms of the number of high redshift clusters included,
we note that the uncertainties on the scaling relation parameters are still rather large, and a combination 
of better measurements of individual cluster properties and a larger sample are required to make further 
progress. Future studies will look at implementing the Bayesian method described by \cite{Kelly2007} to account for intrinsic scatter and measurement errors and looking at possible selection effects.

\section*{Acknowledgements}

SW and MH acknowledge financial support from the National Research Foundation and SKA South 
Africa. This research has made use of the NASA/IPAC Extragalactic Database (NED) which is operated by the Jet 
Propulsion Laboratory, California Institute of Technology, under contract with the National Aeronautics and 
Space Administration. Based on observations obtained at the Gemini Observatory, which is operated by the 
Association of Universities for Research in Astronomy, Inc., under a cooperative agreement with the NSF on 
behalf of the Gemini partnership: the National Science Foundation (United States), the National Research 
Council (Canada), CONICYT (Chile), the Australian Research Council (Australia), Minist\'{e}rio da Ci\^{e}ncia, 
Tecnologia e Inova\c{c}\~{a}o (Brazil) and Ministerio de Ciencia, Tecnolog\'{i}a e Innovaci\'{o}n Productiva 
(Argentina).
PAT acknowledges support from the Science and Technology Facilities Council (grant number 
ST/L000652/1). 
JPS gratefully acknowledges support from a Hintze Research Fellowship.
STK acknowledges support from STFC (grant number ST/L000768/1).

\bibliographystyle{mnras}
\bibliography{references.bib}

\appendix

\section{Observations log}

\begin{table*}
\caption{Spectroscopic observations log. For all observations the R400 grating and the OG515 filter was used.}
  \label{obslog}
 \begin{tabular*}{\textwidth}{c @{\extracolsep{\fill}} ccccccc}
  \hline

   Cluster Name & & Mask     & Slits         & Airmass Range & Observation Date & Frames(s) & Seeing ($\prime 
\prime $) \\
 \hline
 XMMXCS J005656.6-274031.9	&	&	GS2012B--Q--011--03	&	33	&	1.22	&	
2012--10--19	&	1 x 1830	&		\\
\ditto	&	&	\ditto	&		&	1.48 -- 1.76	&	2012--10--16	&	2 x 
1830	&		\\
\ditto	&	&	\ditto	&		&	1.01 -- 1.04	&	2012--10--15	&	3 x 
1830	&	0.76 -- 0.80	\\
\ditto	&	&	GS2012B--Q--011--04	&	35	&	1.05 -- 1.35 	&	2012--11--14	
&	4 x 1830	&		\\
XMMXCS J015241.1-133855.9	&	&	GS2011B--Q--050--01	&	33	&	1.05 -- 1.21 	
&	2011--12--02	&	6 x 1830	&		\\
\ditto	&	&	GS2011B--Q--050--02	&	34	&	1.05 -- 1.65	&	2011--12--03	
&	6 x 1830	&		\\
XMMXCS J021734.7-051326.9	&	&	GS2012B--Q--011--06	&	34	&	1.14 -- 1.48	
&	2012--12--05	&	4 x 1830	&		\\
XMMXCS J025006.4-310400.8	&	&	GS2012B--Q--011--09	&	32	&	1.11 -- 1.20	
&	2012--11--24	&	2 x 1830	&		\\
\ditto	&	&	\ditto	&		&	1.01 -- 1.19	&	2012--11--21	&	4 x 
1830	&		\\
\ditto	&	&	GS2010B--Q--046--06	&	35	&	1.06 -- 1.12	&	2010--11--14	
&	2 x 1830	&	0.50 -- 0.60	\\
\ditto	&	&	\ditto	&		&	1.06 -- 1.44	&	2010--11--13	&	5 x 
1830, 1 x 762	&	1	\\
XMMXCS J030205.1-000003.6	&	&	GS2011B--Q--050--03	&	32	&	1.17	&	
2011--12--01	&	1 x 1830	&		\\
\ditto	&	&	\ditto	&		&	1.18	&	2011--11--20	&	1 x 1098	
&		\\
\ditto	&	&	\ditto	&		&	1.17 -- 1.45	&	2011--11--18	&	4 x 
1830	&		\\
\ditto	&	&	GS2011B--Q--050--04	&	32	&	1.32	&	2011--12--31	&	
1 x 1098	&		\\
\ditto	&	&	\ditto	&		&	1.23 -- 1.74	&	2011--12--30	&	4 x 
1830	&	0.85 -- 1.40	\\
\ditto	&	&	GS2011B--Q--050--05	&	33	&	1.27 -- 1.57	&	2012--01--17	
&	2 x 1830	&	0.7	\\
XMMXCS J095940.7+023113.4	&	&	GS2010B--Q--046--02	&	35	&	1.19 -- 1.23	
&	2011--01--09	&	3 x 1830	&		\\
\ditto	&	&	\ditto	&		&	1.19 -- 1.25	&	2011--01--08	&	4 x 
1830	&		\\
\ditto	&	&	GS--2012A--Q--46--01	&	35	&	1.19 -- 1.29	&	2012--03--18	
&	4 x 1830	&		\\
\ditto	&	&	\ditto	&		&	1.19 -- 1.23	&	2012--03--02	&	2 x 
1830	&		\\
\ditto	&	&	GS--2012A--Q--46--02	&	34	&	1.20 -- 1.35	&	2012--03--27	
&	3 x 1830	&		\\
\ditto	&	&	\ditto	&		&	1.23 -- 1.46	&	2012--03--23	&	3 x 
1830	&	0.8	\\
\ditto	&	&	GS--2012A--Q--46--03	&	34	&	1.21 -- 1.54	&	2012--03--22	
&	6 x 1830	&	0.65 -- 0.70	\\
XMMXCS J112349.3+052956.8	&	&	GS--2012A--Q--46--05	&	33	&	1.23 -- 1.33	
&	2012--04--22	&	5 x 1830	&		\\
\ditto	&	&	\ditto	&		&	1.47	&	2012--04--21	&	1 x 1830	
&		\\
\ditto	&	&	GS--2012A--Q--46--06	&	32	&	1.25 -- 1.65	&	2012--05--15	
&	4 x 1830	&	0.63 -- 0.76	\\
\ditto	&	&	\ditto	&		&	1.45 -- 1.66	&	2012--04--22	&	2 x 
1830	&		\\
\ditto	&	&	GS2010B--Q--046--03	&	33	&	1.26	&	2011--01--31	&	
1 x 1830	&		\\
\ditto	&	&	\ditto	&		&	1.35 -- 1.64	&	2011--01--29	&	2 x 
1525, 1 x 975	&		\\
\ditto	&	&	\ditto	&		&	1.23 -- 1.24	&	2011--01--27	&	2 x 
1830	&		\\
XMMXCS J113602.9--032943.2	&	&	GS--2012A--Q--46--07	&	36	&	1.14	&	
2012--05--24	&	1 x 1830	&		\\
\ditto	&	&	\ditto	&		&	1.12	&	2012--05--23	&	1 x 1830	
&		\\
\ditto	&	&	\ditto	&		&	1.12 -- 1.16	&	2012--05--20	&	3 x 
1830	&		\\
\ditto	&	&	\ditto	&		&	1.12	&	2012--05--19	&	1 x 1830	
&		\\
\ditto	&	&	GS--2012A--Q--46--08	&	33	&	1.48 -- 1.76	&	2012--07--15	
&	2 x 1830	&		\\
\ditto	&	&	\ditto	&		&	1.41 -- 1.80	&	2012--07--11	&	3 x 
1830	&	0.50 -- 0.70	\\
\ditto	&	&	\ditto	&		&	1.5	&	2012--07--10	&	1 x 1830	
&		\\
XMMXCS J134305.1-000056.8	&	&	GS--2012A--Q--46--10	&	36	&	1.16 -- 1.23	
&	2012--05--24	&	4 x 1830	&		\\
\ditto	&	&	\ditto	&		&	1.24	&	2012--05--21	&	1 x 1830	
&		\\
\ditto	&	&	GS--2012A--Q--46--11	&	34	&	1.25	&	2012--07--10	&	
1 x 1830	&		\\
\ditto	&	&	\ditto	&		&	1.16 -- 1.19	&	2012--07--09	&	2 x 
1830	&		\\
\ditto	&	&	\ditto	&		&	1.2	&	2012--07--06	&	1 x 1830	
&		\\
\ditto	&	&	\ditto	&		&	1.54 -- 1.84	&	2012--06--22	&	2 x 
1830	&		\\
XMMXCS J145009.3+090428.8	&	&	GN2012A--Q--070--05	&	32	&	1.02 -- 1.05	
&	2012--07--09	&	2 x 1800	&	1.15	\\
\ditto	&	&	\ditto	&		&	1.11 -- 1.62	&	2012--06--26	&	4 x 
1800	&	0.84 -- 0.98	\\
\ditto	&	&	GN2012A--Q--070--06	&	34	&	1.02 -- 1.04	&	2012--07--07	
&	2 x 1800	&		\\
\ditto	&	&	\ditto	&		&	1.09 -- 1.17	&	2012--07--06	&	2 x 
1800	&		\\
\ditto	&	&	\ditto	&		&	1.48 -- 1.79	&	2012--06--27	&	2 x 
1800	&		\\
\ditto	&	&	GN2012A--Q--070--07	&	33	&	1.22 -- 1.59	&	2012--07--22	
&	3 x 1800	&		\\
\ditto	&	&	\ditto	&		&	1.04 -- 1.16	&	2012--07--08	&	3 x 
1800	&	1	\\
XMMXCS J215221.0--273022.6	&	&	GS2010B--Q--046--04	&	36	&	1.14 -- 1.24	
&	2010--11--12	&	2 x 1830	&		\\
\ditto	&	&	\ditto	&		&	1.02 -- 1.21	&	2010--09--14	&	4 x 
1830	&		\\
\ditto	&	&	GS2011B--Q--050--06	&	34	&	1.07 -- 1.15	&	2011--10--05	
&	2 x 1830	&		\\
\ditto	&	&	\ditto	&		&	1.12 -- 1.56	&	2011--09--18	&	4 x 
1830	&	0.60 -- 1.00	\\
\ditto	&	&	GS2011B--Q--050--07	&	34	&	1.00 -- 1.10	&	2011--10--24	
&	4 x 1830	&		\\
\ditto	&	&	\ditto	&		&	1.05 -- 1.12	&	2011--10--16	&	2 x 
1830	&		\\
XMMXCS J230247.7+084355.9	&	&	GN2012A--Q--070--10	&	34	&	1.37	&	
2012--08--08	&	1 x 1800	&	0.60 -- 0.68	\\
\ditto	&	&	\ditto	&		&	1.02 -- 1.11	&	2012--07--30	&	5 x 
1800	&	0.43 -- 0.86	\\
\ditto	&	&	GN2012A--Q--070--11	&	33	&	1.18 -- 1.31	&	2012--08--13	
&	2 x 1800	&		\\
\ditto	&	&	\ditto	&		&	1.02 -- 1.08	&	2012--08--09	&	3 x 
1800	&	0.60 -- 0.68	\\
\ditto	&	&	\ditto	&		&	1.19	&	2012--08--08	&	1 x 1800	
&	1	\\
\hline
\end{tabular*}
\end{table*}

\clearpage

\section{Redshift catalogue}
\label{RedshiftCatalogue}
Tables of galaxy redshifts measured in each cluster to appear in the online version of the article.
%
%
%
%
%
%
%
%
%
%
%

\begin{table*}
\centering

\begin{tabular*}{\textwidth}{l @{\extracolsep{\fill}} lllll}
  \hline
   Name     & RA(J2000)         & Dec(J2000)  & z & Reference\\
 \hline
GALEXASC J000046.26-251058.3	&	00$^\mathrm{h}$00$^\mathrm{m}$46.26$^\mathrm{s}$	&	-25$^\circ$10$^\prime$59.9$^{\prime \prime}$	&	0.0821	&	\citet{Colless2003}	\\
GALEXASC J000057.30-250623.9	&	00$^\mathrm{h}$00$^\mathrm{m}$57.23$^\mathrm{s}$	&	-25$^\circ$06$^\prime$25.9$^{\prime \prime}$	&	0.0821	&	\citet{Colless2003}	\\
2dFGRS S134Z161	&	23$^\mathrm{h}$59$^\mathrm{m}$32.60$^\mathrm{s}$	&	-25$^\circ$13$^\prime$02.2$^{\prime \prime}$	&	0.0823	&	\citet{Colless2003}	\\
GALEXASC J235957.50-250120.7	&	23$^\mathrm{h}$59$^\mathrm{m}$57.51$^\mathrm{s}$	&	-25$^\circ$01$^\prime$20.5$^{\prime \prime}$	&	0.0823	&	\citet{Colless2003}	\\
2dFGRS S134Z139	&	00$^\mathrm{h}$00$^\mathrm{m}$25.25$^\mathrm{s}$	&	-25$^\circ$20$^\prime$48.2$^{\prime \prime}$	&	0.0834	&	\citet{Colless2003}	\\
2MASX J00002408-2510371	&	00$^\mathrm{h}$00$^\mathrm{m}$24.10$^\mathrm{s}$	&	-25$^\circ$10$^\prime$38.0$^{\prime \prime}$	&	0.0837	&	\citet{Colless2003}	\\
2MASX J00004212-2512080	&	00$^\mathrm{h}$00$^\mathrm{m}$42.12$^\mathrm{s}$	&	-25$^\circ$12$^\prime$08.1$^{\prime \prime}$	&	0.0837	&	\citet{Jones2009}	\\
\hline

\end{tabular*}
\caption{XMMXCSJ000013.9-251052.1: Column 1 gives the name of the galaxy, column 2 and 3 give the right ascension and declination respectively, and column 4 gives the redshift of the galaxy. Column 5 gives the reference from which the redshift was taken.}
\label{Table1}
\end{table*}

\begin{table*}
\centering

   \begin{tabular*}{\textwidth}{l @{\extracolsep{\fill}} lllll}
   \hline
   Name     & RA(J2000)         & Dec(J2000)  & z & Reference\\
 \hline
GALEXMSC J003407.05-432228.7	&	00$^\mathrm{h}$34$^\mathrm{m}$06.97$^\mathrm{s}$	&	-43$^\circ$22$^\prime$30.2$^{\prime \prime}$	&	0.3877	&	\citet{Feruglio2008}	\\
SWIRE3 J003407.30-432152.8	&	00$^\mathrm{h}$34$^\mathrm{m}$07.29$^\mathrm{s}$	&	-43$^\circ$21$^\prime$52.9$^{\prime \prime}$	&	0.3880	&	\citet{Sacchi2009}	\\
SWIRE3 J003400.83-432122.5	&	00$^\mathrm{h}$34$^\mathrm{m}$00.77$^\mathrm{s}$	&	-43$^\circ$21$^\prime$22.5$^{\prime \prime}$	&	0.3890	&	\citet{Sacchi2009}	\\
SWIRE3 J003408.87-432244.6	&	00$^\mathrm{h}$34$^\mathrm{m}$08.87$^\mathrm{s}$	&	-43$^\circ$22$^\prime$44.7$^{\prime \prime}$	&	0.3890	&	\citet{Sacchi2009}	\\
SWIRE3 J003405.07-431742.3	&	00$^\mathrm{h}$34$^\mathrm{m}$05.06$^\mathrm{s}$	&	-43$^\circ$17$^\prime$42.3$^{\prime \prime}$	&	0.3910	&	\citet{Sacchi2009}	\\
IRAC 236852	&	00$^\mathrm{h}$34$^\mathrm{m}$27.40$^\mathrm{s}$	&	-43$^\circ$18$^\prime$59.6$^{\prime \prime}$	&	0.3925	&	\citet{Feruglio2008}	\\
SWIRE3 J003425.16-431912.1	&	00$^\mathrm{h}$34$^\mathrm{m}$25.15$^\mathrm{s}$	&	-43$^\circ$19$^\prime$12.2$^{\prime \prime}$	&	0.3930	&	\citet{Sacchi2009}	\\
\hline

\end{tabular*}
\caption{XMMXCSJ003430.1-431905.6: All columns are as explained in Table \ref{Table1}.}

\end{table*}

\begin{table*}
\centering

   \begin{tabular*}{\textwidth}{l @{\extracolsep{\fill}} lllll}
  \hline
   Name     & RA(J2000)         & Dec(J2000)  & z & Reference\\
 \hline
Cl 0053-37:[CHM98] 12	&	00$^\mathrm{h}$55$^\mathrm{m}$55.56$^\mathrm{s}$	&	-37$^{\circ}$32$^{\prime}$19.7$^{\prime \prime}$	&	0.1613	&	\citet{Guzzo2009}	\\
Cl 0053-37:[CHM98] 14	&	00$^\mathrm{h}$55$^\mathrm{m}$57.74$^\mathrm{s}$	&	-37$^{\circ}$32$^{\prime}$22.6$^{\prime \prime}$	&	0.1613	&	\citet{Guzzo2009}	\\
Cl 0053-37:[CHM98] 08	&	00$^\mathrm{h}$55$^\mathrm{m}$58.90$^\mathrm{s}$	&	-37$^{\circ}$33$^{\prime}$00.7$^{\prime \prime}$	&	0.1617	&	\citet{Cappi1998}	\\
(GSB2009) J005552.68-373143.0	&	00$^\mathrm{h}$55$^\mathrm{m}$52.68$^\mathrm{s}$	&	-37$^{\circ}$31$^{\prime}$43.0$^{\prime \prime}$	&	0.1626	&	\citet{Guzzo2009}	\\
2MASX J00555916-3732186	&	00$^\mathrm{h}$55$^\mathrm{m}$59.11$^\mathrm{s}$	&	-37$^{\circ}$32$^{\prime}$18.2$^{\prime \prime}$	&	0.1631	&	\citet{Cappi1998}	\\
APMUKS(BJ) B005339.96-374945.2	&	00$^\mathrm{h}$56$^\mathrm{m}$01.56$^\mathrm{s}$	&	-37$^{\circ}$33$^{\prime}$32.0$^{\prime \prime}$	&	0.1633	&	\citet{Guzzo2009}	\\
Cl 0053-37:[CHM98] 15	&	00$^\mathrm{h}$55$^\mathrm{m}$59.32$^\mathrm{s}$	&	-37$^{\circ}$32$^{\prime}$35.2$^{\prime \prime}$	&	0.1636	&	\citet{Cappi1998}	\\
\hline

\end{tabular*}
\caption{XMMXCSJ005603.0-373248.0: All columns are as explained in Table \ref{Table1}.}

\end{table*}

\begin{table*}
\centering

   \begin{tabular*}{\textwidth}{l @{\extracolsep{\fill}} lllll}
  \hline
   Name     & RA(J2000)         & Dec(J2000)  & z & Reference\\
 \hline
2MASX J01530967+0102002	&	01$^\mathrm{h}$53$^\mathrm{m}$09.69$^\mathrm{s}$	&	+01$^\circ$02$^\prime$00.3$^{\prime \prime}$	&	0.0579	&	\citet{Berlind2006}	\\
2MASX J01531414+0103443	&	01$^\mathrm{h}$53$^\mathrm{m}$14.16$^\mathrm{s}$	&	+01$^\circ$03$^\prime$44.4$^{\prime \prime}$	&	0.0579	&	\citet{SDSS2003}	\\
2MASX J01531014+0101302	&	01$^\mathrm{h}$53$^\mathrm{m}$10.13$^\mathrm{s}$	&	+01$^\circ$01$^\prime$30.5$^{\prime \prime}$	&	0.0586	&	\citet{SDSS2003}	\\
2MASX J01530528+0059092	&	01$^\mathrm{h}$53$^\mathrm{m}$05.29$^\mathrm{s}$	&	+00$^\circ$59$^\prime$09.2$^{\prime \prime}$	&	0.0588	&	\citet{SDSS2002}	\\
2MASX J01531034+0105562	&	01$^\mathrm{h}$53$^\mathrm{m}$10.37$^\mathrm{s}$	&	+01$^\circ$05$^\prime$56.7$^{\prime \prime}$	&	0.0590	&	\citet{SDSS2003}	\\
2MASX J01530827+0101502	&	01$^\mathrm{h}$53$^\mathrm{m}$08.29$^\mathrm{s}$	&	+01$^\circ$01$^\prime$50.7$^{\prime \prime}$	&	0.0594	&	\citet{Berlind2006}	\\
SDSS J015251.27+010705.4	&	01$^\mathrm{h}$52$^\mathrm{m}$51.28$^\mathrm{s}$	&	+01$^\circ$07$^\prime$05.4$^{\prime \prime}$	&	0.0595	&	\citet{SDSS2002}	\\
\hline

\end{tabular*}
\caption{XMMXCSJ015315.0+010214.2: All columns are as explained in Table \ref{Table1}.}

\end{table*}

\begin{table*}
\centering

   \begin{tabular*}{\textwidth}{l @{\extracolsep{\fill}} lllll}
  \hline
   Name     & RA(J2000)         & Dec(J2000)  & z & Reference\\
 \hline
(MLF2006) J072049.93+710856.6	&	07$^\mathrm{h}$20$^\mathrm{m}$49.93$^\mathrm{s}$	&	+71$^\circ$08$^{\prime}$56.6$^{\prime \prime}$	&	0.2270	&	\citet{Mulchaey2006}	\\
1RXS J072049.0+710724	&	07$^\mathrm{h}$20$^\mathrm{m}$47.35$^\mathrm{s}$	&	+71$^\circ$07$^{\prime}$15.4$^{\prime \prime}$	&	0.2274	&	\citet{Jeltema2007}	\\
(MLF2006) J072103.45+710906.0	&	07$^\mathrm{h}$21$^\mathrm{m}$03.45$^\mathrm{s}$	&	+71$^\circ$09$^{\prime}$06.0$^{\prime \prime}$	&	0.2274	&	\citet{Mulchaey2006}	\\
(MLF2006) J072044.89+710845.5	&	07$^\mathrm{h}$20$^\mathrm{m}$44.89$^\mathrm{s}$	&	+71$^\circ$08$^{\prime}$45.5$^{\prime \prime}$	&	0.2275	&	\citet{Mulchaey2006}	\\
(MLF2006) J072039.75+710950.8	&	07$^\mathrm{h}$20$^\mathrm{m}$39.75$^\mathrm{s}$	&	+71$^\circ$09$^{\prime}$50.8$^{\prime \prime}$	&	0.2277	&	\citet{Mulchaey2006}	\\
(MLF2006) J072046.96+710922.8	&	07$^\mathrm{h}$20$^\mathrm{m}$46.96$^\mathrm{s}$	&	+71$^\circ$09$^{\prime}$22.8$^{\prime \prime}$	&	0.2285	&	\citet{Jeltema2007}	\\
(MLF2006) J072059.93+710932.2	&	07$^\mathrm{h}$20$^\mathrm{m}$59.93$^\mathrm{s}$	&	+71$^\circ$09$^{\prime}$32.2$^{\prime \prime}$	&	0.2288	&	\citet{Mulchaey2006}	\\
\hline

\end{tabular*}
\caption{XMMXCSJ072054.3+710900.5: All columns are as explained in Table \ref{Table1}.}

\end{table*}

\begin{table*}
\centering

   \begin{tabular*}{\textwidth}{l @{\extracolsep{\fill}} lllll}
  \hline
   Name     & RA(J2000)         & Dec(J2000)  & z & Reference\\
 \hline

(VMF98) 049:[BBS2002] 021	&	08$^\mathrm{h}$19$^\mathrm{m}$42.12$^\mathrm{s}$	&	+70$^\circ$54$^\prime$32.1$^{\prime \prime}$	&	0.2272	&	\citet{Balogh2002}	\\
(VMF98) 049:[BBS2002] 040	&	08$^\mathrm{h}$19$^\mathrm{m}$11.99$^\mathrm{s}$	&	+70$^\circ$55$^\prime$43.8$^{\prime \prime}$	&	0.2274	&	\citet{Balogh2002}	\\
(VMF98) 049:[BBS2002] 027	&	08$^\mathrm{h}$19$^\mathrm{m}$01.38$^\mathrm{s}$	&	+70$^\circ$54$^\prime$51.6$^{\prime \prime}$	&	0.2277	&	\citet{Balogh2002}	\\
(VMF98) 049:[BBS2002] 002	&	08$^\mathrm{h}$18$^\mathrm{m}$48.72$^\mathrm{s}$	&	+70$^\circ$52$^\prime$55.8$^{\prime \prime}$	&	0.2278	&	\citet{Balogh2002}	\\
(VMF98) 049:[BBS2002] 026	&	08$^\mathrm{h}$19$^\mathrm{m}$22.86$^\mathrm{s}$	&	+70$^\circ$54$^\prime$44.9$^{\prime \prime}$	&	0.2280	&	\citet{Balogh2002}	\\
(VMF98) 049:[BBS2002] 033	&	08$^\mathrm{h}$19$^\mathrm{m}$17.75$^\mathrm{s}$	&	+70$^\circ$55$^\prime$01.0$^{\prime \prime}$	&	0.2289	&	\citet{Balogh2002}	\\
(VMF98) 049:[BBS2002] 013	&	08$^\mathrm{h}$19$^\mathrm{m}$46.92$^\mathrm{s}$	&	+70$^\circ$53$^\prime$53.9$^{\prime \prime}$	&	0.2294	&	\citet{Balogh2002}	\\
\hline

\end{tabular*}
\caption{XMMXCSJ081918.6+705457.5: All columns are as explained in Table \ref{Table1}.}

\end{table*}

\begin{table*}
\centering

   \begin{tabular*}{\textwidth}{l @{\extracolsep{\fill}} lllll}
  \hline
   Name     & RA(J2000)         & Dec(J2000)  & z & Reference\\
 \hline
SDSS J094400.28+164011.1	&	09$^\mathrm{h}$44$^\mathrm{m}$00.29$^\mathrm{s}$	&	+16$^\circ$40$^\prime$11.2$^{\prime \prime}$	&	0.2487	&	\citet{Verdugo2008}	\\
SDSS J094334.06+164035.7	&	09$^\mathrm{h}$43$^\mathrm{m}$34.06$^\mathrm{s}$	&	+16$^\circ$40$^\prime$35.7$^{\prime \prime}$	&	0.2500	&	\citet{Verdugo2008}	\\
SDSS J094353.35+163958.8	&	09$^\mathrm{h}$43$^\mathrm{m}$53.35$^\mathrm{s}$	&	+16$^\circ$39$^\prime$58.8$^{\prime \prime}$	&	0.2511	&	\citet{Verdugo2008}	\\
SDSS J094345.51+164130.4	&	09$^\mathrm{h}$43$^\mathrm{m}$45.52$^\mathrm{s}$	&	+16$^\circ$41$^\prime$30.5$^{\prime \prime}$	&	0.2514	&	\citet{Verdugo2008}	\\
SDSS J094353.48+164022.7	&	09$^\mathrm{h}$43$^\mathrm{m}$53.49$^\mathrm{s}$	&	+16$^\circ$40$^\prime$22.7$^{\prime \prime}$	&	0.2516	&	\citet{Verdugo2008}	\\
SDSS J094358.03+164116.5	&	09$^\mathrm{h}$43$^\mathrm{m}$58.03$^\mathrm{s}$	&	+16$^\circ$41$^\prime$16.6$^{\prime \prime}$	&	0.2520	&	\citet{Verdugo2008}	\\
2MASX J09435839+1641091	&	09$^\mathrm{h}$43$^\mathrm{m}$58.33$^\mathrm{s}$	&	+16$^\circ$41$^\prime$09.1$^{\prime \prime}$	&	0.2527	&	\citet{Verdugo2008}	\\
\hline

\end{tabular*}
\caption{XMMXCSJ094358.2+164120.7: All columns are as explained in Table \ref{Table1}.}

\end{table*}

\begin{table*}
\centering

   \begin{tabular*}{\textwidth}{l @{\extracolsep{\fill}} lllll}
  \hline
   Name     & RA(J2000)         & Dec(J2000)  & z & Reference\\
 \hline
SDSS J100027.86+251750.9	&	10$^\mathrm{h}$00$^\mathrm{m}$27.87$^\mathrm{s}$	&	+25$^\circ$17$^\prime$51.0$^{\prime \prime}$	&	0.0495	&	\citet{LeFevre1995}	\\
2MASX J09590595+2512006	&	09$^\mathrm{h}$59$^\mathrm{m}$05.97$^\mathrm{s}$	&	+25$^\circ$12$^\prime$01.0$^{\prime \prime}$	&	0.0511	&	\citet{Adelman2008}	\\
2MASX J09585969+2518276	&	09$^\mathrm{h}$58$^\mathrm{m}$59.70$^\mathrm{s}$	&	+25$^\circ$18$^\prime$27.8$^{\prime \prime}$	&	0.0518	&	\citet{Adelman2008}	\\
SDSS J095848.25+251116.9	&	09$^\mathrm{h}$58$^\mathrm{m}$48.25$^\mathrm{s}$	&	+25$^\circ$11$^\prime$16.9$^{\prime \prime}$	&	0.0518	&	\citet{Adelman2008}	\\
2MASX J09594076+2459301	&	09$^\mathrm{h}$59$^\mathrm{m}$40.76$^\mathrm{s}$	&	+24$^\circ$59$^\prime$30.5$^{\prime \prime}$	&	0.0519	&	\citet{Adelman2008}	\\
2MASX J10001346+2450242	&	10$^\mathrm{h}$00$^\mathrm{m}$13.49$^\mathrm{s}$	&	+24$^\circ$50$^\prime$24.1$^{\prime \prime}$	&	0.0519	&	\citet{Adelman2008}	\\
UGC 05361	&	09$^\mathrm{h}$59$^\mathrm{m}$00.86$^\mathrm{s}$	&	+25$^\circ$12$^\prime$08.2$^{\prime \prime}$	&	0.0519	&	
\citet{Huchra2012}	\\
\hline

\end{tabular*}
\caption{XMMXCSJ095957.6+251629.0: All columns are as explained in Table \ref{Table1}.}

\end{table*}

\begin{table*}
\centering

   \begin{tabular*}{\textwidth}{l @{\extracolsep{\fill}} lllll}
  \hline
   Name     & RA(J2000)         & Dec(J2000)  & z & Reference\\
 \hline
SDSS J100052.66+013941.1	&	10$^\mathrm{h}$00$^\mathrm{m}$52.67$^\mathrm{s}$	&	+01$^\circ$39$^\prime$41.2$^{\prime \prime}$	&	0.2166	&	\citet{Lilly2007}	\\
zCOSMOS 805589	&	10$^\mathrm{h}$00$^\mathrm{m}$33.07$^\mathrm{s}$	&	+01$^\circ$41$^\prime$04.8$^{\prime \prime}$	&	0.2179	&	\citet{Lilly2007}	\\
SDSS J100057.84+013637.1	&	10$^\mathrm{h}$00$^\mathrm{m}$57.84$^\mathrm{s}$	&	+01$^\circ$36$^\prime$37.2$^{\prime \prime}$	&	0.2190	&	\citet{Lilly2007}	\\
2MASX J10004852+0139137	&	10$^\mathrm{h}$00$^\mathrm{m}$48.54$^\mathrm{s}$	&	+01$^\circ$39$^\prime$13.9$^{\prime \prime}$	&	0.2190	&	\citet{Colless2003b}	\\
SDSS J100037.77+014013.9	&	10$^\mathrm{h}$00$^\mathrm{m}$37.77$^\mathrm{s}$	&	+01$^\circ$40$^\prime$14.0$^{\prime \prime}$	&	0.2194	&	\citet{Lilly2007}	\\
SDSS J100057.85+013752.1	&	10$^\mathrm{h}$00$^\mathrm{m}$57.85$^\mathrm{s}$	&	+01$^\circ$37$^\prime$52.2$^{\prime \prime}$	&	0.2195	&	\citet{Lilly2007}	\\
SDSS J100045.29+013847.4	&	10$^\mathrm{h}$00$^\mathrm{m}$45.30$^\mathrm{s}$	&	+01$^\circ$38$^\prime$47.4$^{\prime \prime}$	&	0.2205	&	\citet{SDSS2004}	\\
\hline

\end{tabular*}
\caption{
XMMXCSJ100047.4+013926.9: All columns are as explained in Table \ref{Table1}.}

\end{table*}

\begin{table*}
\centering

   \begin{tabular*}{\textwidth}{l @{\extracolsep{\fill}} lllll}
  \hline
   Name     & RA(J2000)         & Dec(J2000)  & z & Reference\\
 \hline
SDSS J100125.41+023145.1	&	10$^\mathrm{h}$01$^\mathrm{m}$25.42$^\mathrm{s}$	&	+02$^\circ$31$^\prime$45.2$^{\prime \prime}$	&	0.1210	&	\cite{SDSS2004}	\\
SDSS J100131.47+022556.7	&	10$^\mathrm{h}$01$^\mathrm{m}$31.48$^\mathrm{s}$	&	+02$^\circ$25$^\prime$56.7$^{\prime \prime}$	&	0.1211	&	\cite{Trump2007}	\\
SDSS J100157.95+022746.1	&	10$^\mathrm{h}$01$^\mathrm{m}$57.96$^\mathrm{s}$	&	+02$^\circ$27$^\prime$46.1$^{\prime \prime}$	&	0.1213	&	\cite{Lilly2007}	\\
SDSS J100120.63+022601.5	&	10$^\mathrm{h}$01$^\mathrm{m}$20.63$^\mathrm{s}$	&	+02$^\circ$26$^\prime$01.6$^{\prime \prime}$	&	0.1216	&	\cite{Lilly2007}	\\
GALEXMSC J100120.08+022019.9	&	10$^\mathrm{h}$01$^\mathrm{m}$20.07$^\mathrm{s}$	&	+02$^\circ$20$^\prime$20.0$^{\prime \prime}$	&	0.1224	&	\cite{Lilly2007}	\\
SDSS J100131.65+022001.3	&	10$^\mathrm{h}$01$^\mathrm{m}$31.66$^\mathrm{s}$	&	+02$^\circ$20$^\prime$01.3$^{\prime \prime}$	&	0.1224	&	\cite{Lilly2007}	\\
SDSS J100135.32+022831.2	&	10$^\mathrm{h}$01$^\mathrm{m}$35.33$^\mathrm{s}$	&	+02$^\circ$28$^\prime$31.2$^{\prime \prime}$	&	0.1225	&	\cite{Lilly2007}	\\
\hline

\end{tabular*}
\caption{XMMXCSJ100141.7+022539.8: All columns are as explained in Table \ref{Table1}.}

\end{table*}

\clearpage

\begin{table*}
\centering

   \begin{tabular*}{\textwidth}{l @{\extracolsep{\fill}} lllll}
  \hline
   Name     & RA(J2000)         & Dec(J2000)  & z & Reference\\
 \hline
2MASX J10403601+4002107	&	10$^\mathrm{h}$40$^\mathrm{m}$35.98$^\mathrm{s}$	&	+40$^\circ$02$^\prime$10.8$^{\prime \prime}$	&	0.1331	&	\citet{SDSS2006}	\\
SDSS J104045.34+395448.5	&	10$^\mathrm{h}$40$^\mathrm{m}$45.35$^\mathrm{s}$	&	+39$^\circ$54$^\prime$48.6$^{\prime \prime}$	&	0.1341	&	\citet{Almeida2011}	\\
B3 1037+401	&	10$^\mathrm{h}$40$^\mathrm{m}$44.45$^\mathrm{s}$	&	+39$^\circ$54$^\prime$54.9$^{\prime \prime}$	&	0.1341	&	\citet{SDSS2006}	\\
SDSS J104126.57+395845.5	&	10$^\mathrm{h}$41$^\mathrm{m}$26.58$^\mathrm{s}$	&	+39$^\circ$58$^\prime$45.5$^{\prime \prime}$	&	0.1352	&	\citet{Miller2002}	\\
SDSS J104043.44+395705.3	&	10$^\mathrm{h}$40$^\mathrm{m}$43.44$^\mathrm{s}$	&	+39$^\circ$57$^\prime$05.3$^{\prime \prime}$	&	0.1357	&	\citet{SDSS2006}	\\
2MASX J10403391+4003497	&	10$^\mathrm{h}$40$^\mathrm{m}$33.96$^\mathrm{s}$	&	+40$^\circ$03$^\prime$49.5$^{\prime \prime}$	&	0.1369	&	\citet{Miller2002}	\\
2MASX J10405360+4001417	&	10$^\mathrm{h}$40$^\mathrm{m}$53.62$^\mathrm{s}$	&	+40$^\circ$01$^\prime$42.0$^{\prime \prime}$	&	0.1374	&	\citet{SDSS2006}	\\
\hline

\end{tabular*}
\caption{XMMXCSJ104044.4+395710.4: All columns are as explained in Table \ref{Table1}.}

\end{table*}

\begin{table*}
\centering

   \begin{tabular*}{\textwidth}{l @{\extracolsep{\fill}} lllll}
  \hline
   Name     & RA(J2000)         & Dec(J2000)  & z & Reference\\
 \hline
SDSS J111512.70+531930.5	&	11$^\mathrm{h}$15$^\mathrm{m}$12.71$^\mathrm{s}$	&	+53$^\circ$19$^\prime$30.6$^{\prime \prime}$	&	0.4586	&	\citet{Bayliss2011}	\\
SDSS J111520.84+532100.9	&	11$^\mathrm{h}$15$^\mathrm{m}$20.84$^\mathrm{s}$	&	+53$^\circ$21$^\prime$01.0$^{\prime \prime}$	&	0.4601	&	\citet{Bayliss2011}	\\
SDSS J111517.30+532115.2	&	11$^\mathrm{h}$15$^\mathrm{m}$17.30$^\mathrm{s}$	&	+53$^\circ$21$^\prime$15.3$^{\prime \prime}$	&	0.4639	&	\citet{Bayliss2011}	\\
SDSS J111517.85+531949.3	&	11$^\mathrm{h}$15$^\mathrm{m}$17.85$^\mathrm{s}$	&	+53$^\circ$19$^\prime$49.4$^{\prime \prime}$	&	0.4640	&	\citet{Bayliss2011}	\\
SDSS J111507.37+531955.7	&	11$^\mathrm{h}$15$^\mathrm{m}$07.38$^\mathrm{s}$	&	+53$^\circ$19$^\prime$55.8$^{\prime \prime}$	&	0.4642	&	\citet{Bayliss2011}	\\
SDSS J111505.53+532042.4	&	11$^\mathrm{h}$15$^\mathrm{m}$05.54$^\mathrm{s}$	&	+53$^\circ$20$^\prime$42.4$^{\prime \prime}$	&	0.4654	&	\citet{Bayliss2011}	\\
SDSS J111514.49+531948.8	&	11$^\mathrm{h}$15$^\mathrm{m}$14.49$^\mathrm{s}$	&	+53$^\circ$19$^\prime$48.9$^{\prime \prime}$	&	0.4654	&	\citet{Bayliss2011}	\\
\hline

\end{tabular*}
\caption{
XMMXCSJ111515.6+531949.5: All columns are as explained in Table \ref{Table1}.}

\end{table*}

\begin{table*}
\centering

   \begin{tabular*}{\textwidth}{l @{\extracolsep{\fill}} lllll}
  \hline
   Name     & RA(J2000)         & Dec(J2000)  & z & Reference\\
 \hline
2MASX J11511361+5507364	&	11$^\mathrm{h}$51$^\mathrm{m}$13.65$^\mathrm{s}$	&	+55$^\circ$07$^\prime$36.5$^{\prime \prime}$	&	0.0757	&	\citet{SDSS2005}	\\
2MASX J11513633+5505433	&	11$^\mathrm{h}$51$^\mathrm{m}$36.33$^\mathrm{s}$	&	+55$^\circ$05$^\prime$43.3$^{\prime \prime}$	&	0.0781	&	\citet{SDSS2005}	\\
SDSS J115109.52+550645.0	&	11$^\mathrm{h}$51$^\mathrm{m}$09.52$^\mathrm{s}$	&	+55$^\circ$06$^\prime$45.1$^{\prime \prime}$	&	0.0782	&	\citet{SDSS2005}	\\
SDSS J115107.14+550839.1	&	11$^\mathrm{h}$51$^\mathrm{m}$07.14$^\mathrm{s}$	&	+55$^\circ$08$^\prime$39.3$^{\prime \prime}$	&	0.0783	&	\citet{SDSS2005}	\\
2MASX J11510461+5510363	&	11$^\mathrm{h}$51$^\mathrm{m}$04.57$^\mathrm{s}$	&	+55$^\circ$10$^\prime$36.0$^{\prime \prime}$	&	0.0787	&	\citet{SDSS2005}	\\
2MASX J11505351+5512012	&	11$^\mathrm{h}$50$^\mathrm{m}$53.45$^\mathrm{s}$	&	+55$^\circ$12$^\prime$01.3$^{\prime \prime}$	&	0.0793	&	\citet{SDSS2005}	\\
2MASX J11511349+5506594	&	11$^\mathrm{h}$51$^\mathrm{m}$13.47$^\mathrm{s}$	&	+55$^\circ$06$^\prime$59.2$^{\prime \prime}$	&	0.0794	&	\citet{SDSS2005}	\\
\hline

\end{tabular*}
\caption{XMMXCSJ115112.0+550655.5: All columns are as explained in Table \ref{Table1}.}

\end{table*}

\begin{table*}
\centering

   \begin{tabular*}{\textwidth}{l @{\extracolsep{\fill}} lllll}
  \hline
   Name     & RA(J2000)         & Dec(J2000)  & z & Reference\\
 \hline
SDSS J123140.79+413626.3	&	12:31$^\mathrm{m}$40.80$^\mathrm{s}$	&	+41$^\circ$36$^\prime$26.4$^{\prime \prime}$	&	0.1708	&	\citet{SDSS2006}	\\
2MASX J12320414+4133324	&	12:32$^\mathrm{m}$04.12$^\mathrm{s}$	&	+41$^\circ$33$^\prime$32.4$^{\prime \prime}$	&	0.1713	&	\citet{SDSS2006}	\\
2MASX J12315661+4133582	&	12:31$^\mathrm{m}$56.59$^\mathrm{s}$	&	+41$^\circ$33$^\prime$58.5$^{\prime \prime}$	&	0.1719	&	\citet{SDSS2006}	\\
2MASX J12320054+4133322	&	12:32$^\mathrm{m}$00.56$^\mathrm{s}$	&	+41$^\circ$33$^\prime$31.9$^{\prime \prime}$	&	0.1727	&	\citet{SDSS2006}	\\
2MASX J12320463+4133362	&	12:32$^\mathrm{m}$04.68$^\mathrm{s}$	&	+41$^\circ$33$^\prime$36.0$^{\prime \prime}$	&	0.1732	&	\citet{SDSS2007}	\\
2MASX J12315296+4133182	&	12:31$^\mathrm{m}$52.98$^\mathrm{s}$	&	+41$^\circ$33$^\prime$18.7$^{\prime \prime}$	&	0.1739	&	\citet{SDSS2006}	\\
2MASX J12314144+4139361	&	12:31$^\mathrm{m}$41.46$^\mathrm{s}$	&	+41$^\circ$39$^\prime$35.9$^{\prime \prime}$	&	0.1750	&	\citet{SDSS2006}	\\
SDSS J123151.89+413434.7	&	12:31$^\mathrm{m}$51.90$^\mathrm{s}$	&	+41$^\circ$34$^\prime$34.7$^{\prime \prime}$	&	0.1750	&	\citet{SDSS2006}	\\
2MASX J12314386+4137341	&	12:31$^\mathrm{m}$43.90$^\mathrm{s}$	&	+41$^\circ$37$^\prime$33.9$^{\prime \prime}$	&	0.1757	&	\citet{SDSS2006}	\\
SDSS J123136.04+414314.0	&	12:31$^\mathrm{m}$36.05$^\mathrm{s}$	&	+41$^\circ$43$^\prime$14.0$^{\prime \prime}$	&	0.1763	&	\citet{SDSS2006}	\\

\hline

\end{tabular*}
\caption{XMMXCSJ123144.4+413732.0: All columns are as explained in Table \ref{Table1}.}

\end{table*}

\begin{table*}
\centering

   \begin{tabular*}{\textwidth}{l @{\extracolsep{\fill}} lllll}
  \hline
   Name     & RA(J2000)         & Dec(J2000)  & z & Reference\\
 \hline
SDSS J151623.79+000156.4	&	15$^\mathrm{h}$16$^\mathrm{m}$23.79$^\mathrm{s}$	&	+00$^\circ$01$^\prime$56.5$^{\prime \prime}$	&	0.1126	&	\citet{SDSS2002}	\\
SDSS J151626.35-000228.8	&	15$^\mathrm{h}$16$^\mathrm{m}$26.35$^\mathrm{s}$	&	-00$^\circ$02$^\prime$28.9$^{\prime \prime}$	&	0.1142	&	\citet{SDSS2002}	\\
2MASX J15161900+0005067	&	15$^\mathrm{h}$16$^\mathrm{m}$19.00$^\mathrm{s}$	&	+00$^\circ$05$^\prime$06.6$^{\prime \prime}$	&	0.1147	&	\citet{SDSS2003}	\\
2MASX J15162348+0006093	&	15$^\mathrm{h}$16$^\mathrm{m}$23.49$^\mathrm{s}$	&	+00$^\circ$06$^\prime$09.1$^{\prime \prime}$	&	0.1154	&	\citet{SDSS2003}	\\
2MASX J15161563+0005291	&	15$^\mathrm{h}$16$^\mathrm{m}$15.63$^\mathrm{s}$	&	+00$^\circ$05$^\prime$29.0$^{\prime \prime}$	&	0.1158	&	\citet{SDSS2003}	\\
2MASX J15162547+0003377	&	15$^\mathrm{h}$16$^\mathrm{m}$25.45$^\mathrm{s}$	&	+00$^\circ$03$^\prime$37.6$^{\prime \prime}$	&	0.1167	&	\citet{SDSS2003}	\\
SDSS J151626.17+000707.4	&	15$^\mathrm{h}$16$^\mathrm{m}$26.18$^\mathrm{s}$	&	+00$^\circ$07$^\prime$07.5$^{\prime \prime}$	&	0.1167	&	\citet{SDSS2002}	\\
\hline

\end{tabular*}
\caption{XMMXCSJ151618.6+000531.3: All columns are as explained in Table \ref{Table1}.}

\end{table*}

\begin{table*}
\centering

  \begin{tabular*}{\textwidth}{l @{\extracolsep{\fill}} lllll}
  \hline
   Name     & RA(J2000)         & Dec(J2000)  & z & Reference\\
 \hline
SDSS J161126.49+541619.2	&	16$^\mathrm{h}$11$^\mathrm{m}$26.49$^\mathrm{s}$	&	+54$^\circ$16$^\prime$19.2$^{\prime \prime}$	&	0.3310	&	\citet{Trichas2010}	\\
SDSS J161131.39+541333.1	&	16$^\mathrm{h}$11$^\mathrm{m}$31.39$^\mathrm{s}$	&	+54$^\circ$13$^\prime$33.2$^{\prime \prime}$	&	0.3330	&	\citet{Trichas2010}	\\
SDSS J161147.78+541132.2	&	16$^\mathrm{h}$11$^\mathrm{m}$47.78$^\mathrm{s}$	&	+54$^\circ$11$^\prime$32.3$^{\prime \prime}$	&	0.3340	&	\citet{Trichas2010}	\\
ELAIS06S(R) J161127.5+541422	&	16$^\mathrm{h}$11$^\mathrm{m}$27.53$^\mathrm{s}$	&	+54$^\circ$14$^\prime$22.7$^{\prime \prime}$	&	0.3350	&	\citet{Trichas2010}	\\
SDSS J161122.54+541706.5	&	16$^\mathrm{h}$11$^\mathrm{m}$22.54$^\mathrm{s}$	&	+54$^\circ$17$^\prime$06.6$^{\prime \prime}$	&	0.3360	&	\citet{Trichas2010}	\\
SDSS J161134.91+541404.6	&	16$^\mathrm{h}$11$^\mathrm{m}$34.92$^\mathrm{s}$	&	+54$^\circ$14$^\prime$04.7$^{\prime \prime}$	&	0.3360	&	\citet{Trichas2010}	\\
SDSS J161135.93+541634.5	&	16$^\mathrm{h}$11$^\mathrm{m}$35.94$^\mathrm{s}$	&	+54$^\circ$16$^\prime$34.5$^{\prime \prime}$	&	0.3380	&	\citet{Trichas2010}	\\
\hline

\end{tabular*}
\caption{XMMXCSJ161132.7+541628.3: All columns are as explained in Table \ref{Table1}.}

\end{table*}

\begin{table*}
\centering

   \begin{tabular*}{\textwidth}{l @{\extracolsep{\fill}} lllll}
  \hline
   Name     & RA(J2000)         & Dec(J2000)  & z & Reference\\
 \hline
SDSS J163057.87+243412.6	&	16$^\mathrm{h}$30$^\mathrm{m}$57.88$^\mathrm{s}$	&	+24$^\circ$34$^\prime$12.7$^{\prime \prime}$	&	0.0560	&	\citet{SDSS2006}	\\
CGCG 138-016	&	16$^\mathrm{h}$28$^\mathrm{m}$36.04$^\mathrm{s}$	&	+24$^\circ$33$^\prime$19.8$^{\prime \prime}$	&	0.0571	&	\citet{Freudling1992}	\\
SDSS J162854.19+243138.5	&	16$^\mathrm{h}$28$^\mathrm{m}$54.20$^\mathrm{s}$	&	+24$^\circ$31$^\prime$38.6$^{\prime \prime}$	&	0.0574	&	\citet{SDSS2006}	\\
2MASX J16305968+2437076	&	16$^\mathrm{h}$30$^\mathrm{m}$59.72$^\mathrm{s}$	&	+24$^\circ$37$^\prime$08.0$^{\prime \prime}$	&	0.0579	&	\citet{SDSS2007}	\\
SDSS J163039.80+243938.2	&	16$^\mathrm{h}$30$^\mathrm{m}$39.80$^\mathrm{s}$	&	+24$^\circ$39$^\prime$38.3$^{\prime \prime}$	&	0.0580	&	\citet{SDSS2006}	\\
2MASX J16310298+2442456	&	16$^\mathrm{h}$31$^\mathrm{m}$03.00$^\mathrm{s}$	&	+24$^\circ$42$^\prime$45.6$^{\prime \prime}$	&	0.0581	&	\citet{SDSS2006}	\\
2MASX J16311876+2442442	&	16$^\mathrm{h}$31$^\mathrm{m}$18.75$^\mathrm{s}$	&	+24$^\circ$42$^\prime$44.2$^{\prime \prime}$	&	0.0582	&	\citet{SDSS2006}	\\
\hline

\end{tabular*}
\caption{XMMXCSJ163015.6+243423.2: All columns are as explained in Table \ref{Table1}.}

\end{table*}

\begin{table*}
\centering

   \begin{tabular*}{\textwidth}{l @{\extracolsep{\fill}} lllll}
  \hline
   Name     & RA(J2000)         & Dec(J2000)  & z & Reference\\
 \hline
(W2011) J339.86728-05.76581	&	22$^\mathrm{h}$39$^\mathrm{m}$28.15$^\mathrm{s}$	&	-05$^\circ$45$^\prime$56.9$^{\prime \prime}$	&	0.2402	&	\citet{Wegner2011}	\\
(W2011) J339.85526-05.78358	&	22$^\mathrm{h}$39$^\mathrm{m}$25.26$^\mathrm{s}$	&	-05$^\circ$47$^\prime$00.9$^{\prime \prime}$	&	0.2405	&	\citet{Wegner2011}	\\
(W2011) J339.85416-05.78444	&	22$^\mathrm{h}$39$^\mathrm{m}$25.00$^\mathrm{s}$	&	-05$^\circ$47$^\prime$04.0$^{\prime \prime}$	&	0.2410	&	\citet{Wegner2011}	\\
(W2011) J339.84311-05.75219	&	22$^\mathrm{h}$39$^\mathrm{m}$22.35$^\mathrm{s}$	&	-05$^\circ$45$^\prime$07.9$^{\prime \prime}$	&	0.2414	&	\citet{Wegner2011}	\\
(W2011) J339.85796-05.66972	&	22$^\mathrm{h}$39$^\mathrm{m}$25.91$^\mathrm{s}$	&	-05$^\circ$40$^\prime$11.0$^{\prime \prime}$	&	0.2421	&	\citet{Wegner2011}	\\
(W2011) J339.91522-05.73214	&	22$^\mathrm{h}$39$^\mathrm{m}$39.65$^\mathrm{s}$	&	-05$^\circ$43$^\prime$55.7$^{\prime \prime}$	&	0.2421	&	\citet{Wegner2011}	\\
(W2011) J339.92899-05.70703	&	22$^\mathrm{h}$39$^\mathrm{m}$42.96$^\mathrm{s}$	&	-05$^\circ$42$^\prime$25.3$^{\prime \prime}$	&	0.2424	&	\citet{Wegner2011}	\\
\hline

\end{tabular*}
\caption{XMMXCSJ223939.3-054327.4: All columns are as explained in Table \ref{Table1}.}

\end{table*}
\clearpage

\begin{table*}
\centering

   \begin{tabular*}{\textwidth}{l @{\extracolsep{\fill}} lllll}
  \hline
   Name     & RA(J2000)         & Dec(J2000)  & z & Reference\\
 \hline
AGC 331543	&	23$^\mathrm{h}$38$^\mathrm{m}$31.41$^\mathrm{s}$	&	+27$^\circ$13$^\prime$48.2$^{\prime \prime}$	&	0.1218	&	\citet{Scodeggio1995}	\\
AGC 331535	&	23$^\mathrm{h}$37$^\mathrm{m}$55.52$^\mathrm{s}$	&	+27$^\circ$08$^\prime$25.8$^{\prime \prime}$	&	0.1222	&	\citet{Scodeggio1995}	\\
2MASX J23380720+2715028	&	23$^\mathrm{h}$38$^\mathrm{m}$07.24$^\mathrm{s}$	&	+27$^\circ$15$^\prime$02.9$^{\prime \prime}$	&	0.1222	&	\citet{Scodeggio1995}	\\
AGC 331529	&	23$^\mathrm{h}$37$^\mathrm{m}$41.99$^\mathrm{s}$	&	+27$^\circ$04$^\prime$37.4$^{\prime \prime}$	&	0.1226	&	\citet{Scodeggio1995}	\\
AGC 331531	&	23$^\mathrm{h}$37$^\mathrm{m}$47.13$^\mathrm{s}$	&	+27$^\circ$09$^\prime$29.5$^{\prime \prime}$	&	0.1226	&	\citet{Scodeggio1995}	\\
2MASX J23375415+2718448	&	23$^\mathrm{h}$37$^\mathrm{m}$54.16$^\mathrm{s}$	&	+27$^\circ$18$^\prime$44.8$^{\prime \prime}$	&	0.1238	&	\citet{Scodeggio1995}	\\
2MASX J23375559+2711128	&	23$^\mathrm{h}$37$^\mathrm{m}$55.59$^\mathrm{s}$	&	+27$^\circ$11$^\prime$13.1$^{\prime \prime}$	&	0.1245	&	\citet{Giovanelli1995}	\\
AGC 331526	&	23$^\mathrm{h}$37$^\mathrm{m}$22.51$^\mathrm{s}$	&	+27$^\circ$06$^\prime$32.6$^{\prime \prime}$	&	0.1246	&	\citet{Scodeggio1995}	\\
2MASX J23380218+2712088	&	23$^\mathrm{h}$38$^\mathrm{m}$02.18$^\mathrm{s}$	&	+27$^\circ$12$^\prime$09.1$^{\prime \prime}$	&	0.1246	&	\citet{Smith2004}	\\
2MASX J23382282+2717516	&	23$^\mathrm{h}$38$^\mathrm{m}$22.83$^\mathrm{s}$	&	+27$^\circ$17$^\prime$51.6$^{\prime \prime}$	&	0.1263	&	\citet{Scodeggio1995}	\\
2MASX J23372726+2707408	&	23$^\mathrm{h}$37$^\mathrm{m}$27.26$^\mathrm{s}$	&	+27$^\circ$07$^\prime$41.0$^{\prime \prime}$	&	0.1264	&	\citet{Scodeggio1995}	\\
BATC J233724.29+271434.14	&	23$^\mathrm{h}$37$^\mathrm{m}$24.10$^\mathrm{s}$	&	+27$^\circ$14$^\prime$32.8$^{\prime \prime}$	&	0.1265	&	\citet{Pinkney1993}	\\

\hline

\end{tabular*}
\caption{XMMXCSJ233757.0+271121.0: All columns are as explained in Table \ref{Table1}.}

\end{table*}

\begin{table*}
\centering

 \begin{tabular*}{\textwidth}{l @{\extracolsep{\fill}} lllllll}
  \hline
   ID & RA(J2000)         & Dec(J2000)  & z & Reference & Member\\
 \hline
 
 1	&	10$^\mathrm{h}$57$^\mathrm{m}$03.77$^\mathrm{s}$	&	-03$^\circ$37$^\prime$37.3$^{\prime \prime}$	&	0.8312	&	\citet{Nastasi2014} \citet{Tran1999}	&	$\surd$	\\
2	&	10$^\mathrm{h}$56$^\mathrm{m}$50.84$^\mathrm{s}$	&	-03$^\circ$37$^\prime$37.4$^{\prime \prime}$	&	0.8390	&	\citet{Nastasi2014} \citet{Tran1999}	&	$\surd$	\\
3	&	10$^\mathrm{h}$57$^\mathrm{m}$04.10$^\mathrm{s}$	&	-03$^\circ$37$^\prime$37.3$^{\prime \prime}$	&	0.8329	&	\citet{Nastasi2014} \citet{Tran1999}	&	$\surd$	\\
4	&	10$^\mathrm{h}$56$^\mathrm{m}$54.31$^\mathrm{s}$	&	-03$^\circ$37$^\prime$37.4$^{\prime \prime}$	&	0.8380	&	\citet{Nastasi2014} \citet{Tran1999}	&	$\surd$	\\
5	&	10$^\mathrm{h}$56$^\mathrm{m}$51.77$^\mathrm{s}$	&	-03$^\circ$37$^\prime$37.4$^{\prime \prime}$	&	0.8371	&	\citet{Nastasi2014} \citet{Tran1999}	&	$\surd$	\\
6	&	10$^\mathrm{h}$56$^\mathrm{m}$58.44$^\mathrm{s}$	&	-03$^\circ$37$^\prime$37.4$^{\prime \prime}$	&	0.8353	&	\citet{Nastasi2014} \citet{Tran1999}	&	$\surd$	\\
7	&	10$^\mathrm{h}$56$^\mathrm{m}$56.37$^\mathrm{s}$	&	-03$^\circ$37$^\prime$37.4$^{\prime \prime}$	&	0.8317	&	\citet{Nastasi2014} \citet{Tran1999}	&	$\surd$	\\
\hline

\end{tabular*}
\caption{XMMXCSJ105659.5-033728.0: Columns 2 - 5 are as explained in Table \ref{Table1}. Column 1 gives an arbitrary ID for each galaxy and column 6 shows whether or not the galaxy was included as a member for the determination of the velocity dispersion.}
\label{NastasiTable}
\end{table*}

\begin{table*}
\centering

 \begin{tabular*}{\textwidth}{l @{\extracolsep{\fill}} lllllll}
  \hline
 ID & RA(J2000)         & Dec(J2000)  & z & Reference & Member\\
 \hline
 
 1	&	11$^\mathrm{h}$40$^\mathrm{m}$22.31$^\mathrm{s}$	&	66$^\circ$08$^\prime$15.1$^{\prime \prime}$	&	0.7844	&	\citet{Nastasi2014} \citet{Donahue1999}	&	$\surd$	\\
2	&	11$^\mathrm{h}$40$^\mathrm{m}$22.01$^\mathrm{s}$	&	66$^\circ$08$^\prime$12.1$^{\prime \prime}$	&	0.7903	&	\citet{Nastasi2014} \citet{Donahue1999}	&	$\surd$	\\
3	&	11$^\mathrm{h}$40$^\mathrm{m}$22.61$^\mathrm{s}$	&	66$^\circ$08$^\prime$17.1$^{\prime \prime}$	&	0.7844	&	\citet{Nastasi2014} \citet{Donahue1999}	&	$\surd$	\\
4	&	11$^\mathrm{h}$40$^\mathrm{m}$22.01$^\mathrm{s}$	&	66$^\circ$08$^\prime$19.0$^{\prime \prime}$	&	0.7927	&	\citet{Nastasi2014} \citet{Donahue1999}	&	$\surd$	\\
5	&	11$^\mathrm{h}$40$^\mathrm{m}$21.41$^\mathrm{s}$	&	66$^\circ$08$^\prime$20.1$^{\prime \prime}$	&	0.7826	&	\citet{Nastasi2014} \citet{Donahue1999}	&	$\surd$	\\
6	&	11$^\mathrm{h}$40$^\mathrm{m}$17.70$^\mathrm{s}$	&	66$^\circ$08$^\prime$23.0$^{\prime \prime}$	&	0.7808	&	\citet{Nastasi2014} \citet{Donahue1999}	&	$\surd$	\\
7	&	11$^\mathrm{h}$40$^\mathrm{m}$16.91$^\mathrm{s}$	&	66$^\circ$08$^\prime$25.1$^{\prime \prime}$	&	0.7909	&	\citet{Nastasi2014} \citet{Donahue1999}	&	$\surd$	\\
\hline

\end{tabular*}
\caption{XMMXCSJ114023.0+660819.0: All columns are as explained in Table \ref{NastasiTable}.}

\end{table*}

\begin{table*}
\centering

 \begin{tabular*}{\textwidth}{l @{\extracolsep{\fill}} lllllll}
  \hline
 ID & RA(J2000)         & Dec(J2000)  & z & Reference & Member\\
 \hline

\hline
1	&	18$^\mathrm{h}$21$^\mathrm{m}$16.61$^\mathrm{s}$	&	68$^\circ$30$^\prime$25.0$^{\prime \prime}$	&	0.8151	&	\citet{Nastasi2014} \citet{Gioia2004}	&	$\surd$	\\
2	&	18$^\mathrm{h}$21$^\mathrm{m}$21.50$^\mathrm{s}$	&	68$^\circ$29$^\prime$57.1$^{\prime \prime}$	&	0.8225	&	\citet{Nastasi2014} \citet{Gioia2004}	&	$\surd$	\\
3	&	18$^\mathrm{h}$21$^\mathrm{m}$26.21$^\mathrm{s}$	&	68$^\circ$29$^\prime$31.0$^{\prime \prime}$	&	0.8206	&	\citet{Nastasi2014} \citet{Gioia2004}	&	$\surd$	\\
4	&	18$^\mathrm{h}$21$^\mathrm{m}$30.80$^\mathrm{s}$	&	68$^\circ$29$^\prime$29.1$^{\prime \prime}$	&	0.8204	&	\citet{Nastasi2014} \citet{Gioia2004}	&	$\surd$	\\
5	&	18$^\mathrm{h}$21$^\mathrm{m}$27.90$^\mathrm{s}$	&	68$^\circ$28$^\prime$43.0$^{\prime \prime}$	&	0.8229	&	\citet{Nastasi2014} \citet{Gioia2004}	&	$\surd$	\\
6	&	18$^\mathrm{h}$21$^\mathrm{m}$33.80$^\mathrm{s}$	&	68$^\circ$28$^\prime$40.1$^{\prime \prime}$	&	0.8235	&	\citet{Nastasi2014} \citet{Gioia2004}	&	$\surd$	\\
7	&	18$^\mathrm{h}$21$^\mathrm{m}$32.60$^\mathrm{s}$	&	68$^\circ$27$^\prime$54.1$^{\prime \prime}$	&	0.8167	&	\citet{Nastasi2014} \citet{Gioia2004}	&	$\surd$	\\
\hline
\end{tabular*}
\caption{XMMXCSJ182132.9+682755.0: All columns are as explained in Table \ref{NastasiTable}.}

\end{table*}

\begin{table*}
\centering

 \begin{tabular*}{\textwidth}{l @{\extracolsep{\fill}} llllccc}
  \hline
   ID & Mask     & RA(J2000)         & Dec(J2000)  & z & Quality & Identification & Member\\
 \hline
1	&	GS-2012B-Q-011-03	&	00$^\mathrm{h}$56$^\mathrm{m}$46.52$^\mathrm{s}$	&	-27$^\circ$41$^\prime$57.9$^{\prime \prime}$	&	0.6625	&	3	&	X	&		\\
2	&	GS-2012B-Q-011-03	&	00$^\mathrm{h}$56$^\mathrm{m}$47.54$^\mathrm{s}$	&	-27$^\circ$41$^\prime$53.6$^{\prime \prime}$	&	0.6294	&	2	&	X	&		\\
3	&	GS-2012B-Q-011-03	&	00$^\mathrm{h}$56$^\mathrm{m}$47.56$^\mathrm{s}$	&	-27$^\circ$41$^\prime$03.2$^{\prime \prime}$	&	0.5864	&	3	&	X	&		\\
5	&	GS-2012B-Q-011-03	&	00$^\mathrm{h}$56$^\mathrm{m}$47.54$^\mathrm{s}$	&	-27$^\circ$41$^\prime$19.7$^{\prime \prime}$	&	0.5604	&	3	&	X	&		\\
6	&	GS-2012B-Q-011-03	&	00$^\mathrm{h}$56$^\mathrm{m}$54.47$^\mathrm{s}$	&	-27$^\circ$41$^\prime$52.9$^{\prime \prime}$	&	0.5604	&	3	&	X	&	$\surd$	\\
7	&	GS-2012B-Q-011-03	&	00$^\mathrm{h}$56$^\mathrm{m}$52.73$^\mathrm{s}$	&	-27$^\circ$41$^\prime$46.0$^{\prime \prime}$	&	0.5744	&	3	&	X	&		\\
9	&	GS-2012B-Q-011-03	&	00$^\mathrm{h}$56$^\mathrm{m}$53.45$^\mathrm{s}$	&	-27$^\circ$41$^\prime$28.0$^{\prime \prime}$	&	0.5754	&	3	&	X	&		\\
\hline

\end{tabular*}
\caption{XMMXCSJ005656.6-274031.9:: Column 1 gives an ID for each galaxy in the mask given in column 2. Column 3 and 4 give the right ascension and declination respectively. Column 5 gives the redshift, column 6 gives the quality of the redshift according to the following scheme: $Q=3$ corresponds to two or more strongly detected features; $Q=2$ refers to one strongly detected or two weakly detected features; $Q=1$ one weakly detected feature and $Q=0$ when no features could be identified. Column 7 shows whether the redshift was found via visible inspection (V) or cross-correlation (X). Column 8 shows whether or not the galaxy was classified as a member and used in the calculation of the velocity dispersion.}
\label{geminiTable}
\end{table*}

\begin{table*}
\centering

\begin{tabular*}{\textwidth}{l @{\extracolsep{\fill}} llllccc}
  \hline
   ID & Mask     & RA(J2000)         & Dec(J2000)  & z & Quality& Identification & Member\\
 \hline1	&	\citet{Nastasi2014}	&	01$^\mathrm{h}$52$^\mathrm{m}$41.30$^\mathrm{s}$	&	-13$^\circ$38$^\prime$55.0$^{\prime \prime}$	&	0.8307	&	2	&		&	$\surd$	\\
3	&	\citet{Nastasi2014}	&	01$^\mathrm{h}$52$^\mathrm{m}$41.11$^\mathrm{s}$	&	-13$^\circ$39$^\prime$07.9$^{\prime \prime}$	&	0.8303	&	3	&		&	$\surd$	\\
4	&	\citet{Nastasi2014}	&	01$^\mathrm{h}$52$^\mathrm{m}$42.71$^\mathrm{s}$	&	-13$^\circ$38$^\prime$59.1$^{\prime \prime}$	&	0.8261	&	4	&		&	$\surd$	\\
5	&	\citet{Nastasi2014}	&	01$^\mathrm{h}$52$^\mathrm{m}$39.41$^\mathrm{s}$	&	-13$^\circ$38$^\prime$55.8$^{\prime \prime}$	&	0.8233	&	3	&		&	$\surd$	\\
6	&	\citet{Nastasi2014}	&	01$^\mathrm{h}$52$^\mathrm{m}$40.70$^\mathrm{s}$	&	-13$^\circ$39$^\prime$21.0$^{\prime \prime}$	&	0.8261	&	2	&		&		\\
7	&	\citet{Nastasi2014}	&	01$^\mathrm{h}$52$^\mathrm{m}$40.21$^\mathrm{s}$	&	-13$^\circ$39$^\prime$24.1$^{\prime \prime}$	&	0.8307	&	3	&		&	$\surd$	\\
8	&	\citet{Nastasi2014}	&	01$^\mathrm{h}$52$^\mathrm{m}$43.20$^\mathrm{s}$	&	-13$^\circ$38$^\prime$29.5$^{\prime \prime}$	&	0.8268	&	3	&		&	$\surd$	\\
\hline

\end{tabular*}
\caption{XMMXCSJ015241.1-133855.9: All columns are as explained in Table \ref{geminiTable}.}

\end{table*}

\begin{table*}
\centering

 \begin{tabular*}{\textwidth}{l @{\extracolsep{\fill}} llllccc}
  \hline
   ID & Mask     & RA(J2000)         & Dec(J2000)  & z & Quality& Identification & Member\\
 \hline
2	&	GS-2012B-Q-011-06	&	02$^\mathrm{h}$17$^\mathrm{m}$30.50$^\mathrm{s}$	&	-05$^\circ$11$^\prime$31.4$^{\prime \prime}$	&	0.5674	&	3	&	X	&		\\
3	&	GS-2012B-Q-011-06	&	02$^\mathrm{h}$17$^\mathrm{m}$28.40$^\mathrm{s}$	&	-05$^\circ$12$^\prime$03.6$^{\prime \prime}$	&	0.9216	&	3	&	X	&		\\
6	&	GS-2012B-Q-011-06	&	02$^\mathrm{h}$17$^\mathrm{m}$27.60$^\mathrm{s}$	&	-05$^\circ$13$^\prime$60.0$^{\prime \prime}$	&	0.6424	&	3	&	X	&	$\surd$	\\
8	&	GS-2012B-Q-011-06	&	02$^\mathrm{h}$17$^\mathrm{m}$34.35$^\mathrm{s}$	&	-05$^\circ$12$^\prime$43.8$^{\prime \prime}$	&	0.6474	&	3	&	X	&	$\surd$	\\
10	&	GS-2012B-Q-011-06	&	02$^\mathrm{h}$17$^\mathrm{m}$33.72$^\mathrm{s}$	&	-05$^\circ$12$^\prime$51.6$^{\prime \prime}$	&	0.6408	&	3	&	X	&	$\surd$	\\
11	&	GS-2012B-Q-011-06	&	02$^\mathrm{h}$17$^\mathrm{m}$35.84$^\mathrm{s}$	&	-05$^\circ$12$^\prime$58.9$^{\prime \prime}$	&	0.8246	&	3	&	X	&		\\
12	&	GS-2012B-Q-011-06	&	02$^\mathrm{h}$17$^\mathrm{m}$30.52$^\mathrm{s}$	&	-05$^\circ$13$^\prime$21.4$^{\prime \prime}$	&	0.8336	&	2	&	X	&		\\
\hline
\end{tabular*}
\caption{XMMXCSJ021734.7-051326.9: All columns are as explained in Table \ref{geminiTable}.}

\end{table*}

\begin{table*}
\centering

 \begin{tabular*}{\textwidth}{l @{\extracolsep{\fill}} llllccc}
  \hline
   ID & Mask     & RA(J2000)         & Dec(J2000)  & z & Quality& Identification & Member\\
 \hline
 
1	&	GS-2010B-Q-46-06	&	02$^\mathrm{h}$50$^\mathrm{m}$22.92$^\mathrm{s}$	&	-31$^\circ$03$^\prime$53.0$^{\prime \prime}$	&	0.8337	&	3	&	X	&		\\
8	&	GS-2010B-Q-46-06	&	02$^\mathrm{h}$50$^\mathrm{m}$15.25$^\mathrm{s}$	&	-31$^\circ$03$^\prime$33.5$^{\prime \prime}$	&	0.7263	&	3	&	X	&		\\
9	&	GS-2010B-Q-46-06	&	02$^\mathrm{h}$50$^\mathrm{m}$13.27$^\mathrm{s}$	&	-31$^\circ$03$^\prime$35.7$^{\prime \prime}$	&	0.6168	&	3	&	X	&		\\
10	&	GS-2010B-Q-46-06	&	02$^\mathrm{h}$50$^\mathrm{m}$12.05$^\mathrm{s}$	&	-31$^\circ$03$^\prime$08.0$^{\prime \prime}$	&	0.7146	&	3	&	X	&		\\
14	&	GS-2010B-Q-46-06	&	02$^\mathrm{h}$50$^\mathrm{m}$06.63$^\mathrm{s}$	&	-31$^\circ$03$^\prime$13.7$^{\prime \prime}$	&	0.8496	&	3	&	X	&		\\
15	&	GS-2010B-Q-46-06	&	02$^\mathrm{h}$50$^\mathrm{m}$08.70$^\mathrm{s}$	&	-31$^\circ$03$^\prime$49.7$^{\prime \prime}$	&	0.9052	&	3	&	X	&	$\surd$	\\
16	&	GS-2010B-Q-46-06	&	02$^\mathrm{h}$50$^\mathrm{m}$03.78$^\mathrm{s}$	&	-31$^\circ$03$^\prime$51.5$^{\prime \prime}$	&	0.3533	&	3	&	X	&		\\
\hline

\end{tabular*}
\caption{XMMXCSJ025006.4-310400.8: All columns are as explained in Table \ref{geminiTable}.}

\end{table*}

\begin{table*}
\centering

 \begin{tabular*}{\textwidth}{l @{\extracolsep{\fill}} llllccc}
  \hline
   ID & Mask     & RA(J2000)         & Dec(J2000)  & z & Quality& Identification & Member\\
 \hline
 
1	&	GS-2011B-Q-050-03	&	03$^\mathrm{h}$02$^\mathrm{m}$13.84$^\mathrm{s}$	&	-00$^\circ$00$^\prime$41.6$^{\prime \prime}$	&	0.4988	&	3	&	X	&		\\
7	&	GS-2011B-Q-050-03	&	03$^\mathrm{h}$02$^\mathrm{m}$13.04$^\mathrm{s}$	&	00$^\circ$00$^\prime$41.5$^{\prime \prime}$	&	0.5958	&	3	&	X	&		\\
9	&	GS-2011B-Q-050-03	&	03$^\mathrm{h}$02$^\mathrm{m}$06.74$^\mathrm{s}$	&	-00$^\circ$00$^\prime$06.1$^{\prime \prime}$	&	0.6432	&	3	&	X	&	$\surd$	\\
11	&	GS-2011B-Q-050-03	&	03$^\mathrm{h}$02$^\mathrm{m}$07.86$^\mathrm{s}$	&	00$^\circ$00$^\prime$21.2$^{\prime \prime}$	&	0.6425	&	3	&	X	&	$\surd$	\\
12	&	GS-2011B-Q-050-03	&	03$^\mathrm{h}$02$^\mathrm{m}$06.20$^\mathrm{s}$	&	-00$^\circ$00$^\prime$04.0$^{\prime \prime}$	&	0.6451	&	3	&	X	&	$\surd$	\\
13	&	GS-2011B-Q-050-03	&	03$^\mathrm{h}$02$^\mathrm{m}$08.78$^\mathrm{s}$	&	00$^\circ$00$^\prime$39.9$^{\prime \prime}$	&	0.6486	&	3	&	X	&	$\surd$	\\
15	&	GS-2011B-Q-050-03	&	03$^\mathrm{h}$02$^\mathrm{m}$05.64$^\mathrm{s}$	&	-00$^\circ$00$^\prime$00.5$^{\prime \prime}$	&	0.6452	&	3	&	X	&	$\surd$	\\
\hline

\end{tabular*}
\caption{XMMXCSJ030205.1-000003.6: All columns are as explained in Table \ref{geminiTable}.}

\end{table*}

\begin{table*}
\centering

 \begin{tabular*}{\textwidth}{l @{\extracolsep{\fill}} llllccc}
  \hline
   ID & Mask     & RA(J2000)         & Dec(J2000)  & z & Quality & Identification & Member\\
 \hline
2	&	GS-2010B-Q-046-01	&	09$^\mathrm{h}$59$^\mathrm{m}$41.78$^\mathrm{s}$	&	02$^\circ$33$^\prime$39.1$^{\prime \prime}$	&	0.6964	&	3	&	X	&		\\
3	&	GS-2010B-Q-046-01	&	09$^\mathrm{h}$59$^\mathrm{m}$34.47$^\mathrm{s}$	&	02$^\circ$33$^\prime$47.0$^{\prime \prime}$	&	0.6766	&	3	&	X	&		\\
4	&	GS-2010B-Q-046-01	&	09$^\mathrm{h}$59$^\mathrm{m}$43.01$^\mathrm{s}$	&	02$^\circ$32$^\prime$32.0$^{\prime \prime}$	&	0.6963	&	3	&	X	&		\\
5	&	GS-2010B-Q-046-01	&	09$^\mathrm{h}$59$^\mathrm{m}$42.91$^\mathrm{s}$	&	02$^\circ$31$^\prime$49.7$^{\prime \prime}$	&	0.7268	&	3	&	X	&	$\surd$	\\
7	&	GS-2010B-Q-046-01	&	09$^\mathrm{h}$59$^\mathrm{m}$39.41$^\mathrm{s}$	&	02$^\circ$33$^\prime$00.1$^{\prime \prime}$	&	0.4701	&	3	&	X	&		\\
8	&	GS-2010B-Q-046-01	&	09$^\mathrm{h}$59$^\mathrm{m}$39.35$^\mathrm{s}$	&	02$^\circ$32$^\prime$39.2$^{\prime \prime}$	&	0.6778	&	3	&	X	&		\\
9	&	GS-2010B-Q-046-01	&	09$^\mathrm{h}$59$^\mathrm{m}$38.13$^\mathrm{s}$	&	02$^\circ$32$^\prime$11.0$^{\prime \prime}$	&	0.7296	&	3	&	X	&	$\surd$	\\
\hline

\end{tabular*}
\caption{XMMXCSJ095940.7+023113.4: All columns are as explained in Table \ref{geminiTable}.}

\end{table*}

\begin{table*}
\centering

 \begin{tabular*}{\textwidth}{l @{\extracolsep{\fill}} llllccc}
  \hline
   ID & Mask     & RA(J2000)         & Dec(J2000)  & z & Quality & Identification & Member\\
 \hline

2	&	GS-2010B-Q-046-03	&	11$^\mathrm{h}$24$^\mathrm{m}$00.44$^\mathrm{s}$	&	05$^\circ$27$^\prime$13.8$^{\prime \prime}$	&	0.4726	&	3	&	X	&		\\
4	&	GS-2010B-Q-046-03	&	11$^\mathrm{h}$24$^\mathrm{m}$00.65$^\mathrm{s}$	&	05$^\circ$27$^\prime$42.1$^{\prime \prime}$	&	0.3752	&	3	&	X	&		\\
5	&	GS-2010B-Q-046-03	&	11$^\mathrm{h}$24$^\mathrm{m}$00.16$^\mathrm{s}$	&	05$^\circ$27$^\prime$43.0$^{\prime \prime}$	&	0.1554	&	3	&	X	&		\\
6	&	GS-2010B-Q-046-03	&	11$^\mathrm{h}$24$^\mathrm{m}$02.58$^\mathrm{s}$	&	05$^\circ$28$^\prime$43.8$^{\prime \prime}$	&	0.3942	&	3	&	X	&		\\
11	&	GS-2010B-Q-046-03	&	11$^\mathrm{h}$23$^\mathrm{m}$58.50$^\mathrm{s}$	&	05$^\circ$28$^\prime$10.8$^{\prime \prime}$	&	0.6544	&	3	&	X	&		\\
12	&	GS-2010B-Q-046-03	&	11$^\mathrm{h}$23$^\mathrm{m}$59.15$^\mathrm{s}$	&	05$^\circ$28$^\prime$07.0$^{\prime \prime}$	&	0.5945	&	3	&	X	&		\\
13	&	GS-2010B-Q-046-03	&	11$^\mathrm{h}$23$^\mathrm{m}$57.11$^\mathrm{s}$	&	05$^\circ$28$^\prime$34.9$^{\prime \prime}$	&	0.6514	&	3	&	X	&	$\surd$	\\
\end{tabular*}
\caption{XMMXCSJ112349.4+052955.1: All columns are as explained in Table \ref{geminiTable}.}

\end{table*}

\begin{table*}
\centering

 \begin{tabular*}{\textwidth}{l @{\extracolsep{\fill}} llllccc}
  \hline
   ID & Mask     & RA(J2000)         & Dec(J2000)  & z & Quality & Identification & Member\\
 \hline
3	&	GS-2012A-Q-046-07	&	11$^\mathrm{h}$35$^\mathrm{m}$58.38$^\mathrm{s}$	&	-03$^\circ$31$^\prime$15.8$^{\prime \prime}$	&	1.0306	&	3	&	X	&		\\
5	&	GS-2012A-Q-046-07	&	11$^\mathrm{h}$35$^\mathrm{m}$59.26$^\mathrm{s}$	&	-03$^\circ$31$^\prime$33.9$^{\prime \prime}$	&	0.8323	&	3	&	X	&	$\surd$	\\
6	&	GS-2012A-Q-046-07	&	11$^\mathrm{h}$35$^\mathrm{m}$59.51$^\mathrm{s}$	&	-03$^\circ$32$^\prime$04.2$^{\prime \prime}$	&	0.8353	&	3	&	X	&		\\
8	&	GS-2012A-Q-046-07	&	11$^\mathrm{h}$36$^\mathrm{m}$01.95$^\mathrm{s}$	&	-03$^\circ$31$^\prime$10.1$^{\prime \prime}$	&	0.8255	&	3	&	X	&	$\surd$	\\
9	&	GS-2012A-Q-046-07	&	11$^\mathrm{h}$36$^\mathrm{m}$02.76$^\mathrm{s}$	&	-03$^\circ$31$^\prime$04.6$^{\prime \prime}$	&	0.8841	&	3	&	X	&		\\
11	&	GS-2012A-Q-046-07	&	11$^\mathrm{h}$36$^\mathrm{m}$01.15$^\mathrm{s}$	&	-03$^\circ$29$^\prime$33.8$^{\prime \prime}$	&	0.8215	&	3	&	X	&	$\surd$	\\
12	&	GS-2012A-Q-046-07	&	11$^\mathrm{h}$36$^\mathrm{m}$04.68$^\mathrm{s}$	&	-03$^\circ$30$^\prime$39.0$^{\prime \prime}$	&	0.8266	&	3	&	X	&	$\surd$	\\
\hline

\end{tabular*}
\caption{XMMXCSJ113602.9–032943.2: All columns are as explained in Table \ref{geminiTable}.}

\end{table*}

\begin{table*}
\centering

 \begin{tabular*}{\textwidth}{l @{\extracolsep{\fill}} llllccc}
  \hline
   ID & Mask     & RA(J2000)         & Dec(J2000)  & z & Quality &Identification & Member\\
 \hline
4	&	GS-2012A-Q-046-10	&	13$^\mathrm{h}$42$^\mathrm{m}$58.07$^\mathrm{s}$	&	-00$^\circ$02$^\prime$43.9$^{\prime \prime}$	&	0.8343	&	3	&	X	&		\\
5	&	GS-2012A-Q-046-10	&	13$^\mathrm{h}$42$^\mathrm{m}$59.51$^\mathrm{s}$	&	00$^\circ$00$^\prime$08.0$^{\prime \prime}$	&	0.6925	&	3	&	X	&	$\surd$	\\
6	&	GS-2012A-Q-046-10	&	13$^\mathrm{h}$42$^\mathrm{m}$56.97$^\mathrm{s}$	&	-00$^\circ$00$^\prime$47.4$^{\prime \prime}$	&	0.4383	&	3	&	X	&		\\
7	&	GS-2012A-Q-046-10	&	13$^\mathrm{h}$42$^\mathrm{m}$58.82$^\mathrm{s}$	&	-00$^\circ$00$^\prime$53.6$^{\prime \prime}$	&	0.8342	&	3	&	X	&		\\
8	&	GS-2012A-Q-046-10	&	13$^\mathrm{h}$43$^\mathrm{m}$02.03$^\mathrm{s}$	&	-00$^\circ$01$^\prime$17.4$^{\prime \prime}$	&	0.6829	&	3	&	X	&	$\surd$	\\
9	&	GS-2012A-Q-046-10	&	13$^\mathrm{h}$42$^\mathrm{m}$59.93$^\mathrm{s}$	&	-00$^\circ$01$^\prime$20.0$^{\prime \prime}$	&	0.5510	&	3	&	X	&		\\
10	&	GS-2012A-Q-046-10	&	13$^\mathrm{h}$43$^\mathrm{m}$01.63$^\mathrm{s}$	&	-00$^\circ$01$^\prime$41.7$^{\prime \prime}$	&	0.6833	&	3	&	X	&	$\surd$	\\
\hline

\end{tabular*}
\caption{XMMXCSJ134305.1-000056.8: All columns are as explained in Table \ref{geminiTable}.}

\end{table*}

\begin{table*}
\centering

 \begin{tabular*}{\textwidth}{l @{\extracolsep{\fill}} llllccc}
  \hline
   ID & Mask     & RA(J2000)         & Dec(J2000)  & z & Quality & Identification & Member\\
 \hline

1	&	\citet{Nastasi2014}	&	14$^\mathrm{h}$50$^\mathrm{m}$09.30$^\mathrm{s}$	&	09$^\circ$04$^\prime$39.3$^{\prime \prime}$	&	0.6419	&	3	&		&		\\
2	&	\citet{Nastasi2014}	&	14$^\mathrm{h}$50$^\mathrm{m}$09.20$^\mathrm{s}$	&	09$^\circ$04$^\prime$45.1$^{\prime \prime}$	&	0.6425	&	3	&		&		\\
3	&	\citet{Nastasi2014}	&	14$^\mathrm{h}$50$^\mathrm{m}$10.40$^\mathrm{s}$	&	09$^\circ$04$^\prime$23.5$^{\prime \prime}$	&	0.6418	&	2	&		&	$\surd$	\\
4	&	\citet{Nastasi2014}	&	14$^\mathrm{h}$50$^\mathrm{m}$09.50$^\mathrm{s}$	&	09$^\circ$04$^\prime$15.9$^{\prime \prime}$	&	0.6430	&	3	&		&		\\
5	&	\citet{Nastasi2014}	&	14$^\mathrm{h}$50$^\mathrm{m}$07.40$^\mathrm{s}$	&	09$^\circ$04$^\prime$28.8$^{\prime \prime}$	&	0.6429	&	3	&		&	$\surd$	\\
6	&	\citet{Nastasi2014}	&	14$^\mathrm{h}$50$^\mathrm{m}$09.30$^\mathrm{s}$	&	09$^\circ$04$^\prime$06.5$^{\prime \prime}$	&	0.6377	&	3	&		&		\\
7	&	\citet{Nastasi2014}	&	14$^\mathrm{h}$50$^\mathrm{m}$10.40$^\mathrm{s}$	&	09$^\circ$03$^\prime$53.4$^{\prime \prime}$	&	0.6462	&	2	&		&	$\surd$	\\
\hline

\end{tabular*}
\caption{XMMXCSJ145009.3+090428.8: All columns are as explained in Table \ref{geminiTable}.}

\end{table*}

\begin{table*}
\centering

 \begin{tabular*}{\textwidth}{l @{\extracolsep{\fill}} llllccc}
  \hline
   ID & Mask     & RA(J2000)         & Dec(J2000)  & z & Quality&Identification & Member\\
 \hline
5	&	GS-2010B-Q-046-04	&	21$^\mathrm{h}$52$^\mathrm{m}$20.26$^\mathrm{s}$	&	-27$^\circ$28$^\prime$46.6$^{\prime \prime}$	&	0.4662	&	2	&	X	&		\\
8	&	GS-2010B-Q-046-04	&	21$^\mathrm{h}$52$^\mathrm{m}$16.97$^\mathrm{s}$	&	-27$^\circ$28$^\prime$39.8$^{\prime \prime}$	&	0.6330	&	3	&	X	&		\\
12	&	GS-2010B-Q-046-04	&	21$^\mathrm{h}$52$^\mathrm{m}$23.98$^\mathrm{s}$	&	-27$^\circ$30$^\prime$29.9$^{\prime \prime}$	&	0.8217	&	3	&	X	&	$\surd$	\\
13	&	GS-2010B-Q-046-04	&	21$^\mathrm{h}$52$^\mathrm{m}$19.74$^\mathrm{s}$	&	-27$^\circ$30$^\prime$32.1$^{\prime \prime}$	&	0.8262	&	2	&	X	&	$\surd$	\\
14	&	GS-2010B-Q-046-04	&	21$^\mathrm{h}$52$^\mathrm{m}$14.96$^\mathrm{s}$	&	-27$^\circ$30$^\prime$00.0$^{\prime \prime}$	&	0.8293	&	3	&	X	&	$\surd$	\\
17	&	GS-2010B-Q-046-04	&	21$^\mathrm{h}$52$^\mathrm{m}$17.29$^\mathrm{s}$	&	-27$^\circ$29$^\prime$22.2$^{\prime \prime}$	&	0.8297	&	2	&	X	&	$\surd$	\\
19	&	GS-2010B-Q-046-04	&	21$^\mathrm{h}$52$^\mathrm{m}$20.19$^\mathrm{s}$	&	-27$^\circ$29$^\prime$52.9$^{\prime \prime}$	&	0.8281	&	2	&	X	&	$\surd$	\\
\hline

\end{tabular*}
\caption{XMMXCSJ215221.0–273022.6: All columns are as explained in Table \ref{geminiTable}.}

\end{table*}
\clearpage

\begin{table*}
\centering

 \begin{tabular*}{\textwidth}{l @{\extracolsep{\fill}} llllccc}
  \hline
   ID & Mask     & RA(J2000)         & Dec(J2000)  & z & Quality & Identification & Member\\
 \hline
1	&	GN-2012A-Q-070-10	&	23$^\mathrm{h}$02$^\mathrm{m}$39.79$^\mathrm{s}$	&	08$^\circ$42$^\prime$48.1$^{\prime \prime}$	&	0.7148	&	3	&	X	&	$\surd$	\\
2	&	GN-2012A-Q-070-10	&	23$^\mathrm{h}$02$^\mathrm{m}$38.96$^\mathrm{s}$	&	08$^\circ$42$^\prime$33.7$^{\prime \prime}$	&	0.7141	&	3	&	X	&	$\surd$	\\
4	&	GN-2012A-Q-070-10	&	23$^\mathrm{h}$02$^\mathrm{m}$39.70$^\mathrm{s}$	&	08$^\circ$42$^\prime$25.3$^{\prime \prime}$	&	0.7169	&	3	&	X	&	$\surd$	\\
9	&	GN-2012A-Q-070-10	&	23$^\mathrm{h}$02$^\mathrm{m}$42.62$^\mathrm{s}$	&	08$^\circ$43$^\prime$57.4$^{\prime \prime}$	&	0.7235	&	2	&	X	&	$\surd$	\\
11	&	GN-2012A-Q-070-10	&	23$^\mathrm{h}$02$^\mathrm{m}$43.85$^\mathrm{s}$	&	08$^\circ$43$^\prime$55.2$^{\prime \prime}$	&	0.7291	&	3	&	X	&	$\surd$	\\
12	&	GN-2012A-Q-070-10	&	23$^\mathrm{h}$02$^\mathrm{m}$42.24$^\mathrm{s}$	&	08$^\circ$42$^\prime$57.6$^{\prime \prime}$	&	0.7231	&	3	&	X	&	$\surd$	\\
14	&	GN-2012A-Q-070-10	&	23$^\mathrm{h}$02$^\mathrm{m}$43.02$^\mathrm{s}$	&	08$^\circ$42$^\prime$48.4$^{\prime \prime}$	&	0.6239	&	3	&	X	&		\\
\hline

\end{tabular*}
\caption{XMMXCSJ230247.7+084355.9: All columns are as explained in Table \ref{geminiTable}.}

\end{table*}
%

\clearpage

\section{Velocity Histograms}
\label{VelocityHistograms}
Velocity histograms of all the clusters in the sample to appear in the online version of the article.
\begin{figure*}

\includegraphics*[width=1\textwidth]{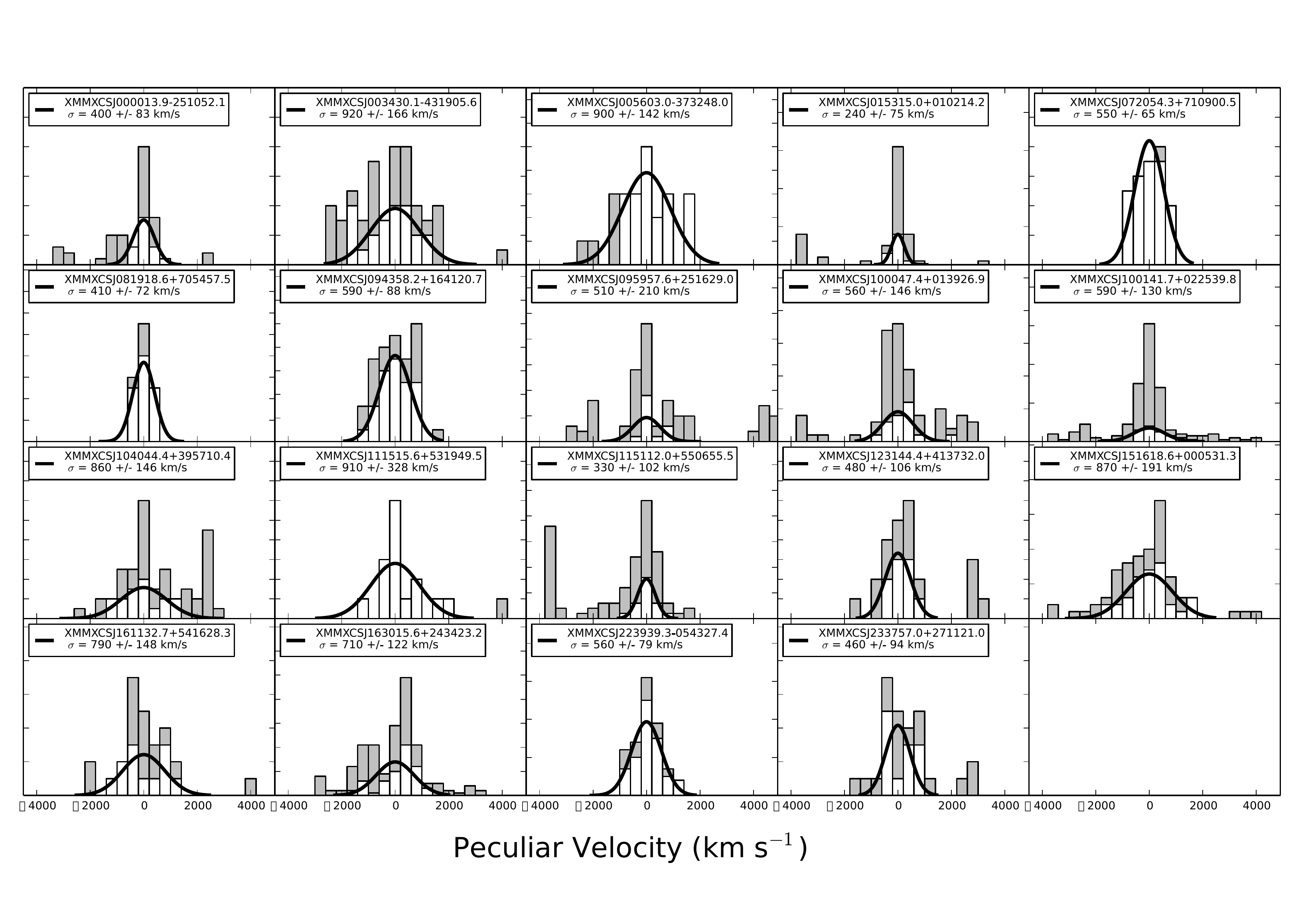}

\caption{We depict a synthetic view of the tables found in Appendix B in the form of histograms for the low redshift sample. The solid grey blocks depict all the galaxies considered as possible members while the white blocks are the final chosen members. The solid black line shows the velocity dispersion calculated using the method described in the paper. }
\label{plot1Appendix}
\end{figure*}
\clearpage
\begin{figure*}

\includegraphics*[width=\textwidth]{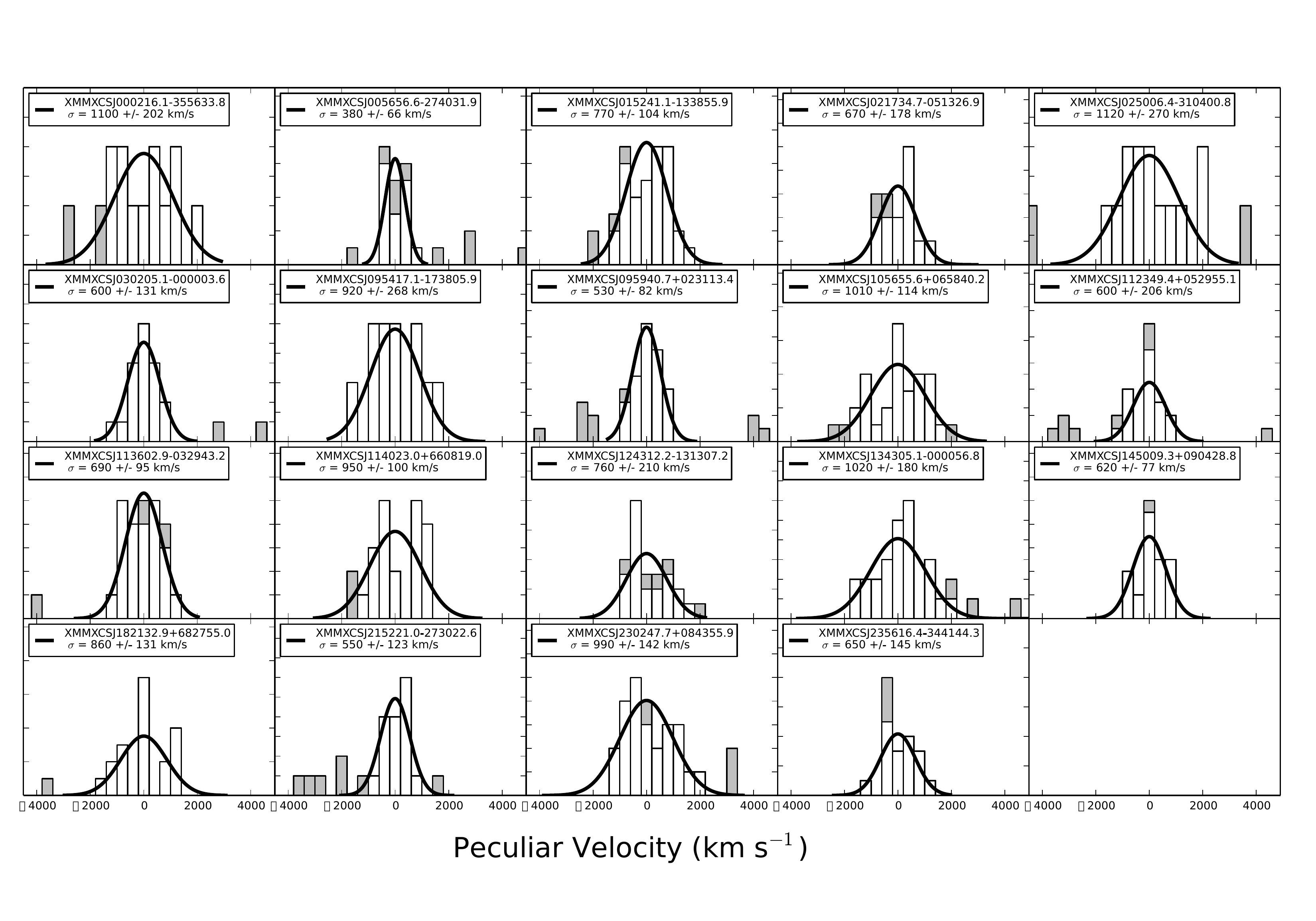}

\caption{We depict a synthetic view of the tables found in Appendix B in the form of histograms for the high redshift sample. All blocks and lines are as in Figure \ref{plot1Appendix}.}
\label{plot2Appendix}
\end{figure*}


\end{document}